\documentclass[runningheads,a4paper]{llncs}

\usepackage[]{algorithm2e}
\usepackage{tabularx}
\usepackage[noend]{algpseudocode}
\usepackage{colortbl}
\usepackage{booktabs}
\usepackage[hyphens]{url}
\usepackage{tabu}
\usepackage{todonotes}
\usepackage{multirow}
\usepackage{xspace}
\usepackage{xcolor}
\usepackage{adjustbox}
\usepackage[normalem]{ulem}
\usepackage{url}
\usepackage{amssymb}
\usepackage{amsmath}
\usepackage{enumitem}

\newcommand{\realspz}{\ensuremath{\mathbb{R}_{\ge 0}}\xspace}

\newcommand{\reals}{\ensuremath{\mathbb{R}}\xspace}
\newcommand{\graph}{\ensuremath{G}\xspace}
\newcommand{\nodes}{\ensuremath{V}\xspace}

\newcommand{\edges}{\ensuremath{E}\xspace}

\newcommand{\eset}{\ensuremath{X}\xspace}

\newcommand{\esetopt}{\ensuremath{X^*}\xspace}

\newcommand{\esim}{\ensuremath{s}\xspace}
\newcommand{\totsim}{\ensuremath{s_{total}}\xspace}
\newcommand{\degree}{\ensuremath{deg}\xspace}
\newcommand{\dns}{\ensuremath{D}\xspace}
\newcommand{\sml}{\ensuremath{S}\xspace}
\newcommand{\neginvd}{\ensuremath{\bar{D}}\xspace}

\newcommand{\uset}{\ensuremath{U}\xspace}
\newcommand{\usetopt}{\ensuremath{U^*}\xspace}
\newcommand{\usetoptc}{\ensuremath{\widebar{U}^*}\xspace}
\newcommand{\flowgr}{\ensuremath{\graph^{\prime}}\xspace}
\newcommand{\flnodes}{\ensuremath{\uset^{\prime}}\xspace}
\newcommand{\fledeges}{\ensuremath{\edges^{\prime}}\xspace}
\newcommand{\flweights}{\ensuremath{w^{\prime}}\xspace}
\newcommand{\optmccost}{\ensuremath{\text{\sf C}^*\!}\xspace}
\newcommand{\copt}{\ensuremath{c^*\!}\xspace}
\newcommand{\Qopt}{\ensuremath{Q^*\!}\xspace}
\newcommand{\snk}{\ensuremath{t}\xspace}
\newcommand{\src}{\ensuremath{s}\xspace}

\newcommand{\pareto}{\ensuremath{\mathcal{P}}\xspace}

\newcommand{\mincut}{\textsc{\large min-cut}\xspace}
\newcommand{\LRES}{\textsc{\large dss}\xspace}
\newcommand{\LRID}{{\textsc{\large dss-inv}}\xspace}

\newcommand{\OLR}{\ensuremath{O_{\mu}\xspace}}
\newcommand{\OID}{\ensuremath{O_{\lambda}\xspace}}
\newcommand{\Qproblem}{\ensuremath{\text{\sc\large q}}\xspace}
\newcommand{\FPproblem}{\ensuremath{\text{\sc\large fp}}\xspace}
\newcommand{\FPalgo}{\ensuremath{\text{\sf\small FP-algo}}\xspace}
\newcommand{\ourmethod}{\ensuremath{\text{\sf\small DenSim}}\xspace}
\newcommand{\bld}{\ensuremath{\text{\sf\small BLDen}}\xspace}
\newcommand{\bls}{\ensuremath{\text{\sf\small BLSim}}\xspace}

\newcommand{\dataset}[1]{\textsl{#1}}
\newcommand{\cs}{\dataset{CS-Aarhus}\xspace}
\newcommand{\air}{\dataset{EU-Air}\xspace}
\newcommand{\concel}{\dataset{Neuronal~C.e.}\xspace}
\newcommand{\gencel}{\dataset{Genetic~C.e.}\xspace}
\newcommand{\genara}{\dataset{Genetic~A.th.}\xspace}

\renewcommand{\algorithmicrequire}{\textbf{Input:}}
\renewcommand{\algorithmicensure}{\textbf{Output:}}
\renewcommand{\algorithmicforall}{\textbf{for each}}

\newcommand{\bigO}{\ensuremath{\mathcal{O}}\xspace}

\newcommand{\spara}[1]{\smallskip\noindent{\bf #1}}

\makeatletter
\newcommand\rel@kern[1]{\kern#1\dimexpr\macc@kerna}
\newcommand\widebar[1]{%
  \begingroup
  \def\mathaccent##1##2{%
    \rel@kern{0.8}%
    \overline{\rel@kern{-0.8}\macc@nucleus\rel@kern{0.2}}%
    \rel@kern{-0.2}%
  }%
  \macc@depth\@ne
  \let\math@bgroup\@empty \let\math@egroup\macc@set@skewchar
  \mathsurround\z@ \frozen@everymath{\mathgroup\macc@group\relax}%
  \macc@set@skewchar\relax
  \let\mathaccentV\macc@nested@a
  \macc@nested@a\relax111{#1}%
  \endgroup
}
\makeatother

\makeatletter
 \def\@textbottom{\vskip \z@ \@plus 1pt}
 \let\@texttop\relax
\makeatother

\newcommand{\alabel}{\ensuremath{\color{yafcolor1!100}a}\xspace}
\newcommand{\blabel}{\ensuremath{\color{yafcolor2!100}b}\xspace}
\newcommand{\clabel}{\ensuremath{\color{yafcolor3!100}c}\xspace}
\newcommand{\dlabel}{\ensuremath{\color{yafcolor4!100}d}\xspace}
\newcommand{\elabel}{\ensuremath{\color{yafcolor5!100}e}\xspace}
\newcommand{\flabel}{\ensuremath{\color{yafcolor7!100}f}\xspace}
\newcommand{\glabel}{\ensuremath{\color{yafcolor8!100}g}\xspace}
\newcommand{\hlabel}{\ensuremath{\color{yafcolor4!100}h}\xspace}
\newcommand{\ilabel}{\ensuremath{\color{yafcolor5!100}i}\xspace}
\newcommand{\jlabel}{\ensuremath{\color{yafcolor7!100}j}\xspace}
\newcommand{\klabel}{\ensuremath{\color{yafcolor8!100}k}\xspace}

\newcommand{\alabelt}{\ensuremath{\color{yafcolor1!50}a}\xspace}
\newcommand{\blabelt}{\ensuremath{\color{yafcolor2!50}b}\xspace}
\newcommand{\clabelt}{\ensuremath{\color{yafcolor3!50}c}\xspace}
\newcommand{\dlabelt}{\ensuremath{\color{yafcolor4!50}d}\xspace}
\newcommand{\elabelt}{\ensuremath{\color{yafcolor5!50}e}\xspace}
\newcommand{\flabelt}{\ensuremath{\color{yafcolor7!50}f}\xspace}
\newcommand{\glabelt}{\ensuremath{\color{yafcolor8!50}g}\xspace}
\newcommand{\hlabelt}{\ensuremath{\color{yafcolor4!50}h}\xspace}
\newcommand{\ilabelt}{\ensuremath{\color{yafcolor5!50}i}\xspace}
\newcommand{\jlabelt}{\ensuremath{\color{yafcolor7!50}j}\xspace}

\newcommand{\tikzscale}{{0.7}}

\usepackage{tikz,pgfplots,pgfplotstable}
\usetikzlibrary{positioning,fit,shapes}
\usetikzlibrary{decorations.pathreplacing}
\usetikzlibrary{arrows}

\pgfdeclarelayer{background}
\pgfdeclarelayer{foreground}
\pgfsetlayers{background,main,foreground}

\makeatletter
\tikzset{multicircle/.style  args={#1, #2}{%
 alias=tmp@name, %
  postaction={%
    insert path={
     \pgfextra{%
     \pgfpointdiff{\pgfpointanchor{\pgf@node@name}{center}}%
                  {\pgfpointanchor{\pgf@node@name}{east}}%
     \pgfmathsetmacro\insiderad{\pgf@x}%
        \fill[white] (\pgf@node@name.center)  circle (\insiderad-\pgflinewidth);%
        \draw[#2] (\pgf@node@name.center)  circle (\insiderad-\pgflinewidth);%
        \fill[#2] (\pgf@node@name.center)  -- ++(0:\insiderad-\pgflinewidth) arc (0:#1:\insiderad-\pgflinewidth)--cycle;%
        }}}}}
\makeatother

\definecolor{yafaxiscolor}{rgb}{0.3, 0.3, 0.3}
\definecolor{yafcolor1}{rgb}{0.4, 0.165, 0.553}
\definecolor{yafcolor2}{rgb}{0.949, 0.482, 0.216}
\definecolor{yafcolor3}{rgb}{0.47, 0.549, 0.306}
\definecolor{yafcolor4}{rgb}{0.925, 0.165, 0.224}
\definecolor{yafcolor5}{rgb}{0.141, 0.345, 0.643}
\definecolor{yafcolor6}{rgb}{0.965, 0.933, 0.267}
\definecolor{yafcolor7}{rgb}{0.627, 0.118, 0.165}
\definecolor{yafcolor8}{rgb}{0.878, 0.475, 0.686}
\definecolor{yafcolor9}{rgb}{0.965, 0.733, 0.767}

\newlength{\yafaxispad}
\newlength{\yaftlpad}
\newlength{\yaflabelpad}
\newlength{\yafaxiswidth}
\newlength{\yafticklen}
\makeatletter
\def\pgfplots@drawtickgridlines@INSTALLCLIP@onorientedsurf#1{}
\makeatother

\newcommand{\yafdrawxaxis}[2]{
  \pgfplotstransformcoordinatex{#1}\let\xmincoord=\pgfmathresult 
  \pgfplotstransformcoordinatex{#2}\let\xmaxcoord=\pgfmathresult 
  \pgfsetlinewidth{\yafaxiswidth} 
  \pgfsetcolor{yafaxiscolor}
  \pgfpathmoveto{\pgfpointadd{\pgfpointadd{\pgfplotspointrelaxisxy{0}{0}}{\pgfqpointxy{\xmincoord}{0}}}{\pgfqpoint{-0.5\yafaxiswidth}{\yafaxispad}}}
  \pgfpathlineto{\pgfpointadd{\pgfpointadd{\pgfplotspointrelaxisxy{0}{0}}{\pgfqpointxy{\xmaxcoord}{0}}}{\pgfqpoint{0.5\yafaxiswidth}{\yafaxispad}}}
  \pgfusepath{stroke}

}
\newcommand{\yafdrawyaxis}[2]{
  \pgfplotstransformcoordinatey{#1}\let\ymincoord=\pgfmathresult 
  \pgfplotstransformcoordinatey{#2}\let\ymaxcoord=\pgfmathresult 
  \pgfsetlinewidth{\yafaxiswidth} 
  \pgfsetcolor{yafaxiscolor}
  \pgfpathmoveto{\pgfpointadd{\pgfpointadd{\pgfplotspointrelaxisxy{0}{0}}{\pgfqpointxy{0}{\ymincoord}}}{\pgfqpoint{\yafaxispad}{-0.5\yafaxiswidth}}}
  \pgfpathlineto{\pgfpointadd{\pgfpointadd{\pgfplotspointrelaxisxy{0}{0}}{\pgfqpointxy{0}{\ymaxcoord}}}{\pgfqpoint{\yafaxispad}{0.5\yafaxiswidth}}}
  \pgfusepath{stroke}
}

\pgfplotscreateplotcyclelist{yaf}{%
{yafcolor1,mark options={scale=0.75},mark=o}, 
{yafcolor2,mark options={scale=0.75},mark=square},
{yafcolor3,mark options={scale=0.75},mark=triangle},
{yafcolor4,mark options={scale=0.75},mark=o},
{yafcolor5,mark options={scale=0.75},mark=o},
{yafcolor6,mark options={scale=0.75},mark=o},
{yafcolor7,mark options={scale=0.75},mark=o},
{yafcolor8,mark options={scale=0.75},mark=o}} 

\pgfplotsset{axis y line=left, axis x line=bottom,
  tick align=outside,
  compat = 1.3,
  tickwidth=\yafticklen,
  clip = false,
  every axis title shift = 0pt,
    x axis line style= {-, line width = 0pt, opacity = 0},
    y axis line style= {-, line width = 0pt, opacity = 0},
    x tick style= {line width = \yafaxiswidth, color=yafaxiscolor, yshift = \yafaxispad},
    y tick style= {line width = \yafaxiswidth, color=yafaxiscolor, xshift = \yafaxispad},
    x tick label style = {font=\scriptsize, yshift = \yaftlpad},
    y tick label style = {font=\scriptsize, xshift = \yaftlpad},
    every axis y label/.style = {at = {(ticklabel cs:0.5)}, rotate=90, anchor=center, font=\scriptsize, yshift = -\yaflabelpad},
    every axis x label/.style = {at = {(ticklabel cs:0.5)}, anchor=center, font=\scriptsize, yshift = \yaflabelpad},
    x tick label style = {font=\scriptsize, yshift = 1pt},
    grid = major,
    major grid style  = {dash pattern = on 1pt off 3 pt},
  every axis plot post/.append style= {line width=\yafaxiswidth} ,
  legend cell align = left,
  legend style = {inner sep = 1pt, cells = {font=\scriptsize}},
  legend image code/.code={%
    \draw[mark repeat=2,mark phase=2,#1] 
    plot coordinates { (0cm,0cm) (0.15cm,0cm) (0.3cm,0cm) };%
  } 
}

\usepackage{cite}
\usepackage{amsmath, amssymb, amsfonts}
\usepackage{graphicx}
\usepackage{xcolor}
\usepackage[misc]{ifsym}

\begin{document}

\title{Mining Dense Subgraphs with Similar Edges}

\author{Polina Rozenshtein\inst{1}$^{\textrm{\Letter}}$ \and
	Giulia Preti\inst{2} \and
	Aristides Gionis\inst{3} \and
	Yannis Velegrakis\inst{4}
}

\authorrunning{P. Rozenshtein et al.}

\institute{Institute of Data Science, National University of Singapore, Singapore \\\email{idspoli@nus.edu.sg} \and
	ISI Foundation, Italy
	\and
	KTH Royal Institute of Technology, Sweden
	\and
	Utrecht University, The Netherlands
	}
%

%
\maketitle   

\begin{abstract}	
	
When searching for interesting structures in graphs,
it is often important to take into account not only the graph connectivity, 
but also the metadata available, such as node and edge labels, or temporal information. 
In this paper we are interested in settings where such metadata is used to define a similarity between edges.
We consider the problem of finding subgraphs 
that are {\em dense} and whose edges are {\em similar} to each other with respect to a given similarity function. 
Depending on the application, this function can be, for example, 
the Jaccard similarity between the edge label sets, or 
the temporal correlation of the edge occurrences in a temporal graph.

We formulate a Lagrangian relaxation-based optimization problem to search
for dense subgraphs with high pairwise edge similarity. 
We design a novel algorithm to solve the problem 
through parametric \mincut~\cite{gallo1989fast, hochbaum2008pseudoflow},
and provide an efficient search scheme to iterate through the values of the Lagrangian multipliers.
Our study is complemented by an evaluation on real-world datasets, 
which demonstrates the usefulness and efficiency of the proposed approach.

\end{abstract}

\section{Introduction}\label{intro}

Searching for densely-connected structures in graphs is a task with numerous applications \cite{angel2012dense,chen2012dense,tsourakakis2013denser, chan2015graph}
and extensive theoretical work \cite{bhaskara2010detecting,goldberg1984finding,khuller2009finding}. 
A densely-connected subset of nodes may represent 
a community in a social network, 
a set of interacting proteins, or
a group of related entities in a knowledge base.
Given the relevance of the problem in different applications, 
a number of measures have been used to capture graph density, 
including
average degree~\cite{khuller2009finding}, 
quasi-cliques~\cite{tsourakakis2013denser}, and
$k$-clique subgraphs~\cite{tsourakakis2015novel}.

Often, however, real-world graphs have attributes associated with their edges, which describe how nodes are related with each other.
This is common in social networks, where we can distinguish multiple types of relationships between individuals
(friends, family, class-mates, work, etc.),
as well as several types of interactions 
(likes, messages, and comments). 
Similarly, a communication network records information that describes the 
communication patterns between its nodes, 
the volume of data exchanged between two nodes, 
or the level of congestion at a given link, as a function of time.

Incorporating this rich information into standard graph-mining tasks, 
such as dense-subgraph mining, 
can provide a better understanding of the graph, and enable the discovery of clear, 
cohesive, and homogeneous groups and patterns~\cite{fang2016effective}.
For instance, a group of hashtags that form a dense subgraph in the Twitter's hashtag co-occurence network becomes more meaningful for a social scientist if those hashtags are also correlated in time, as they likely indicate a recurrent topic of discussion, or an emerging trend.

In this paper, we study a general graph-mining problem 
where the input is a graph $\graph=(\nodes,\edges)$
and a function $\esim: \edges\times\edges\to\realspz$
that measures similarity between pairs of edges.
We do not restrict the choice of edge similarity $\esim$, meaning that
it can be defined using any type of information that is available about the edges.
For example, for a graph with edge labels, the similarity of two edges can be defined as the Jaccard similarity between their label sets, while for a temporal graph, where edges are active in some timestamps and inactive in others, the similarity can be defined as the temporal correlation between the~edge time series.
Given a similarity function, we are interested in finding {\em dense} subgraphs
whose edges are {\em similar} to each other.
Consider the following example.

\begin{figure}[t]
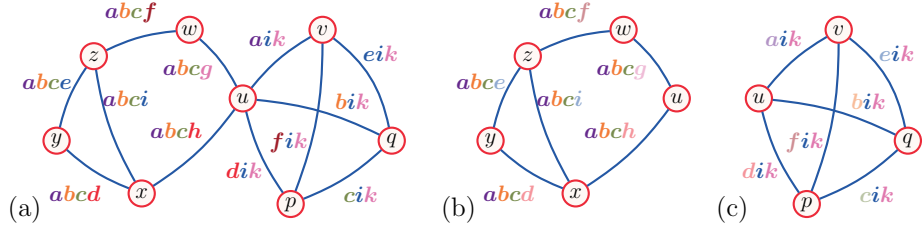

	\centering
	\begin{tabular}{ccc}
		\begin{tikzpicture}[scale=\tikzscale,every node/.style={scale=\tikzscale}]]

\input{tikz-intro-defs}


\node[exnode] (x) at ( 1.1, 0  ) {$x$};
\node[exnode] (y) at (-0.5, 1  ) {$y$};
\node[exnode] (z) at ( 0.2, 2.6) {$z$};
\node[exnode] (w) at ( 2,   3.1) {$w$};
\node[exnode] (u) at ( 3,   1.8) {$u$};
\node[exnode] (v) at ( 4.5, 3.1) {$v$};
\node[exnode] (p) at ( 3.9,-0.2) {$p$};
\node[exnode] (q) at ( 5.8, 1  ) {$q$};

\draw (x) edge[-, exedge, bend left  = 10] node[exlabel, pos = 0.5] {$\boldsymbol{\alabel\blabel\clabel\dlabel}$} (y);
\draw (y) edge[-, exedge, bend left  = 10] node[exlabel, pos = 0.4] {$\boldsymbol{\alabel \blabel \clabel \elabel}$} (z);
\draw (z) edge[-, exedge, bend left  = 10] node[exlabel, pos = 0.7] {$\boldsymbol{\alabel \blabel \clabel \flabel}$} (w);
\draw (u) edge[-, exedge, bend right = 10] node[exlabel, pos = 0.8] {$\boldsymbol{\alabel \blabel \clabel \glabel}$} (w);
\draw (x) edge[-, exedge, bend right = 10] node[exlabel] {$\boldsymbol{\alabel \blabel \clabel \hlabel}$} (u);
\draw (z) edge[-, exedge, bend right = 10] node[exlabel, pos = 0.4] {$\boldsymbol{\alabel \blabel \clabel \ilabel}$} (x);

\draw (u) edge[-, exedge, bend left  = 10] node[exlabel, pos = 0.6] {$\boldsymbol{\alabel \ilabel \klabel}$} (v);
\draw (u) edge[-, exedge, bend left  = 10] node[exlabel, pos = 0.65] {$\boldsymbol{\blabel \ilabel \klabel}$} (q);
\draw (q) edge[-, exedge, bend left  = 10] node[exlabel] {$\boldsymbol{\clabel \ilabel \klabel}$} (p);
\draw (p) edge[-, exedge, bend left  = 10] node[exlabel, pos = 0.5] {$\boldsymbol{\dlabel \ilabel \klabel}$} (u);
\draw (v) edge[-, exedge, bend left  = 20] node[exlabel, pos = 0.4] {$\boldsymbol{\elabel \ilabel \klabel}$} (q);
\draw (p) edge[-, exedge, bend right = 10] node[exlabel, pos = 0.2] {$\boldsymbol{\flabel \ilabel \klabel}$} (v);

\node[fill=white] at (-1.1,-0.3) {\Large (a)};

\end{tikzpicture}%
 \hspace{0.18cm} &
		\begin{tikzpicture}[scale=\tikzscale,every node/.style={scale=\tikzscale}]]

\input{tikz-intro-defs}


\node[exnode] (x) at ( 1.1, 0  ) {$x$};
\node[exnode] (y) at (-0.5, 1  ) {$y$};
\node[exnode] (z) at ( 0.2, 2.6) {$z$};
\node[exnode] (w) at ( 2,   3.1) {$w$};
\node[exnode] (u) at ( 3,   1.8) {$u$};


\draw (x) edge[-, exedge, bend left  = 10] node[exlabel] {$\boldsymbol{\alabel\blabel\clabel\dlabelt}$} (y);
\draw (y) edge[-, exedge, bend left  = 10] node[exlabel, pos = 0.4] {$\boldsymbol{\alabel\blabel\clabel\elabelt}$} (z);
\draw (z) edge[-, exedge, bend left  = 10] node[exlabel, pos = 0.7] {$\boldsymbol{\alabel\blabel\clabel\flabelt}$} (w);
\draw (u) edge[-, exedge, bend right = 10] node[exlabel, pos = 0.8] {$\boldsymbol{\alabel\blabel\clabel\glabelt}$} (w);
\draw (x) edge[-, exedge, bend right = 10] node[exlabel] {$\boldsymbol{\alabel\blabel\clabel\hlabelt}$} (u);
\draw (z) edge[-, exedge, bend right = 10] node[exlabel, pos = 0.4] {$\boldsymbol{\alabel\blabel\clabel\ilabelt}$} (x);


\node[fill=white] at (-1.1,-0.3) {\Large (b)};

\end{tikzpicture}%
 \hspace{0.05cm} &
		\begin{tikzpicture}[scale=\tikzscale,every node/.style={scale=\tikzscale}]]

\input{tikz-intro-defs}



\node[exnode] (u) at ( 3,   1.8) {$u$};
\node[exnode] (v) at ( 4.5, 3.1) {$v$};
\node[exnode] (p) at ( 3.9,-0.2) {$p$};
\node[exnode] (q) at ( 5.8, 1  ) {$q$};


\draw (u) edge[-, exedge, bend left  = 10] node[exlabel, pos = 0.6] {$\boldsymbol{\alabelt\ilabel\klabel}$} (v);
\draw (u) edge[-, exedge, bend left  = 10] node[exlabel, pos = 0.65] {$\boldsymbol{\blabelt\ilabel\klabel}$} (q);
\draw (q) edge[-, exedge, bend left  = 10] node[exlabel] {$\boldsymbol{\clabelt\ilabel\klabel}$} (p);
\draw (p) edge[-, exedge, bend left  = 10] node[exlabel, pos = 0.5] {$\boldsymbol{\dlabelt\ilabel\klabel}$} (u);
\draw (v) edge[-, exedge, bend left  = 20] node[exlabel, pos = 0.4] {$\boldsymbol{\elabelt\ilabel\klabel}$} (q);
\draw (p) edge[-, exedge, bend right = 10] node[exlabel, pos = 0.2] {$\boldsymbol{\flabelt\ilabel\klabel}$} (v);

\node[fill=white] at ( 2.5,   -0.3) {\Large (c)};

\end{tikzpicture}%
\\
	\end{tabular}
	\caption{\label{figure:intro}\small
		(a) Input graph $G$ with edge labels; 
		(b) subgraph $G_B$ of users ${B}=\{x,y,z,w,u\}$, where each edge pair shares $3$ out of $4$ labels;
		(c) clique $G_C$ of users ${C}=\{u,v,p,q\}$, where each edge pair shares $2$ out of $3$ labels.
	}
\end{figure}

\spara{Example.}
As a toy example, 
Figure~\ref{figure:intro}(a) illustrates a portion of a social network, 
where a set of labels is available for each connection, 
describing the topics on which the two users have interacted with each other.
Figures~\ref{figure:intro}(b) and \ref{figure:intro}(c)
highlight two dense subgraphs $G_B$ and $G_C$,
represented by the sets of users $B=\{x,y,z,w,u\}$ and $C=\{u,v,p,q\}$, respectively. 
The graph $G_C$ is denser than $G_B$ ($G_C$ is a clique), meaning that the users in $C$ have interacted more. 
However, the edges of $G_B$ have more labels in common than those of $G_C$ ($3$ out of $4$ per edge pair, 
versus $2$ out of $3$), meaning that the users in $B$ share more topics of interest.  
This example shows that when multiple metrics of interest are taken into consideration, some solutions may optimize some of the metrics, while other solutions may optimize the other metrics. For example, an advertiser may be interested in finding both tighter groups of users and highly similar groups of users, because the first ones have more connections and thus they can influence more other users in the group, while the second ones have more interests in common and thus they are more likely to like similar products. 
\hfill$\Box$

The previous example 
brings an interesting trade-off:
some subgraphs have higher density, while other have higher edge similarity.
This is a typical situation in {\em bi-criteria optimization}~\cite{ehrgott2005multicriteria}.
A common approach to study such problems is by using a Lagrangian relaxation, 
i.e., 
combining the two objectives into a weighted sum and solving the resulting optimization problem for different weights. 
We adopt this approach and combine the density and the edge-similarity of the subgraph induced by an edge set.
Then, we reformulate the problem and design a novel efficient algorithm to solve the relaxation based on 
\emph{parametric minimum cut}~\cite{gallo1989fast, hochbaum2008pseudoflow}.
We explore possible density-similarity trade-offs and provide an efficient search procedure through the values of the Lagrangian multipliers.

We demonstrate experimentally that our method finds efficiently 
a set of solutions on real-world datasets. A wide range 
of the weighting parameter effectively controls the trade-off between similarity and density. 
Additionally, we present a case study where we explore the properties of the discovered subgraphs.

All omitted proofs can be found in the Supplementary Material.

\section{Problem Formulation}\label{problem}

We consider an undirected graph $G=(V,E)$ with node set $V$ and edge set $E$. 
All our algorithms extend to weighted graphs, but
for simplicity of presentation we discuss the unweighted case.
To avoid degenerate cases, we assume that $G$ has at least 2 edges.
We consider subsets of edges and edge-induced subgraphs:
\begin{definition}[Edge-induced subgraph]
Let $\graph=(\nodes,\edges)$ be an undirected graph 
and $\eset$ a subset of edges. 
The subgraph $\graph(\eset)=(\nodes(\eset), \eset)$ of \graph is induced by \eset,
where $\nodes(\eset)$ contains all the nodes that are endpoints of edges in $\eset$.
\end{definition}

We define the density of an edge-induced subgraph as the standard half of average degree or the number of edges divided by the number of nodes \cite{charikar2000greedy,khuller2009finding}:
\begin{definition}[Density] 
\label{definition:edge-set-density}
Given an undirected graph $\graph=(\nodes,\edges)$ and a set of edges $\eset\subseteq \edges$, 
the density of the edge-induced graph $\graph(\eset)=(\nodes(\eset), \eset)$ 
is defined as
\[
\dns(\graph(\eset))=\frac{1}{2}\frac{\sum_{u\in\nodes(\eset)}\degree(u)}{|\nodes(\eset)|} = \frac{|\eset|}{|\nodes(\eset)|},
\]
where $\degree(u)$ denotes the degree of a node $u\in\nodes$.
We refer to $\dns(\graph(\eset))$ as the density of the set of edges $\eset\subseteq \edges$,
and denote it by $\dns(\eset)=\dns(\graph(\eset))$.
\end{definition}

We assume that the graph \graph 
is equipped with a non-negative \emph{edge similarity function} 
$\esim: \edges\times\edges\to\realspz$.
We define the \emph{total edge similarity} of an edge $e$ as 
$\totsim(e, \eset)=\sum_{e_i\in \eset \wedge e \neq e_i} \esim(e, e_i)$.
We then define the \emph{subgraph edge similarity} of an edge-induced subgraph 
as half of the average total edge~similarity:
%
%
\begin{definition}[Subgraph edge similarity]
\label{definition:edge-set-similarity}
The similarity of a set of edges~\eset with at least $2$ edges is defined to be
\[
\sml(\eset) 
= \frac{1}{2}\frac{\sum_{e\in\eset}\totsim(e, \eset)}{|\eset|}
= \frac{1}{|\eset|}\sum_{\{e_i,e_j\}\in \eset^2} \esim(e_i,e_j),
\]
where $\eset^2$ is the set of all the unordered pairs of edges in \eset, 
i.e., $\eset^2 = \{\{e_i, e_j\}\mid e_i,e_j\in \eset \text{ with } e_i\neq e_j \}$. 
If $|\eset| \leq 1$, we set $\sml(\eset)=0$.
\end{definition}

In this paper we look for edge-induced subgraphs 
that have \emph{high density} 
and \emph{high subgraph edge similarity}. 
Note that the more common definition of node-induced subgraphs is not suitable for our problem setting, because a solution to our problem is not defined by a node set. 
Indeed, excluding some edges from a node-induced subgraph may lead to a subgraph, which is less dense, but have edges more similar to each other.

As shown in Figure~\ref{figure:intro}, 
there may not exist solutions that optimize the two objectives at the same time. 
One possible approach is to search for subgraphs whose density and subgraph edge similarity
exceed given thresholds. 
However, setting meaningful thresholds requires domain knowledge, 
which may be expensive to acquire. 
%
Here, we rely on a common approach to cope with bi-criteria optimization problems, 
namely to formulate and solve a Lagrangian relaxation:


\begin{problem}[{\LRES}]
\label{LRES}
Given an undirected graph $\graph=(\nodes,\edges)$ with an edge-similarity function 
$\esim: \edges\times\edges\to\realspz$ and a non-negative real number $\mu\geq0$,
find a subset of edges $\eset\subseteq\edges$, 
that maximizes the objective $\OLR(\eset\mid\mu)=\sml(\eset)+\mu\, \dns(\eset)$.
\end{problem}

\section{Proposed Method}
\label{solution}

\label{section:LR}

We start describing our solution by reformulating the \LRES problem. 
The reformulation will allow us to use efficient algorithmic techniques. 
We alter the \LRES objective by substituting the density term $\dns$ 
with the inverse negated term $-1/\dns$. 
Without loss of generality, 
we require that the solution edge set \eset contains at least one edge.
The resulting problem is the following.

\begin{problem}[\LRID]
\label{LRID}
Given an undirected graph $\graph=(\nodes,\edges)$ and a non-negative real number $\lambda\geq0$,
find a subset of edges $\eset\subseteq\edges$, with $|\eset|\geq 1$,
that maximizes the objective $\OID(\eset\mid \lambda) = \sml(\eset)-\lambda/\dns(\eset)$.
\end{problem}

For shorthand, we denote $-1/\dns$ as $\neginvd$.
We first show that \LRES can be mapped to \LRID, 
so that optimal solutions for the one problem can be found by solving the other, 
with parameters $\mu$ and~$\lambda$ appropriatelly chosen. 
Then, we focus on solving the \LRID problem.

\begin{proposition}
\label{propES-ID}
An edge set $\esetopt$ is an optimal solution for $\LRES$ with parameter~$\mu$ 
if and only if $\esetopt$ is an optimal solution for $\LRID$ with $\lambda=\dns^2(\esetopt)\mu$.	
\end{proposition}

The mapping provided in Proposition~\ref{propES-ID} 
guarantees that a solution to $\LRES$ with a parameter $\mu$ 
can be found by solving $\LRID$ with a corresponding parameter $\lambda$.
A drawback is that to construct an $\LRID$ instance for a given $\LRES$ instance with a fixed $\mu$
we need to know the density of $\LRES$'s solution~$\dns(\esetopt)$. 
However, in general, the Lagrangian multiplier $\mu$ is often not known in advance, 
and the user needs to experiment with several values and select the setting leading 
to an interesting solution. 
In such cases, arguably, there is no difference between experimenting 
with~$\mu$ for \LRES or with $\lambda$ for \LRID.
Furthermore, if the value of $\mu$ is given, 
we will show that our solution provides an efficient framework 
to explore the solution space of $\LRID$ for all possible values of $\lambda$, 
and identify the solutions for the given value of $\mu$.

\subsection{Fractional Programming}
\label{section:fractional-programming}

Following the connection established in the previous section,
our goal is therefore to solve problem \LRID for a given value of $\lambda$.
We use the technique of \emph{fractional programming}, 
based on the work by Gallo et al.~\cite{gallo1989fast}. 
For completeness, we review the technique. 
We first define the \emph{fractional pro\-gram\-ming} (\FPproblem) problem:

\begin{problem}[\FPproblem]
\label{problem:FP}
Given an undirected graph $G=(\nodes,\edges)$, 
and edge set functions $F_1:2^\edges\to\reals$ and $F_2:2^\edges\to\realspz$, find a subset of edges $\eset\subseteq \edges$ 
so that $c(\eset)=\frac{F_1(\eset)}{F_2(\eset)}$ is maximized.
\end{problem}

The following problem, which we call \Qproblem, is closely related to the \FPproblem~problem:
\begin{problem}[\Qproblem]
\label{problem:Q}
Given an undirected graph $G=(\nodes,\edges)$, edge set functions $F_1:2^\edges\to\reals$ and 
$F_2:2^\edges\to\realspz$, and a real number $c\in\reals$, find a subset of edges 
$\eset\subseteq \edges$ so that $Q(\eset\mid c)  = F_1(\eset) - c F_2(\eset)$ is maximized.
\end{problem}

The key result of fractional programming~\cite{gallo1989fast} states that:
\begin{proposition}[Gallo et al.~\cite{gallo1989fast}] 
\label{prop:FP}
A set $\esetopt$ is an optimal solution to an instance of the \FPproblem problem 
with solution value $c(\esetopt)$,
if and only if $\esetopt$ is an optimal solution to the corresponding {\Qproblem} problem
with $c=c(\esetopt)$ and $Q(\esetopt\mid c(\esetopt))=0$.
\end{proposition}

Proposition~\ref{prop:FP} provides the basis for the following iterative algorithm 
(\FPalgo), which finds a solution to \FPproblem by solving a series of instances of 
{\Qproblem} problems~\cite{gallo1989fast}.

\medskip
\noindent
\underline{Algorithm \FPalgo}:
\begin{enumerate}
\item{Select some $\eset_0\subseteq \edges$. Set $c_0 \leftarrow F_1(\eset_0) / F_2(\eset_0)$, and $k\leftarrow 0$.}
\item{Compute $\eset_{k+1}$ by solving the {\Qproblem} problem: 
\[ Q(\eset_{k+1}\mid c_k) \leftarrow \max_{\eset\subseteq \edges}\{F_1(\eset) - c_kF_2(\eset)\}. \]}
\item{If $Q(\eset_{k+1}\mid c_k)=0$, then return $\esetopt \leftarrow \eset_k$.\\Otherwise, 
set $c_{k+1} \leftarrow F_1(\eset_{k+1}) / F_2(\eset_{k+1})$, $k\leftarrow k + 1$, and go to Step (2). }
\end{enumerate}

It can be shown~\cite{gallo1989fast} that the sequence $(c_k)$ generated by \FPalgo is increasing, and that 
if $F_2$ is an integer-valued set function (and we will see that this is our case), 
then the number of iterations of \FPalgo
is bounded by the number of elements in the underlying set, which in our case is the edgeset $\edges$.


\medskip
We formulate \LRID as an instance of \FPproblem. 
As \LRID is parameterized by $\lambda\ge 0$, 
we introduce such parameter in \FPproblem
and set
\[
F_1(\eset\mid\lambda) = \!\sum_{\{e_i,e_j\}\in \eset^2}\!\esim(e_i,e_j) -\lambda |\nodes(\eset)|,
\text{ and }
F_2(\eset) = |\eset|.
\] 
Now, \LRID becomes an instance of \FPproblem and algorithm \FPalgo can be~applied. 
As $F_2(\eset)$ is the number of edges in the solution, 
the algorithm \FPalgo is guaranteed to halt after solving 
$\bigO(|\edges|)$ instances of the~{\Qproblem} problem. 

Each instance of the {\Qproblem} problem at Step (2) of \FPalgo can be solved
efficiently by a parametric preflow/minimum cut algorithm~\cite{gallo1989fast}. 
The construction of the flow graph is presented in the next section.

Since we introduced the parameter $\lambda$, we need to write the objectives of 
\FPproblem and \Qproblem as $c(\eset\mid\lambda)$ and  $Q(\eset\mid c,\lambda)$, respectively, 
but we will omit the dependency on $\lambda$ when it is clear.
We denote the optimal values of \FPproblem and \Qproblem as 
$\copt(\lambda) = c(\esetopt\mid\lambda)$ and $\Qopt(c,\lambda) = Q(\esetopt\mid c,\lambda)$, respectively.

\subsection{Parametric MIN-CUT}
\label{section:parametric-min-cut}

In this section we show how to solve instances of the {\Qproblem} problem
by using a mapping to the \mincut problem.
A similar approach has been used, among others, 
by Goldberg~\cite{goldberg1984finding}
to solve the densest-subgraph problem.
%

\begin{figure}[t]
	\begin{center}
		\begin{tikzpicture}[scale=0.9,every node/.style={scale=0.9}]]


\tikzstyle{exedge} = [draw = yafcolor5!80, thick, text=black!80]
\tikzstyle{direxedge} = [draw = yafcolor3!80, thick, text=black!80]
\tikzstyle{exnode} = [thick, draw = white, fill=white, circle, inner sep = 1pt, text=black, minimum width=0pt]


\node[exnode] (x) at (1.5,  0)   {$x$};
\node[exnode] (y) at (2.1, -1.1)   {$y$};
\node[exnode] (z) at (0.7, -0.9) {$z$};
\node[exnode] (w) at (1.8,  1.1) {$w$};

\draw[-, exedge, bend left  = 10] (x) to (y);
\draw[-, exedge, bend right = 10] (x) to (z);
\draw[-, exedge, bend left  = 10] (y) to (z);
\draw[-, exedge, bend left  = 10] (x) to (w);

\node[fill=white] at (1.4,-2) {\graph};


\node[exnode] (s) at (3, 0) {$s$};

\node[exnode] (xy) at (5,  1.5) {$(x,y)$};
\node[exnode] (xz) at (5,  0.5) {$(x,z)$};
\node[exnode] (yz) at (5, -0.5) {$(y,z)$};
\node[exnode] (xw) at (5, -1.5) {$(x,w)$};

\node[exnode] (x2) at (7,  1.5) {$x$};
\node[exnode] (y2) at (7,  0.5) {$y$};
\node[exnode] (z2) at (7, -0.5) {$z$};
\node[exnode] (w2) at (7, -1.5) {$w$};

\node[exnode] (t) at (8.8, 0) {$t$};

\draw[->, direxedge, bend left  = 20] (s) to (xy);
\draw[->, direxedge, bend left  = 10] (s) to (xz);
\draw[->, direxedge, bend right = 10] (s) to (yz);
\draw[->, direxedge, bend right = 20] (s) to (xw);

\draw[->, direxedge, bend right = 10] (xy) to (xz);
\draw[->, direxedge, bend right = 10] (xz) to (xy);

\draw[->, direxedge, bend right = 10] (xz) to (yz);
\draw[->, direxedge, bend right = 10] (yz) to (xz);

\draw[->, direxedge, bend right = 10] (yz) to (xw);
\draw[->, direxedge, bend right = 10] (xw) to (yz);

\draw[->, direxedge, bend right = 45] (xy) to (yz);
\draw[->, direxedge, bend right = 45] (yz) to (xy);

\draw[->, direxedge, bend right = 45] (xy) to (xw);
\draw[->, direxedge, bend right = 45] (xw) to (xy);

\draw[->, direxedge, bend right = 45] (xz) to (xw);
\draw[->, direxedge, bend right = 45] (xw) to (xz);

\draw[->, direxedge] (xy) to (x2);
\draw[->, direxedge, bend left = 10] (xy) to (y2);

\draw[->, direxedge, bend right = 10] (xz) to (x2);
\draw[->, direxedge, bend left  = 10] (xz) to (z2);

\draw[->, direxedge, bend right = 10] (yz) to (y2);
\draw[->, direxedge] (yz) to (z2);

\draw[->, direxedge, bend right = 10] (xw) to (x2);
\draw[->, direxedge] (xw) to (w2);

\draw[->, direxedge, bend left  = 20] (x2) to (t);
\draw[->, direxedge, bend left  = 10] (y2) to (t);
\draw[->, direxedge, bend right = 10] (z2) to (t);
\draw[->, direxedge, bend right = 20] (w2) to (t);

\node[fill=white] at (6,-2) {\flowgr};

\end{tikzpicture}
		\caption{\label{figure:min-cut}
			A graph \graph (left),
			and the corresponding flow graph \flowgr (right)
			used for solving \Qproblem with  
			parametric \mincut techniques, 
			as described in Section~\ref{section:parametric-min-cut}.
			The edge weights in \flowgr are not shown to avoid clutter.}
	\end{center}
\end{figure}
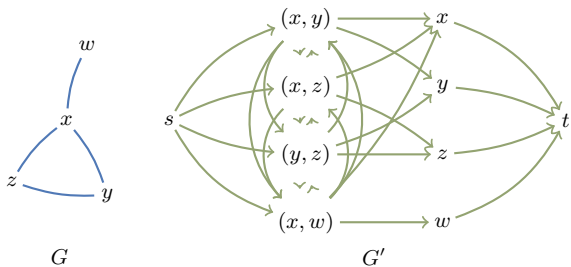

Let the input of \Qproblem be a graph $\graph=(\nodes,\edges)$ 
with edge similarity \sml and parameters $c\in \reals$ and $\lambda\in \realspz$.
We construct the following directed weighed network $\flowgr = (\flnodes, \fledeges, \flweights)$.
The set \flnodes is defined as $\flnodes = \uset_\edges \cup \uset_\nodes \cup \{\src, \snk\}$, 
where $\uset_\edges = \{u_e \mid e\in\edges\}$ contains a node $u_e$ for each edge $e \in \edges$, 
$\uset_\nodes = \{u_v \mid v\in\nodes\}$ contains a node $u_v$ for each node $v \in \nodes$, and
the nodes $\src$ and $\snk$ are additional source and sink nodes. 
The nodes in $\uset_\edges$ are pairwise connected by bi-directional edges 
$(u_{e},u_{d})$ with weight $\frac{1}{2}\sml(e, d)$, whereas the nodes in $\uset_\nodes$ are not connected to each other.
Additionally, there is a directed edge $(u_e,u_v)$ for each $v\in \nodes$ that is an endpoint of $e\in \edges$ with weight $\flweights(u_e,u_v) = +\infty$. 
Finally, the source \src is connected to all the nodes in $\uset_\edges$ by directed edges with 
weight $\flweights(\src, u_e) = \frac{1}{2}\sum_{d\in \edges, d\ne e}\esim(e, d)-c$, 
and each node in $\uset_\nodes$ is connected to \snk 
by a directed edge with weight $\lambda$. 
The construction of $\flowgr$ for a given $\graph$ is clearly polynomial.
An example of the construction of \flowgr
is shown in Figure~\ref{figure:min-cut}.

We now solve the $(\src,\snk)$-\mincut problem on 
the graph $\flowgr = (\flnodes, \fledeges, \flweights)$, 
parameterized with $c$ and $\lambda$.
Let $(\{\src\}\cup\usetopt, \{\snk\}\cup\usetoptc)$ 
be the minimum cut in~\flowgr, 
and let $\optmccost(c,\lambda)$ be its value. 
The next proposition establishes the connection between the optimal values of \mincut on \flowgr and 
the $\Qproblem$ problem on $\graph$, and 
describes how the solution edge set for the \Qproblem problem can be derived from the solution cut set of \mincut.

\begin{proposition}
\label{proposition:q-problem}
The value of the $(\src,\snk)$-\mincut in the graph~$\flowgr = (\flnodes, \fledeges, \flweights)$
for given parameters $c$ and $\lambda$
corresponds to the optimum value for the {\Qproblem} problem
with the same values of~$c$ and $\lambda$. 
The solution edge set $\esetopt$ for {\Qproblem} problem on $\graph$ can be reconstructed from the minimum-cut set $\{\src\}\cup\usetopt\subseteq\flnodes$ in \flowgr as $\esetopt=\usetopt\cap \uset_\edges$.
\end{proposition}

To summarize, in the previous sections we have established the following:

\begin{proposition}
\label{proposition:fp-problem}
An instance of \LRID for a given parameter $\lambda$
can be solved by mapping it to Problem~\FPproblem and applying the \FPalgo.
Problem~\Qproblem in the iterative step of \FPalgo
can be solved by mapping it to the parametric \mincut problem, 
as shown in Proposition~\ref{proposition:q-problem}.
\end{proposition}

Let us evaluate the time and space complexity of the proposed solution.
In \FPalgo we iteratively search for optimal values in the {\Qproblem} problem 
by solving \mincut problems. 
In each iteration, 
only the source link capacities are updated as $c_k$ changes, and, 
as mentioned before, sequence $(c_k)$ grows monotonically. 
This setting can be handled efficiently in the parametric \mincut framework, 
which incrementally updates the solution from the previous iteration. 
The state-of-the-art algorithm for parametric \mincut~\cite{hochbaum2008pseudoflow}
requires $\bigO(mn\log n + kn)$ time and $\bigO(m)$ space 
for a graph with $n$ nodes, $m$ edges, and $k$ updates of edge capacities (iterations in \FPalgo). 
Recall that the number of iterations is bounded by $\bigO(|\edges|)$, and thus,
solving \LRID for a fixed $\lambda$ requires $\bigO(|\edges|^3\log|\edges|)$ time and
$\bigO(|\edges|^2)$ space.

\subsection{$\lambda$-Exploration}

Having discussed how to solve the \LRID problem for a fixed $\lambda$, 
we now introduce a framework to efficiently enumerate the solutions 
for all possible values of~$\lambda$.
The goal is to identify the ranges of values of~$\lambda$ 
that yield identical solutions and exclude them from the search.

First, we show the monotonicity of the optimal solution value
of \LRID, the optimal subgraph similarity and density values with respect to $\lambda$. 

\begin{proposition}
\label{prop:monot}
The optimal solution value of \LRID is a monotonically non-in\-creas\-ing function of $\lambda$. 
The density of the optimal edge set is a mo\-not\-onically non-decreasing function of $\lambda$.
The subgraph edge similarity of the optimal edge set is a mo\-not\-onically non-increasing function of $\lambda$.
\end{proposition}

From the definition of optimality and Proposition~\ref{prop:monot}, it follows that:

\begin{corollary}
	\label{cor}
	Given two solutions $X_1$ and $X_2$ to \LRID for $\lambda_1$ and $\lambda_2$ with $\lambda_1<\lambda_2$,
	either (i) $\sml(X_1)=\sml(X_2)$ and $\dns(X_1)=\dns(X_2)$ or (ii) $\sml(X_1)>\sml(X_2)$ and $\dns(X_1)<\dns(X_2)$.
\end{corollary}

The monotonicity of the optimal values of the objective functions and Corollary~\ref{cor} will guide our exploration of the $\lambda$ ranges. 


Note that the shown monotonic properties are not strict and it is indeed easy to construct an example input graph, where different values of $\lambda$ lead to solutions to \LRID with the same values of subgraph edge similarity and density.
To avoid a redundant search, we would like to solve \LRID for all the values of $\lambda$ that lead to distinct combinations of density and similarity values.
Such redundant values of $\lambda$ can be pruned by observing that when two values $\lambda_1$ and $\lambda_2$ 
give solutions with the same values $\sml_1=\sml_2$ and $\dns_1=\dns_2$, 
then all $\lambda \in [\lambda_1, \lambda_2]$ 
must also lead to the same optimal density and subgraph edge similarity, and thus the interval
$[\lambda_1,\lambda_2]$ can be discarded from further search. This result follows from the monotonicity of the optimal values of density and similarity (Proposition~\ref{prop:monot}).

The proposed approach to search for different values of $\lambda$ is a breadth-first iterative algorithm. 
At the beginning, the set of distinct solutions \pareto is empty, and $\lambda_{\ell}=\lambda_{\min}$ and $\lambda_{u}=\lambda_{\max}$. The algorithm maintains a queue of candidate search intervals $T$, which is initially empty.
To avoid clutter, we denote the solution values of subgraph edge similarity and density $(\sml, \dns)$ for a given $\lambda$ as $t(\lambda)$.

\smallskip
\noindent
\underline{Algorithm $\lambda$-exploration}:
\begin{enumerate}
	\item{Compute set $\eset^*(\lambda_{\ell})$ and add it to~\pareto.}
	\item{Compute set $\eset^*(\lambda_u)$.}
	\item{If $t(\lambda_u)\ne t(\lambda_\ell)$, then push $(\lambda_\ell, \lambda_u, t(\lambda_\ell), t(\lambda_u))$ to the queue $T$ and add $\eset^*(\lambda_u)$ to \pareto.}
	\item{While $Q$ is not empty:
	\begin{enumerate}
	\item{Pop $(\lambda_\ell,\lambda_u, t(\lambda_\ell), t(\lambda_u))$ from $T$.}
	\item{Set $\lambda_m =(\lambda_\ell+\lambda_u)/2$ and compute $\eset^*(\lambda_m)$.}
	\item{If $t(\lambda_m)\ne t(\lambda_\ell)$, then push  $(\lambda_\ell,\lambda_m, t(\lambda_\ell), t(\lambda_m))$ to~$T$.}
	\item{If $t(\lambda_m)\ne t(\lambda_u)$, then push  $(\lambda_m,\lambda_u, t(\lambda_m), t(\lambda_u))$ to~$T$.}
	\item{If $t(\lambda_m)\ne t(\lambda_\ell)$ and $t(\lambda_m)\ne t(\lambda_u)$, add $\eset^*(\lambda_m)$ to~\pareto.}
	\end{enumerate}
	}
\end{enumerate}

To bound the number of calls of $\lambda$-search,
we need to lower bound the difference between two consecutive values of $\lambda$ that lead to two different solutions. This lower bound is given in the next proposition, together with upper and lower bounds for $\lambda$ values.

\begin{proposition}
\label{prop:bounds}
To obtain all the distinct solutions in the $\lambda$-exploration algorithm, 
a lower bound for a value of $\lambda$ is
$\lambda_{min}=s_{min}/2|E|$, 
an upper bound is $\lambda_{max}=s_{max}|E|^2/2$, and 
a lower bound for the difference between two values of $\lambda$ leading to solutions with distinct density and subgraph edge similarity values is $\delta_{\lambda}= s_{min}/2|E|$. 
Here $s_{min} = \min_{\{e_1, e_2\}\in X^2}\esim(e_1, e_2)$ and $s_{max} = \max_{\{e_1, e_2\}\in X^2}\esim(e_1, e_2)$.
\end{proposition}

Given the bounds in Proposition~\ref{prop:bounds}, 
an upper bound on the number of different values of $\lambda$ 
that we need to try is
$I_{\lambda} =(\lambda_{max}-\lambda_{min})/\delta_{\lambda}\leq |E|^3\frac{s_{max}}{s_{min}}$, 
where $s_{max}$ and $s_{min}$ are the largest and the smallest non-zero values of edge similarity between two edges in the input graph. Thus, for a complete exploration of all the possible $\lambda$ values leading to different values of subgraph edge similarity and density of the solution graph, we need $\bigO(|E|^3)$ iterations. This estimate is pessimistic and assumes no subranges of $\lambda$ are pruned during the exploration. As we will see later, on practice the exploration typically requires around $|E|$ number of iterations.

\section{Related Work}\label{related}
In this paper we consider the problem of finding subgraphs that maximize both a density measure and a similarity measure.
The problem of finding dense structures in graphs has been extensively studied in the literature, as it finds applications in many domains such as community detection~\cite{chen2012dense,falih2018community}, event detection~\cite{angel2012dense}, and fraud 
detection~\cite{hooi2016fraudar}.
Existing works have addressed the task of finding the best solution that satisfies the given constraints, such as, the densest subgraph~\cite{goldberg1984finding,hooi2016fraudar}, the densest subgraph of $k$ 
vertices~\cite{bhaskara2010detecting}, the densest 
subgraph in a dual network~\cite{wu2015finding}, or the best $\alpha$-quasi-clique~\cite{tsourakakis2013denser}.
Other works
have aimed at retrieving a set of good solutions, such as top-$k$ densest subgraphs in a graph collection~\cite{valari2012discovery}, $k$ diverse subgraphs with maximum total aggregate density~\cite{balalau2015finding}, or $k$-cores with maximum number of common attributes~\cite{fang2016effective}.
However, these works optimize a single measure, i.e., the density, thus ignoring other properties of the graph, or find a solution that depends on an input query.

There are a few works focusing on edge similarity.
The closest to our work, Boden et al.~\cite{boden2012mining}, considers edge-labeled multilayer graph and looks for vertex sets that are densely connected by edges with similar labels in a subset of the graph layers. They set a pairwise edge similarity threshold for a layer and look for $0.5$ quasi-cliques, which persist for at least 2 layers. 
In contrast, our approach does not require any preset parameters and offers a comprehensive exploration of the space of dense and similar subgraphs.




Motivated by applications in fraud detection, Shin et al.~\cite{hooi2016fraudar} propose a greedy algorithm that detects the $k$ densest blocks in a tensor with $N$ attributes, with guarantees on the approximation.
The framework outputs $k$ blocks by greedy iterative search. Yikun et al.~\cite{yikun2019no} propose a novel model and a measure for dense fraudulent blocks detection. The measure is tailored for the fraud detection in multi-dimensional entity data, such as online product reviews, but could be possibly adapted to capture other types of node and edge similarities. They propose an efficient algorithm, which outputs several graph with some approximation guarantees. 
In contrast to the approaches above, our work offers exploration of \emph{exact} solutions with different trade-offs. 

Multi-objective optimization for interesting graph structures search was studied in the context of frequent pattern 
mining~\cite{shelokar2013mosubdue,carranza2018higher} and graph partitioning~\cite{banos2004new}. 
However, most of the frequent pattern works do not consider the density as an objective function, but depend on the notion of frequency in the graph and cannot be extended to our case. Carranza et al.~\cite{carranza2018higher}, instead, define a conductance measure in terms of an input pattern and then recursively cut the graph into partitions with minimum conductance. The result, however, depends on the input.
Graph partitioning approaches optimize quality functions based on modularity and node/edge attributes, but focus on a complete partition of the input graph~\cite{sanchez2015efficient,combe2015louvain} and do not guarantee to the quality of each individual partition.

%

\section{Experimental Evaluation}
\label{exp}

We evaluate the proposed method using real-world multiplex networks from the CoMuNe lab database%
\footnote{\url{https://comunelab.fbk.eu/data.php}}
and the BioGRID datasets\footnote{\url{https://thebiogrid.org}}.
An implementation of our method is publicly available\footnote{\url{https://github.com/polinapolina/dense-subgraphs-with-similar-edges}}.
In the following, we refer to our approach as \ourmethod.
For the parametric \mincut problem, we use Hochbaum's 
algorithm~\cite{hochbaum2008pseudoflow}
and its open-source C implementation~\cite{chandran2009computational}. 
The experiments are conducted on a Xeon Gold 6148 2.40\,GHz machine.

\spara{Datasets.}
We use the following real-world datasets:
\noindent
\cs is a multiplex social network consisting of five kinds of online and offline relationships 
(Facebook, leisure, work, co-authorship, and lunch) 
between the employees of the Computer Science department at Aarhus University.
\noindent
\air is a multilayer network composed by 37 layers, 
each one corresponding to a different airline operating in Europe.  
\noindent
\concel is the C.\ elegans connectome multiplex that consists of layers corresponding to different synaptic junctions: 
electric, chemical monadic, and polyadic.
\noindent
\gencel and \genara are multiplex networks of genetic interactions for C.\ elegans 
and Arabidopsis thaliana.
Table~\ref{tab:stats} summarizes the main characteristics of the datasets. 

All datasets are multilayer networks 
$G_M=(V,E,\ell)$ with a labeling function 
$\ell : E\to 2^L$, where $L$ is the set of all possible labels. 
We omit edge directionality if it is present in the dataset. 
The edge similarity function is defined as the Jaccard coefficient of the labellings: 
$\esim(e_1, e_2)=|\ell(e_1)\cap\ell(e_2)| / |\ell(e_1)\cup\ell(e_2)|$. 
If the pairwise edge similarity is $0$ for a pair of edges, we do not materialize the corresponding edge in the \mincut problem graph. 

We note that, while the datasets are not large in terms of number of nodes and edges, 
the number of edges co-appearing in at least one layer is significant. 
Furthermore, the subgraphs edge similarity is high.

\begin{table*}[!t]
	\caption{Network characteristics. 
		$|\nodes|$: number of vertices; 
		$|\edges|$: the number of edges; 
		$L$: the number of layers; 
		$|\edges|_{\mathit{avg}}$: number of edges per layer; 
		$|E_{\mathit{mult}}|$: number of edges in the multiplex 
		(same edges on different layers are counted as distinct); 
		$|E_{\mathit{meta}}|$: number of unordered edge pairs co-appeared at least in one layer; 
		$\dns$: density of the network; 
		$\dns_{\mathit{avg}}$: average density across the layers; 
		$\sml$: similarity of the network's edge set $\sml(E)$; 
		$\ell_{\mathit{avg}}$: average participation of an edge to a layer.}
	\label{tab:stats}
	\setlength{\tabcolsep}{0pt}
	\vspace{-2mm}
	\centering
	\begin{tabular*}{\textwidth}{@{\extracolsep{\fill}}l rrrrrrrrrr}
		\toprule 
		Dataset & $|\nodes|$ &$|\edges|$& $L$ & $|\edges|_{\mathit{avg}}$ & $|E_{\mathit{mult}}|$ & 
		$|E_{\mathit{meta}}|$ & $\dns$ & $\dns_{\mathit{avg}}$ & $\sml$ & $\ell_{\mathit{avg}}$\\		
		\midrule
		{\cs}  &  61 & 353 & 5 & 124.00 & 620 & 
		39565 & 5.78 & 2.60 
		& 57.44 & 1.75\\
		{\air}  & 417 & 2953 & 37 & 96.97 & 3588 & 
		360082 & 7.08 &
		1.56 & 94.64 & 1.21\\		
		{\concel}  & 279 & 2290 & 3 & 1036.0 & 5863 & 
		1762756 & 8.20 & 3.86 & 534.70 & 1.35\\
		{\gencel}  & 3879 & 7908 & 6 & 1338.66 & 8182 & 
		17249444 & 2.03 & 1.23 & 2141.88 & 1.01\\
		{\genara} & 6980 & 16713 & 7 & 2499.57 & 18655 & 
		91782863 & 2.39 & 1.18 & 5156.65 & 1.04\\		
		\bottomrule
	\end{tabular*}
\end{table*}

\spara{Baselines.}
We compare \ourmethod with two baselines, \bld and \bls.

Algorithm~\bld optimizes the density directly, but takes into account edge-similarity indirectly. 
It outputs the set of edges of the densest weighted subgraph in the complete graph 
$G_D=(V, E_D, w^D)$, which has the same nodes as the input multiplex network $G_M$.
The edge-weighting function $w^D$ has two components, i.e., $w^D  = (w^D_1,w^D_2)$. 
The weight $w^D_1(u,v)$ captures the graph topology, i.e., $w^D_1(u,v) = 1$ if $(u,v)\in E$ 
for some layer and $0$ otherwise.
The weight $w^D_2(u,v)$ captures the similarity of node activity across layers: 
we first define the node labels as the set of all layers where a node appears in some edge 
$\ell(u) = \cup_{u\in e}\ell(e)$, and then define 
$w^D_2(u,v)$ to be the Jaccard index between the sets $\ell(u)$ and $\ell(v)$.
The final weight of an edge is a weighted sum 
$w^D(u,v) = w^D_1(u,v) + \gamma w^D_2(u,v)$, 
where $\gamma$ regulates the importance of the components. 
By tuning $\gamma$ we can obtain a trade-off between topological density and subgraph edge similarity. 

Algorithm~\bls is the counterpart of \bld, which optimizes the edge-similarity directly, but accounts for density indirectly.
\bls finds the densest weighted subgraph of the complete graph $G_S=(E, E_S, w^S)$ with $w^S = (w^S_1,w^S_2)$. Here weight $w^S_1(e_1,e_2)$ is the edge similarity in the multiplex network $G_M$, 
i.e., $w_1(e_1,e_2) = \esim(e_1,e_2)$ and $w^S_2(e_1,e_2)$ represents the topological information, i.e., 
$w^S_2(e_1,e_2)$ $=$ $1$ if $e_1$ and $e_2$ have a common node in the original graph, and $0$ otherwise.
Again, the final edge weight is a weighted sum $w^S(e_1,e_2) = w^S_1(e_1,e_2) + \gamma w^S_2(e_1,e_2)$. 
When $\gamma = 0$, finding the densest weighted subgraph is equivalent to finding the set of edges in the original graph with the largest similarity. 
We tune $\gamma$ to obtain a trade-off between subgraph edge similarity and topological density. 

As with \ourmethod, we do not materialize 0-weight edges in the baselines. 
Both baselines search for the densest subgraph. 
Similarly to \ourmethod, we use the parametric \mincut framework~{\cite{gallo1989fast}}.

\begin{figure*}[!t]
	\begin{center}
		\setlength{\tabcolsep}{0pt}
		\newlength{\figlength}		
		\setlength{\figlength}{0.25\textwidth}
		\begin{tabular}{c@{\hspace{0.0mm}}c@{\hspace{-1.5mm}}c@{\hspace{-1.5mm}}c@{\hspace{-1.5mm}}c}
		& \hspace*{0.2cm} solution values & Density($\lambda$) & Similarity($\lambda$) &  \#solutions\\
		
		\rotatebox{90}{\hspace*{0.5cm}{\footnotesize \cs}}
		&\includegraphics[width=\figlength	]{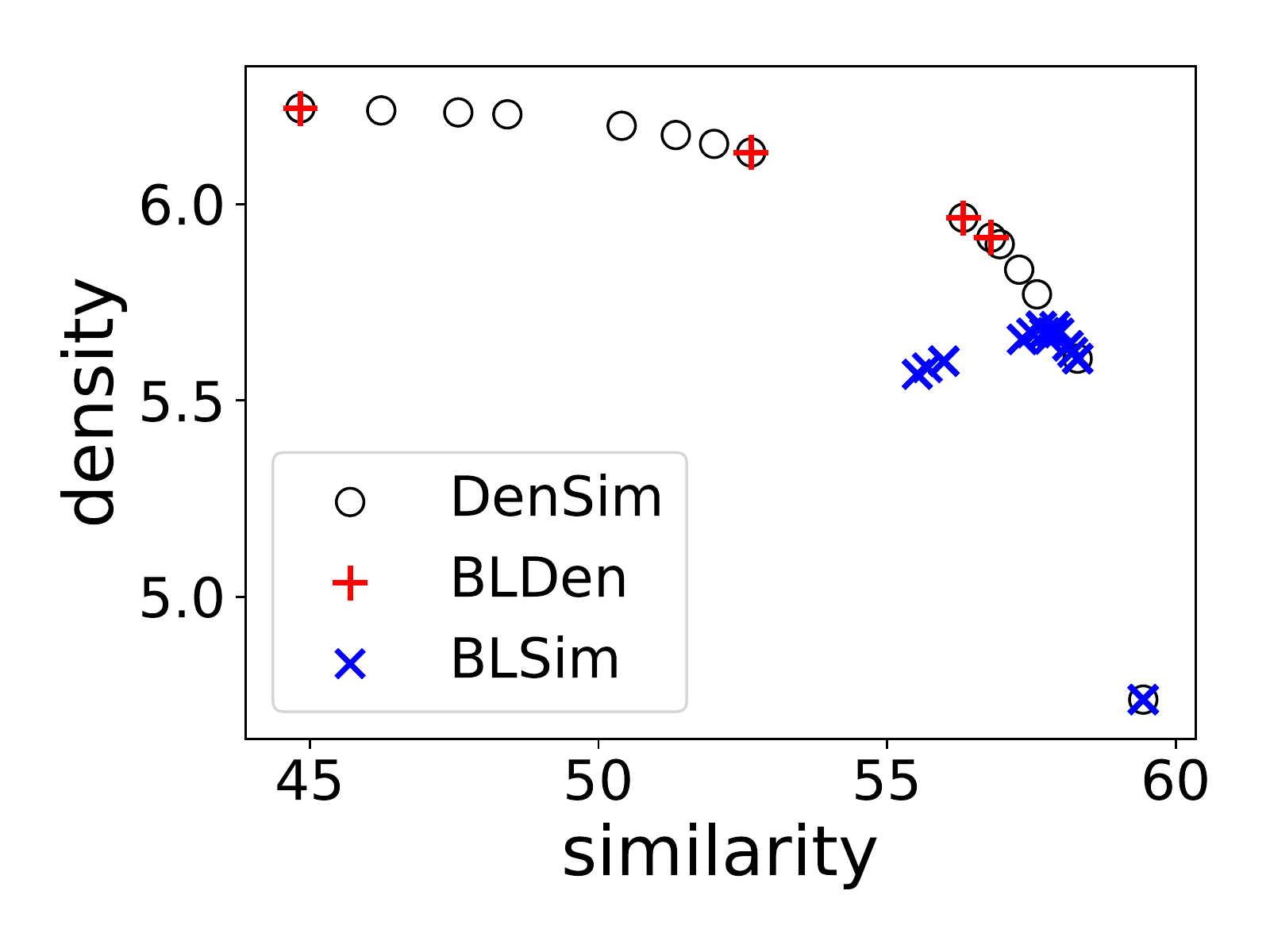} 
		& \includegraphics[width=\figlength	]{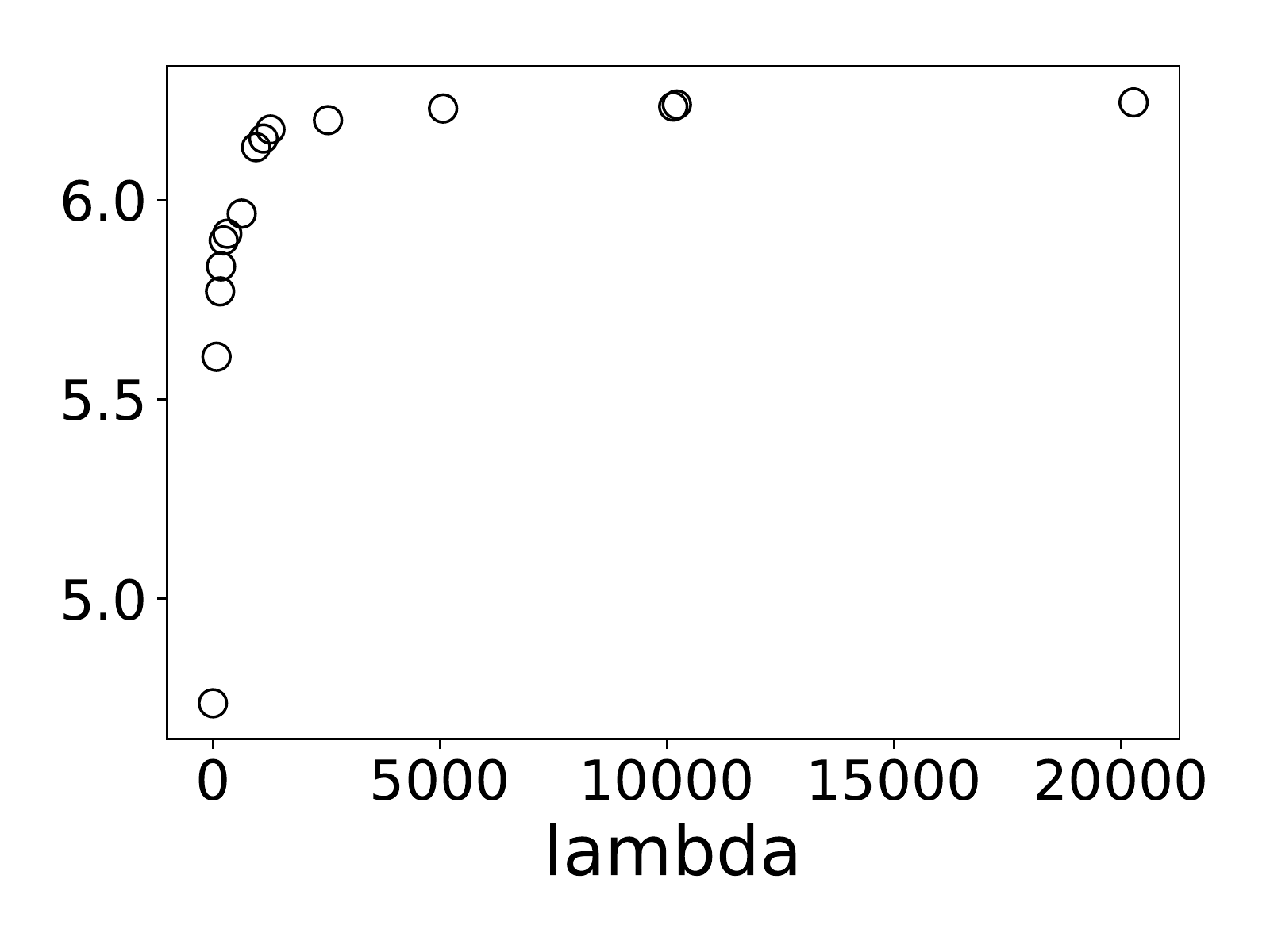}
		& \includegraphics[width=\figlength	]{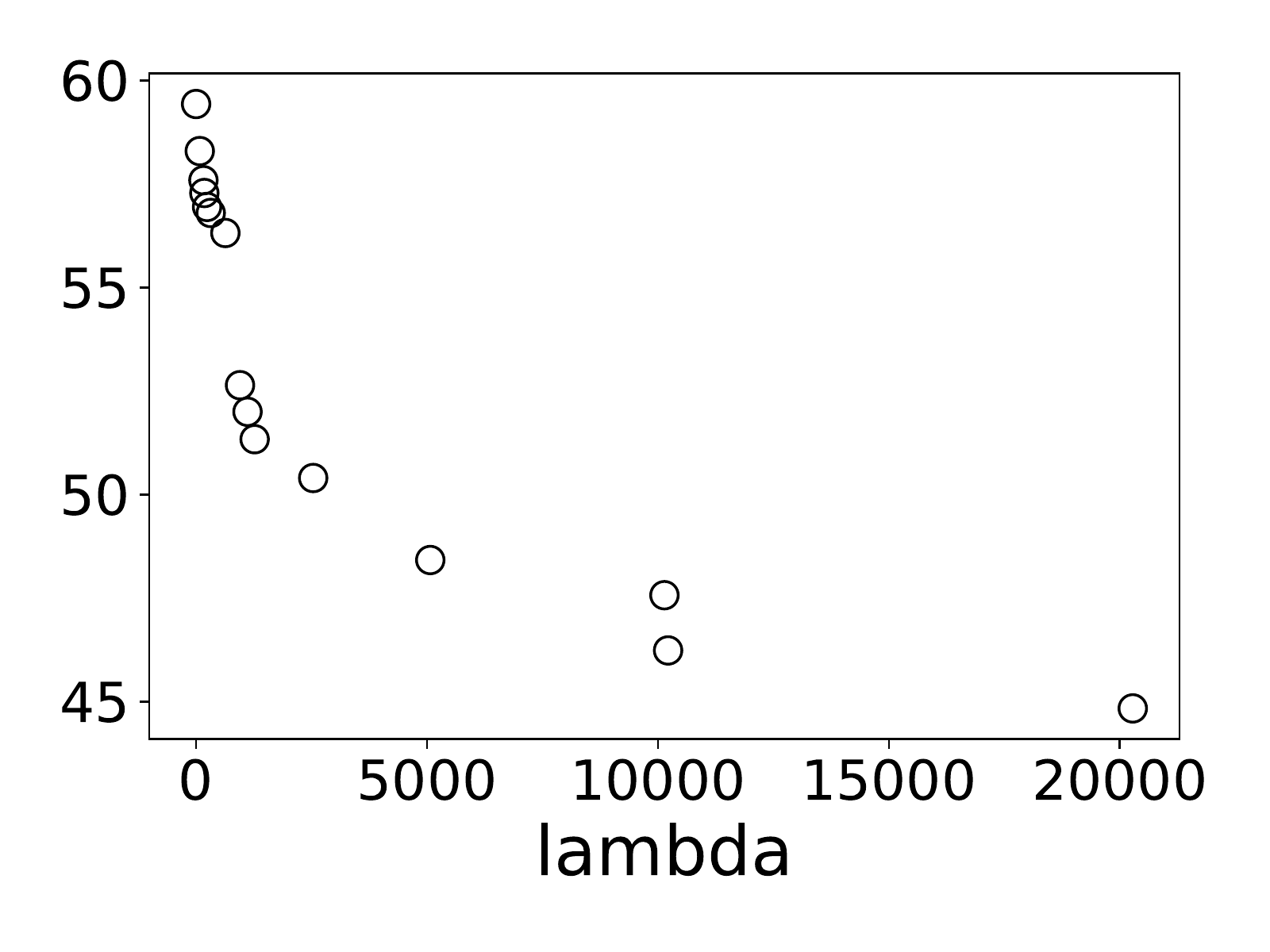}
		& \includegraphics[width=\figlength	]{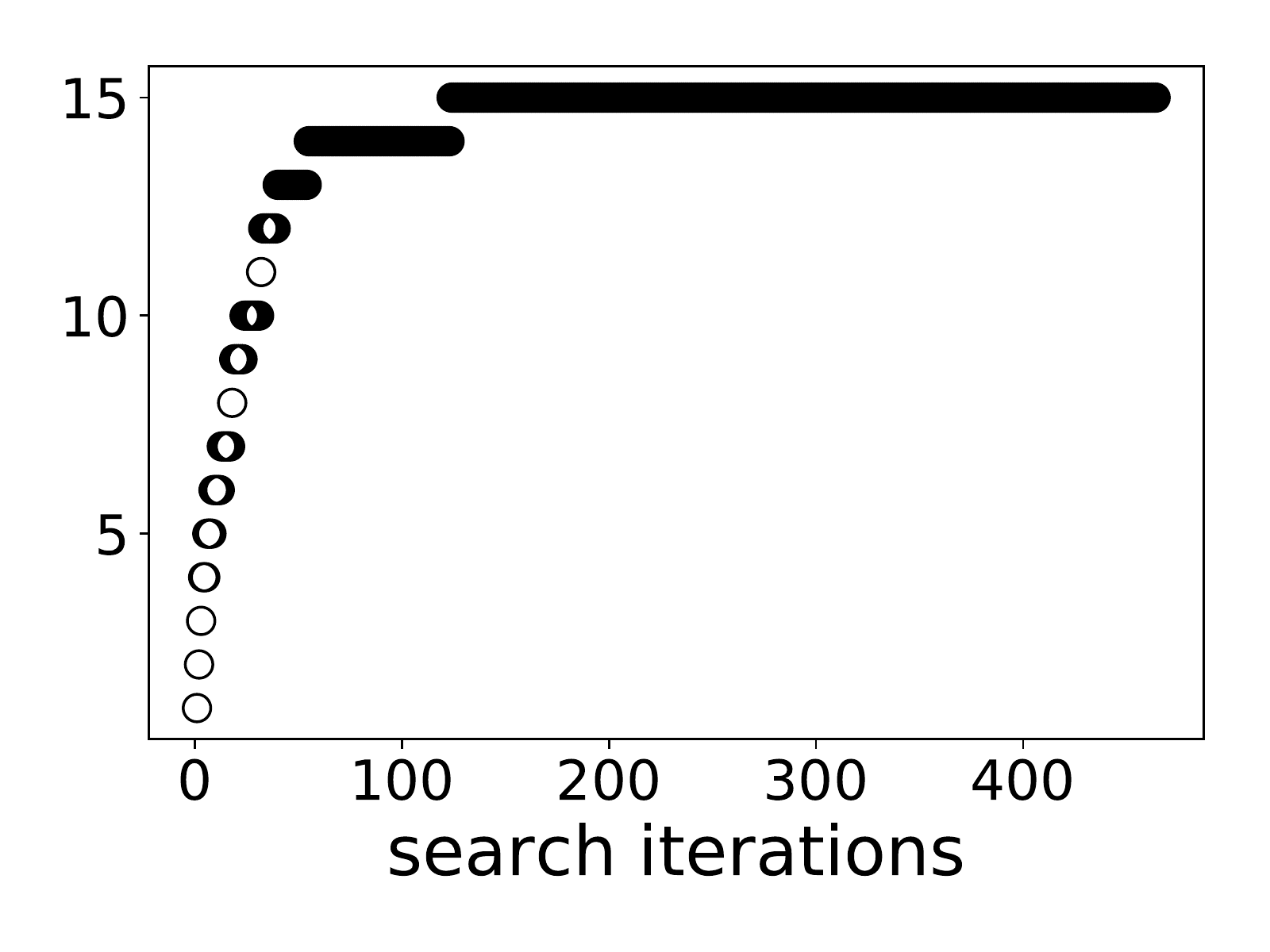} 
		\\[-2mm]
		\rotatebox{90}{\hspace*{0.7cm}{\footnotesize \air}}
		&\includegraphics[width=\figlength	]{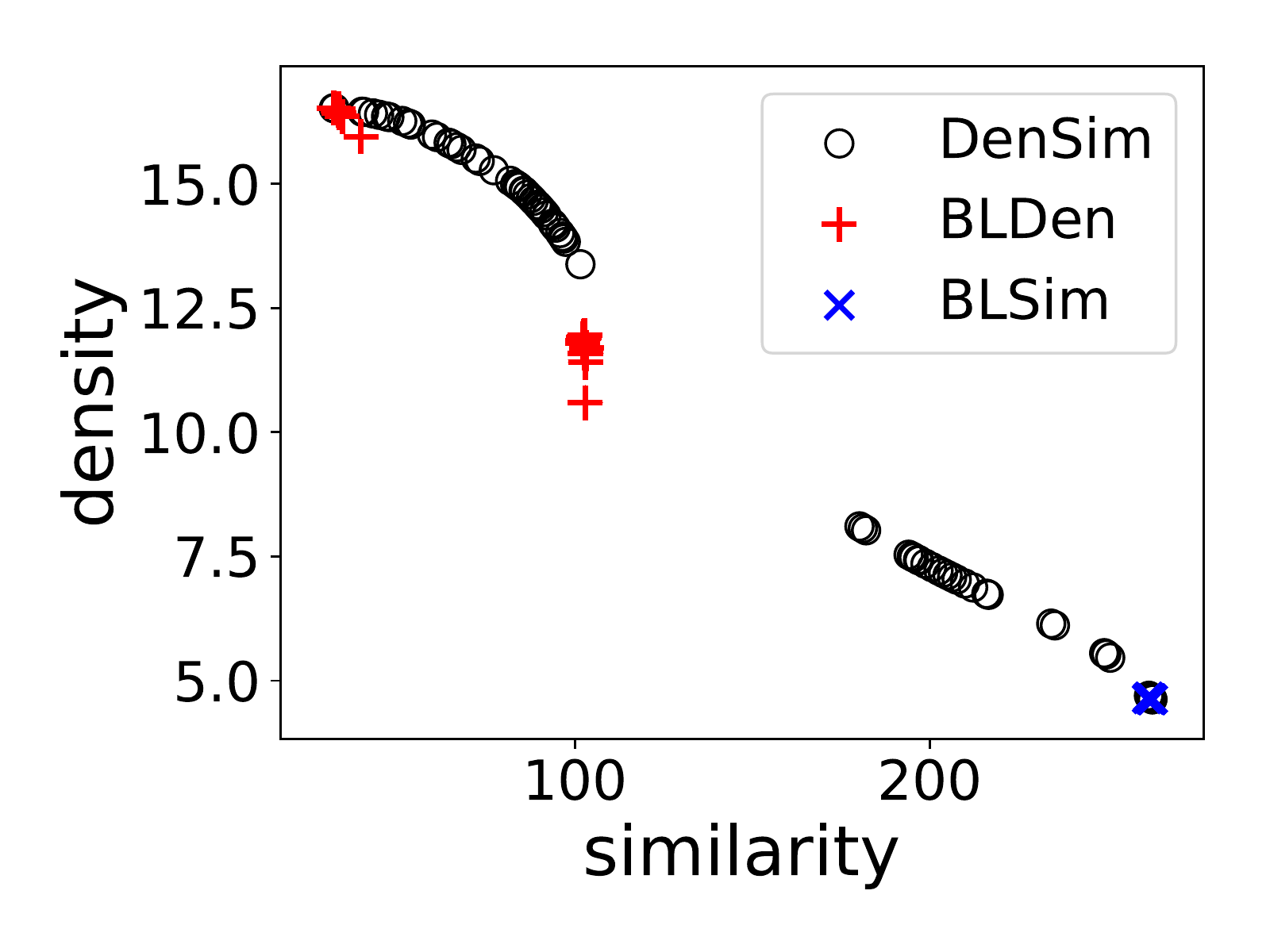}
		& \includegraphics[width=\figlength	]{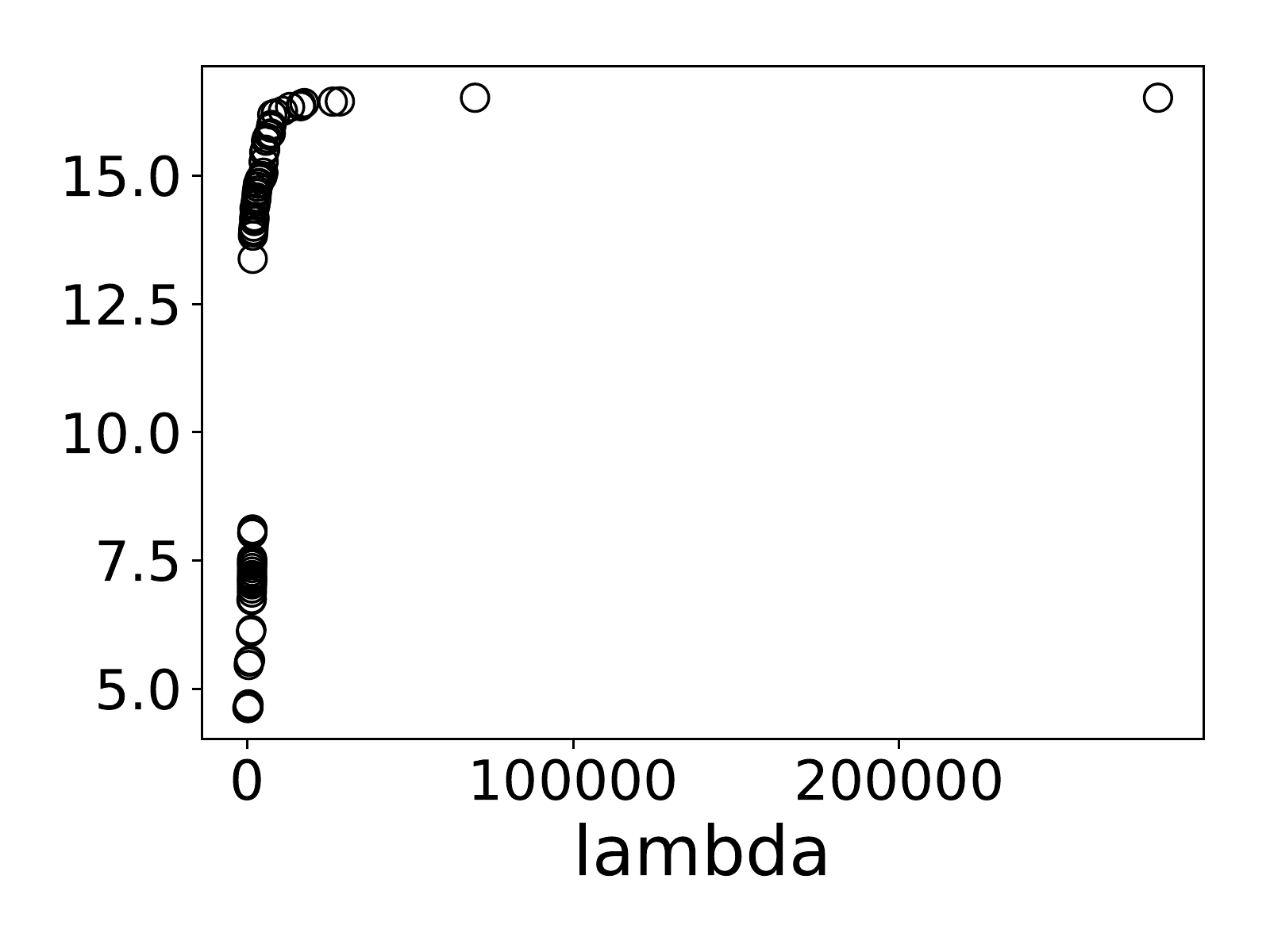}
		& \includegraphics[width=\figlength	]{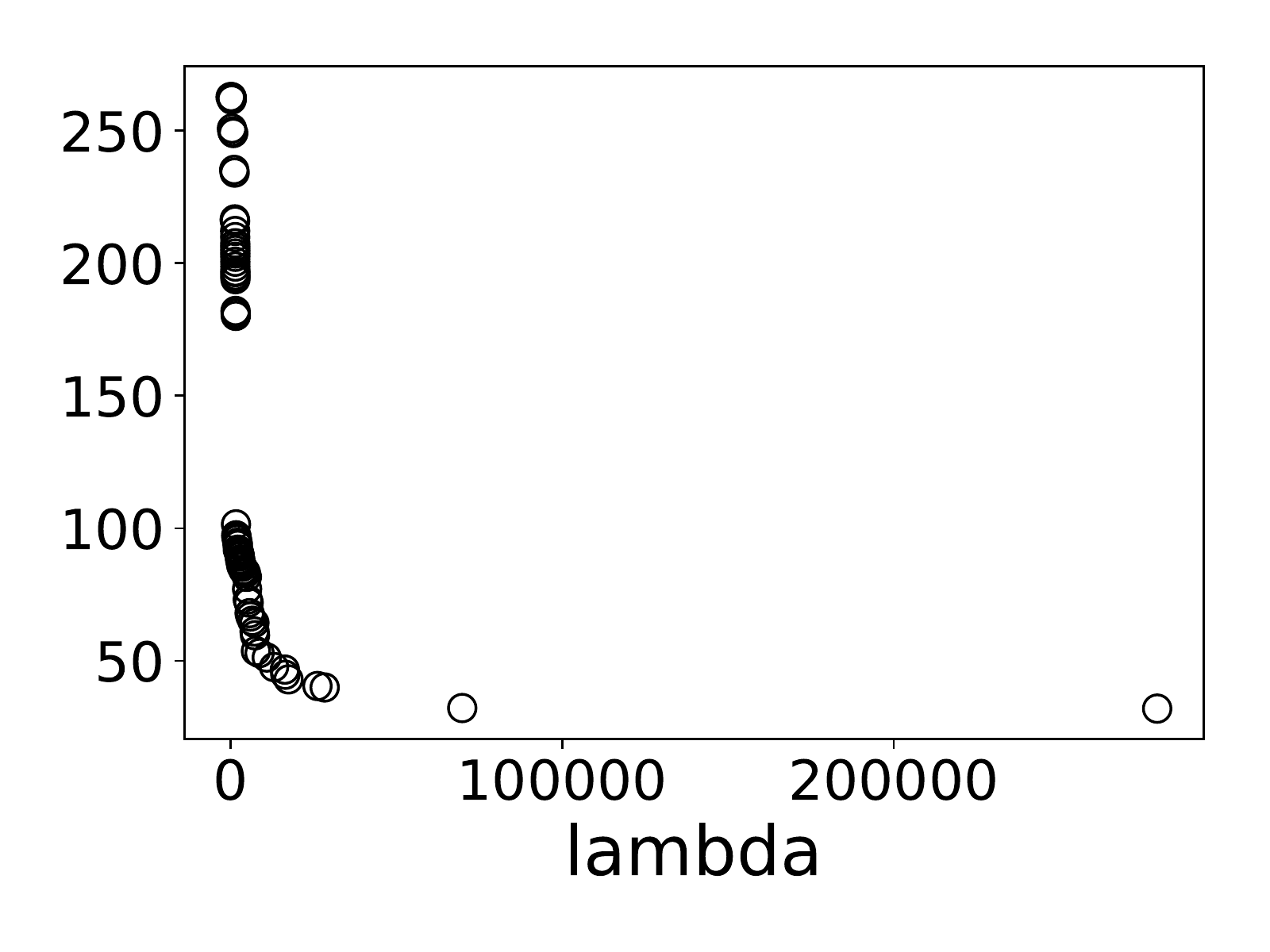}
		& \includegraphics[width=\figlength	]{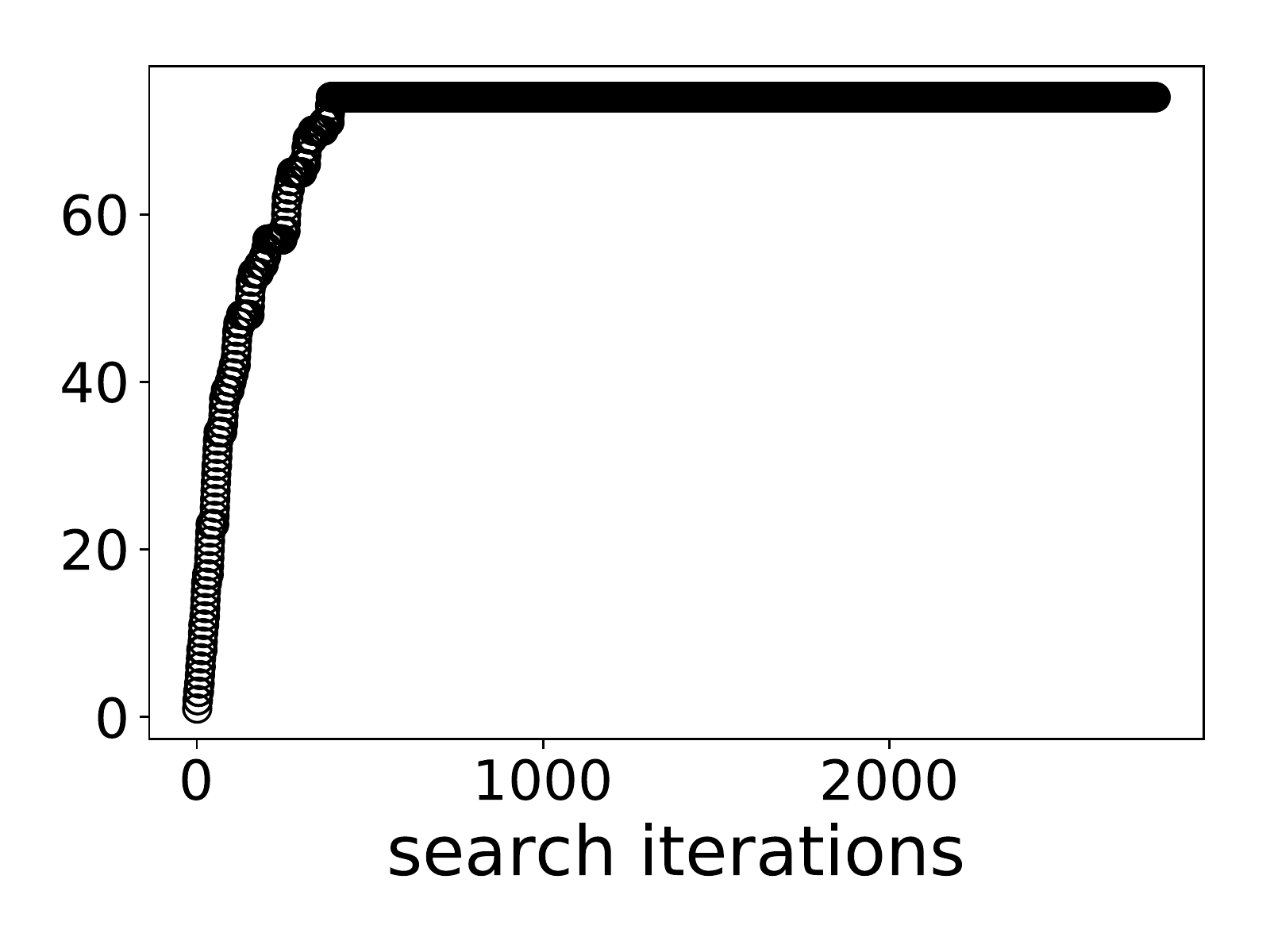}
		\\[-2mm]
		\rotatebox{90}{\hspace*{0.2cm}{\footnotesize \concel}}
		&\includegraphics[width=\figlength	]{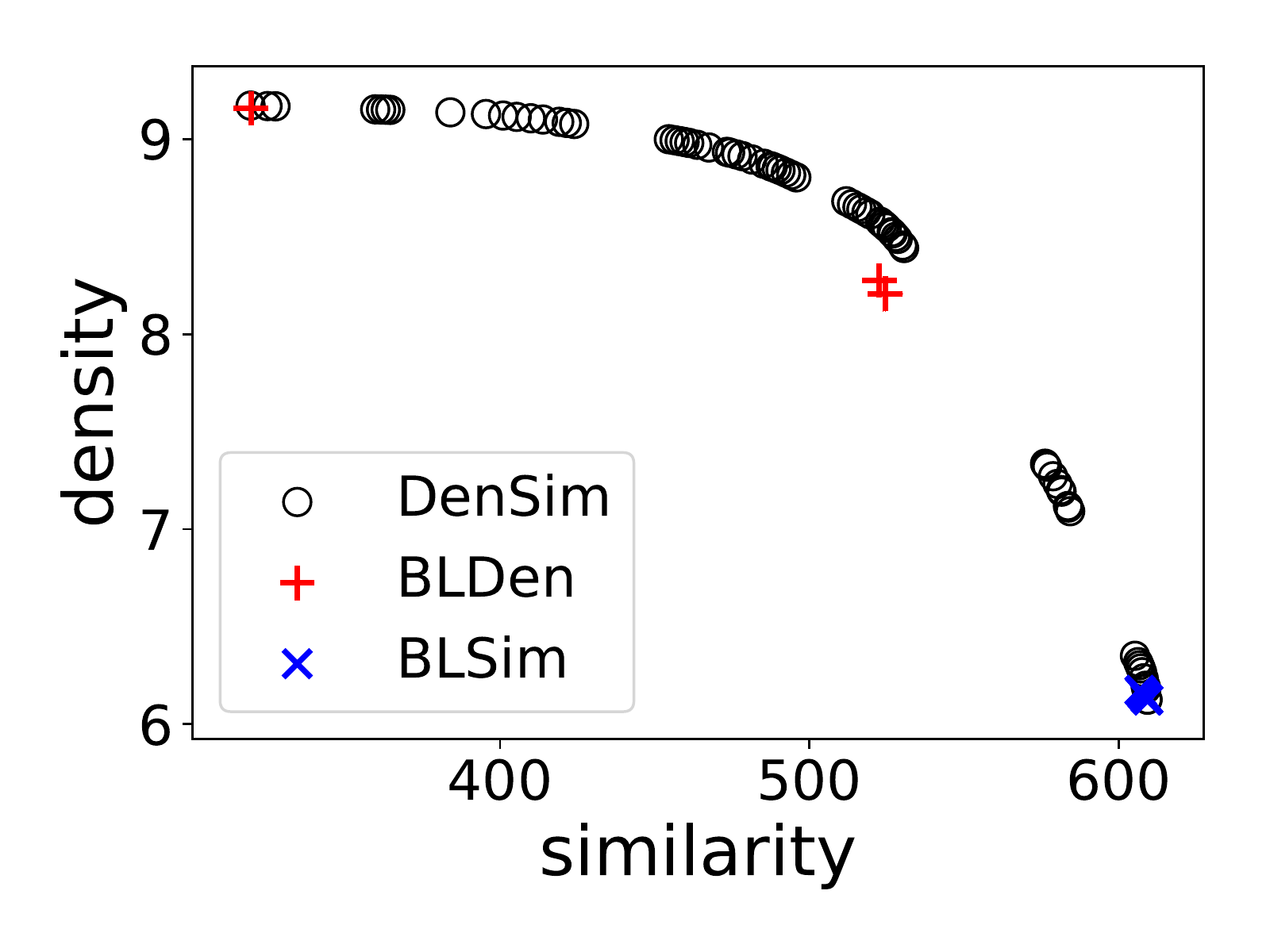}
		& \includegraphics[width=\figlength	]{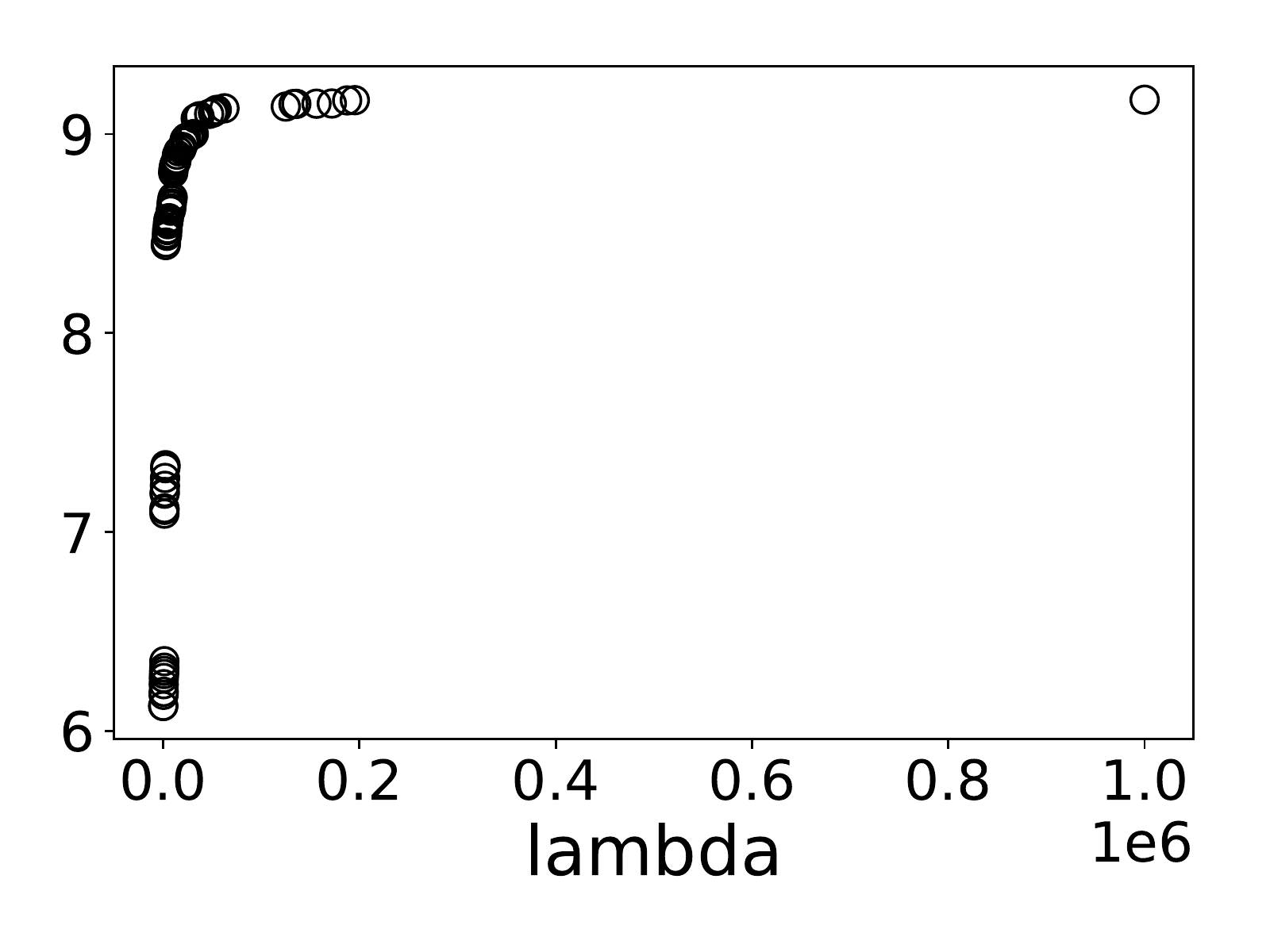}
		& \includegraphics[width=\figlength	]{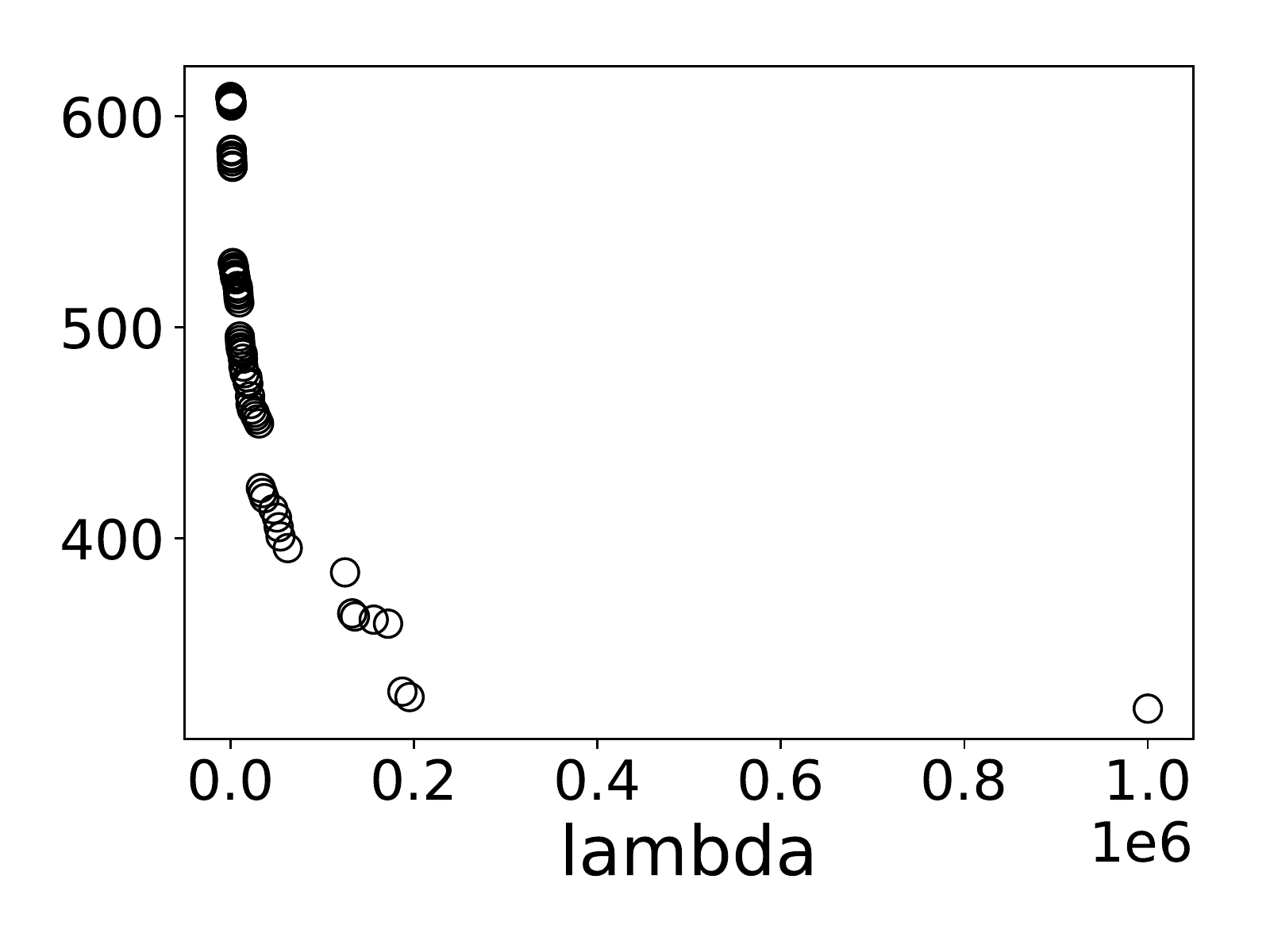}
		& \includegraphics[width=\figlength	]{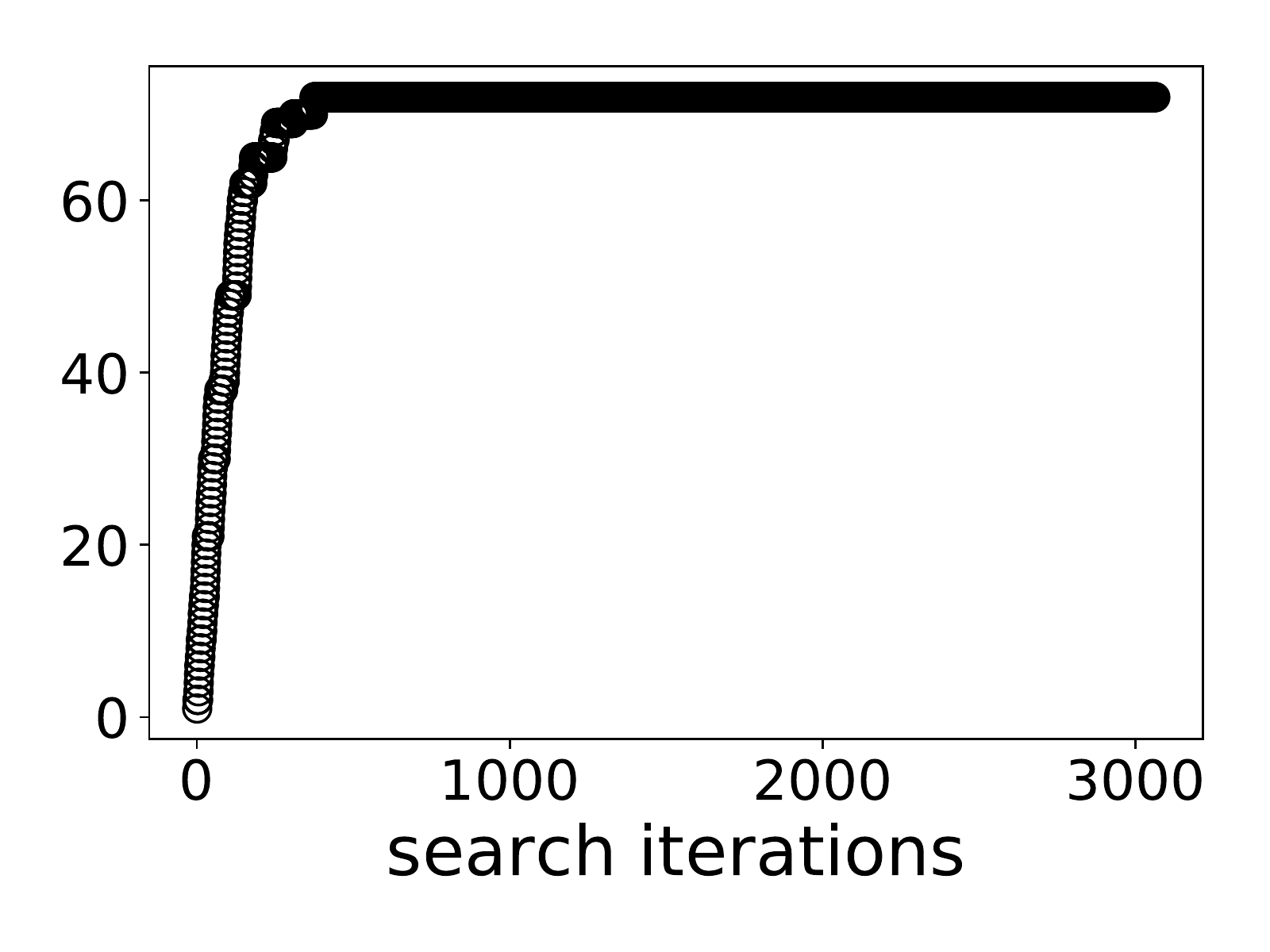} 
		\\[-2mm]
		\rotatebox{90}{\hspace*{0.3cm}{\footnotesize \gencel}}
		&\includegraphics[width=\figlength	]{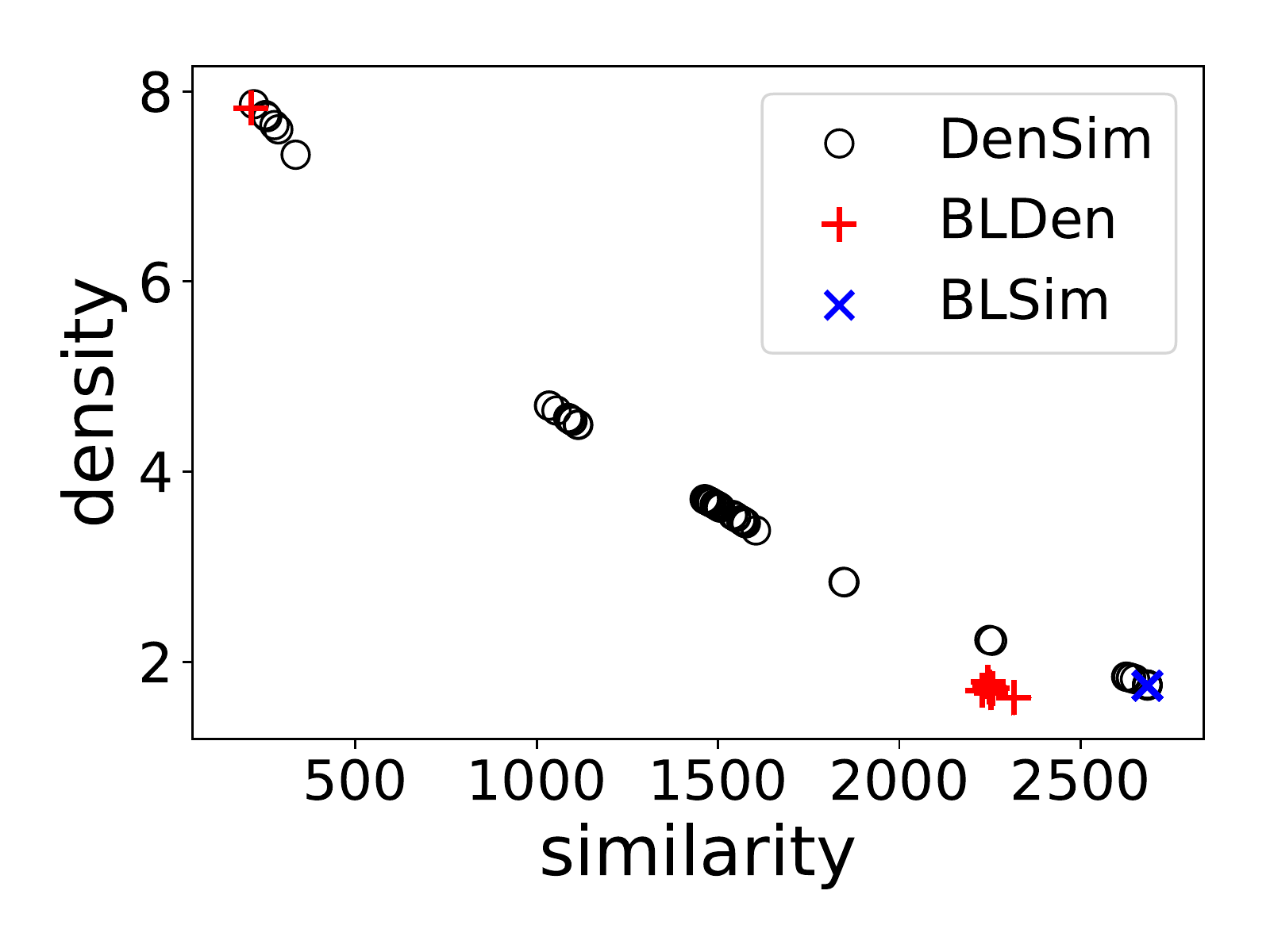}
		& \includegraphics[width=\figlength	]{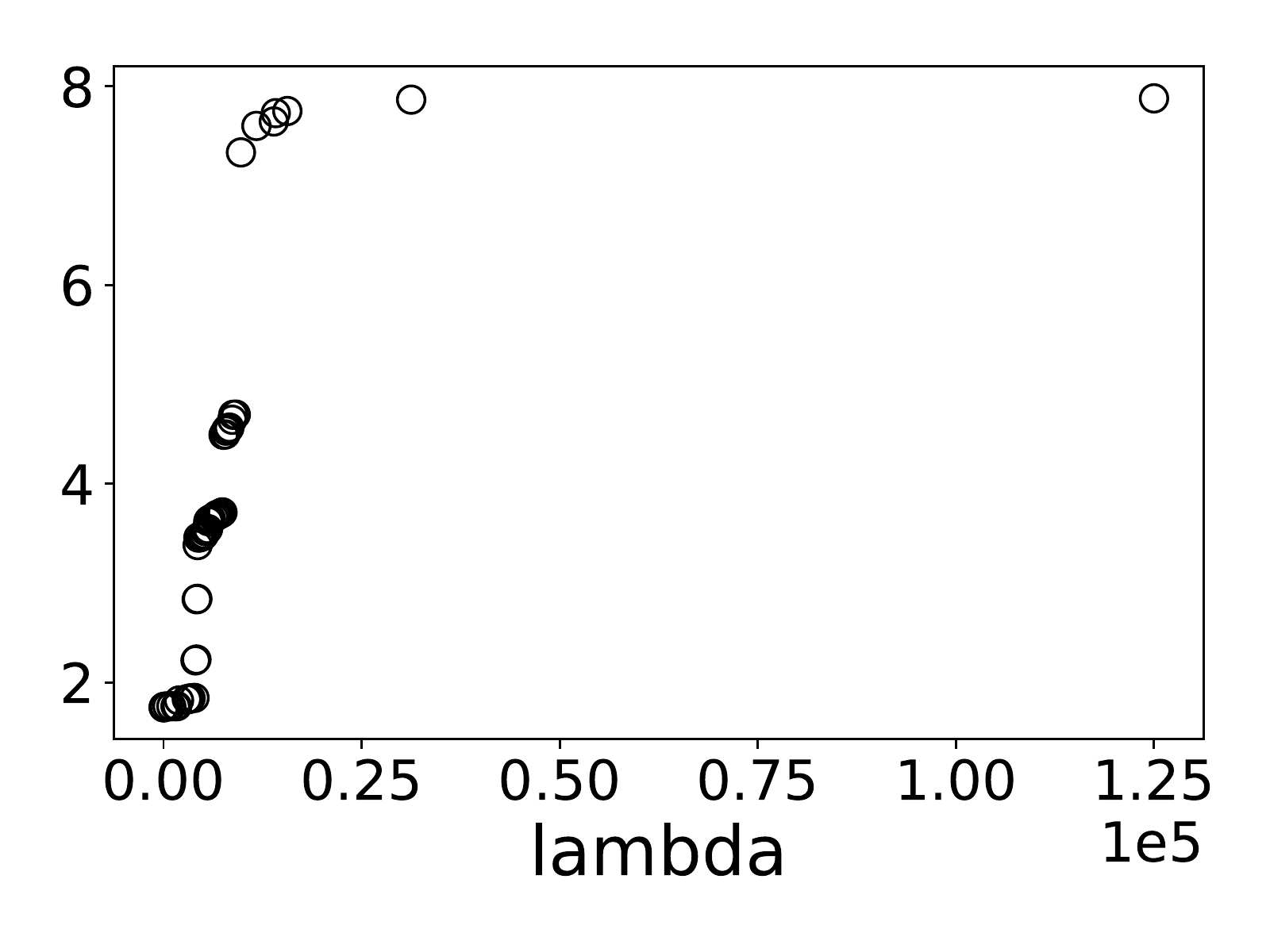}
		& \includegraphics[width=\figlength	]{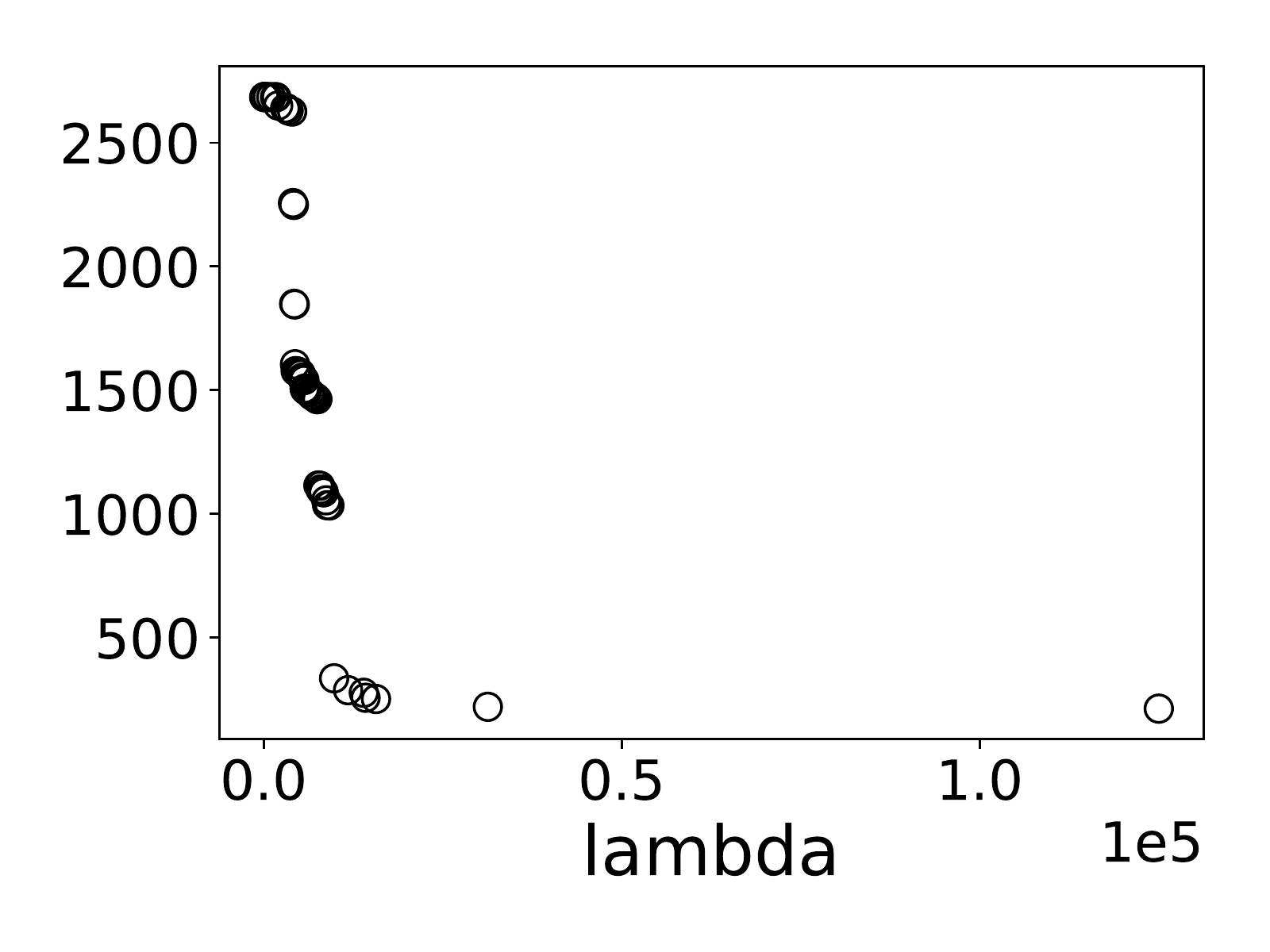}
		& \includegraphics[width=\figlength	]{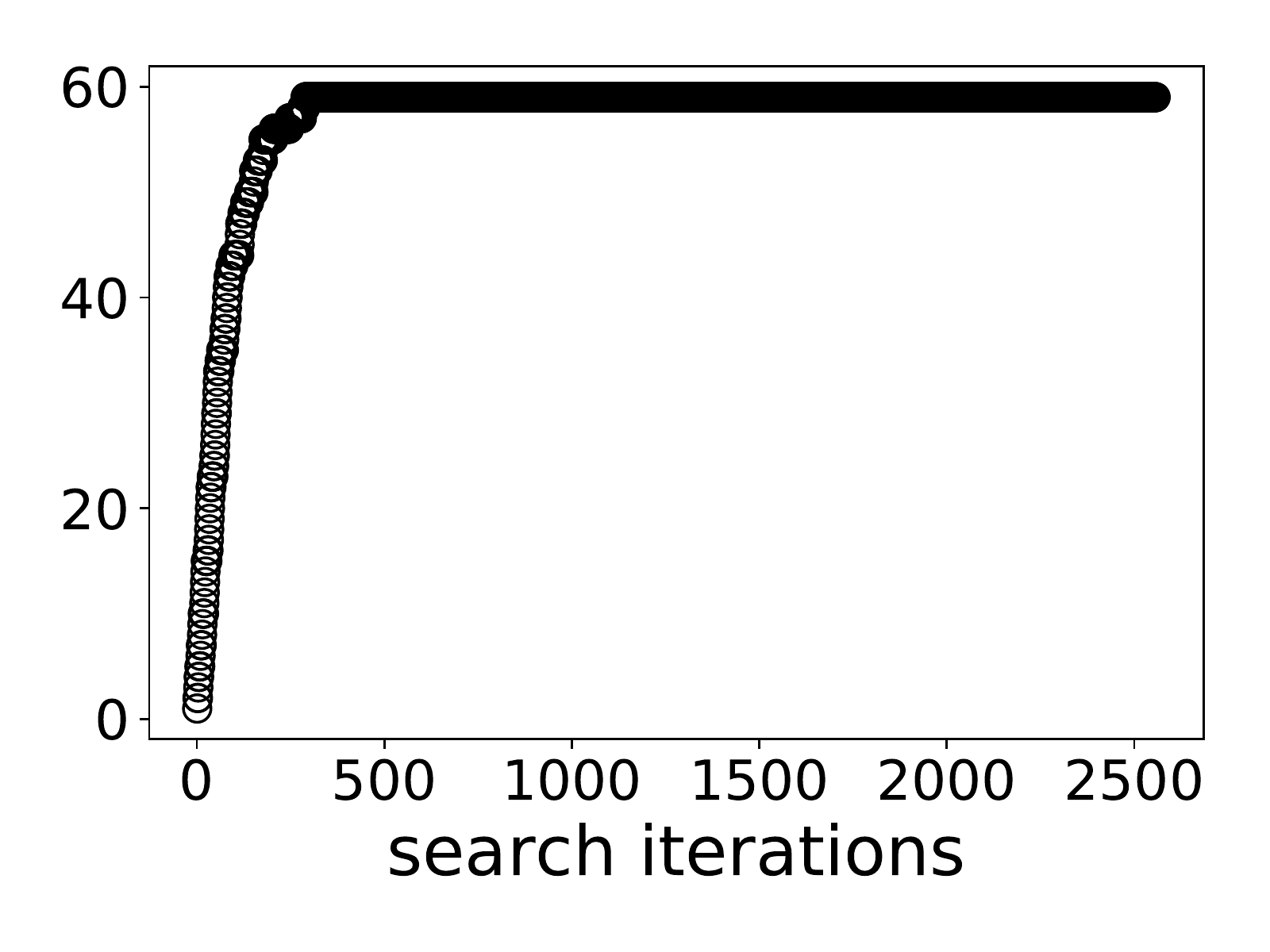} 
		\\[-2mm]
		\rotatebox{90}{\hspace*{0.2cm}{\footnotesize \genara}}
		&\includegraphics[width=\figlength]{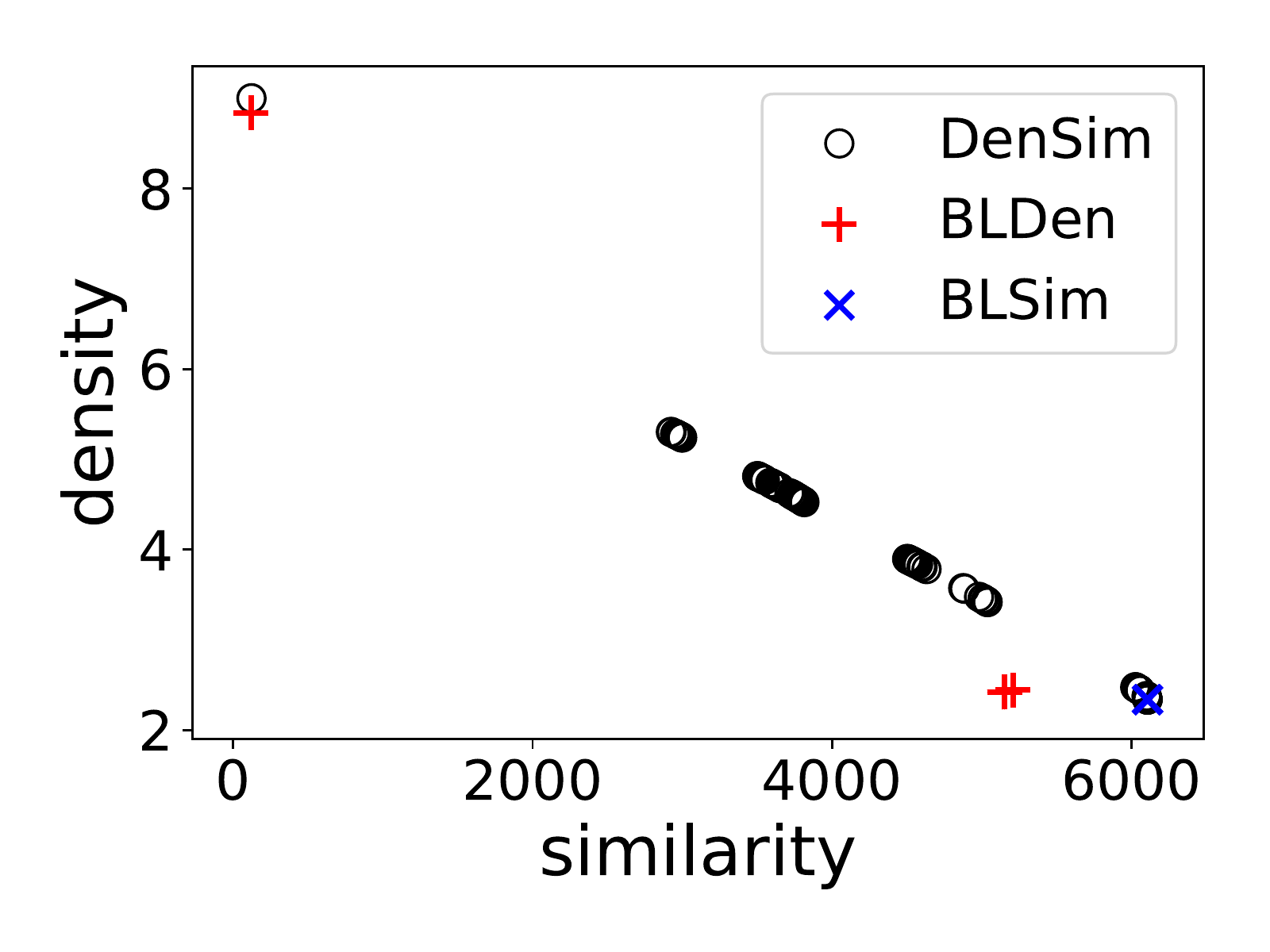}
		&\includegraphics[width=\figlength]{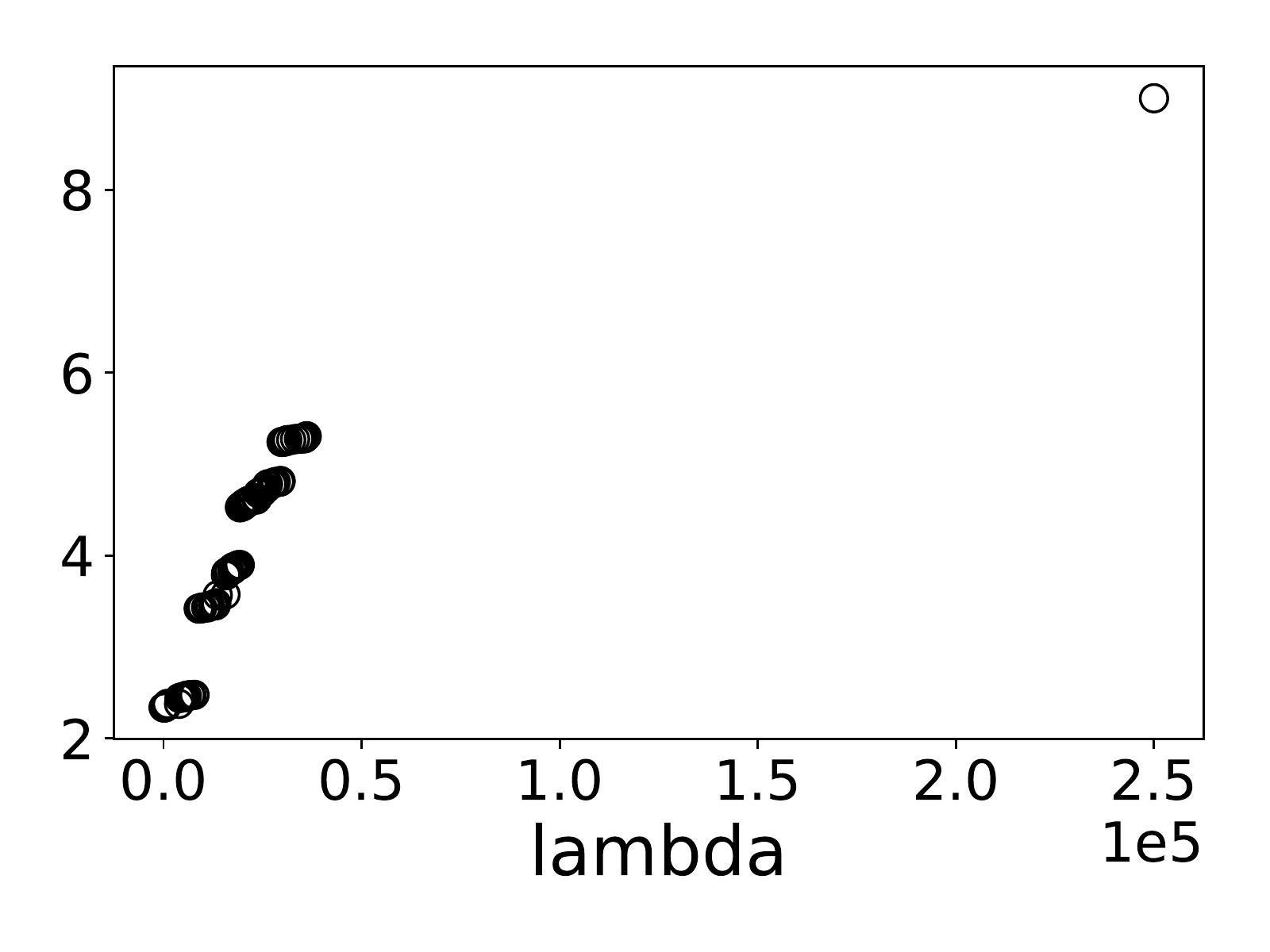}
		&\includegraphics[width=\figlength]{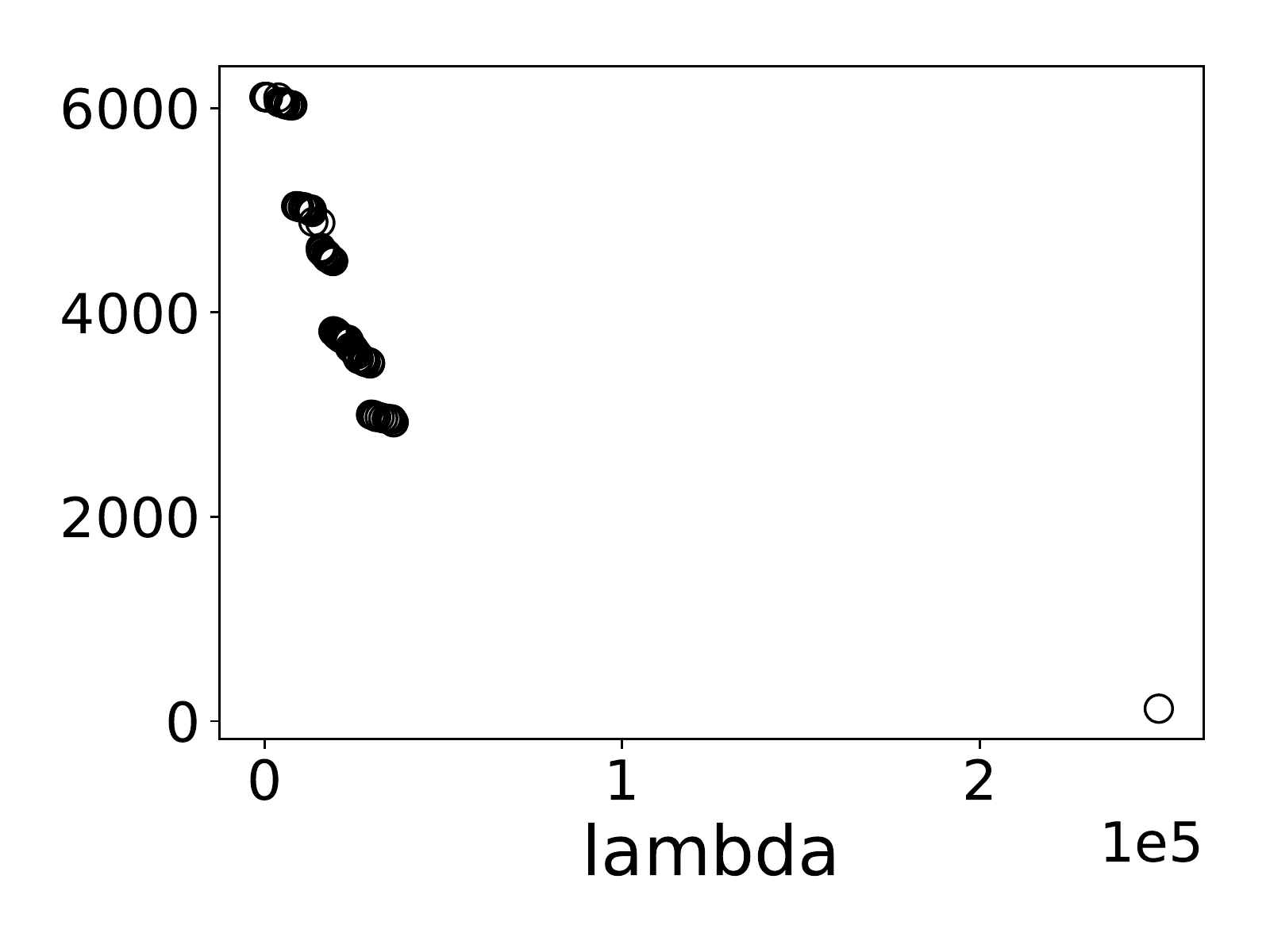}
		&\includegraphics[width=\figlength]{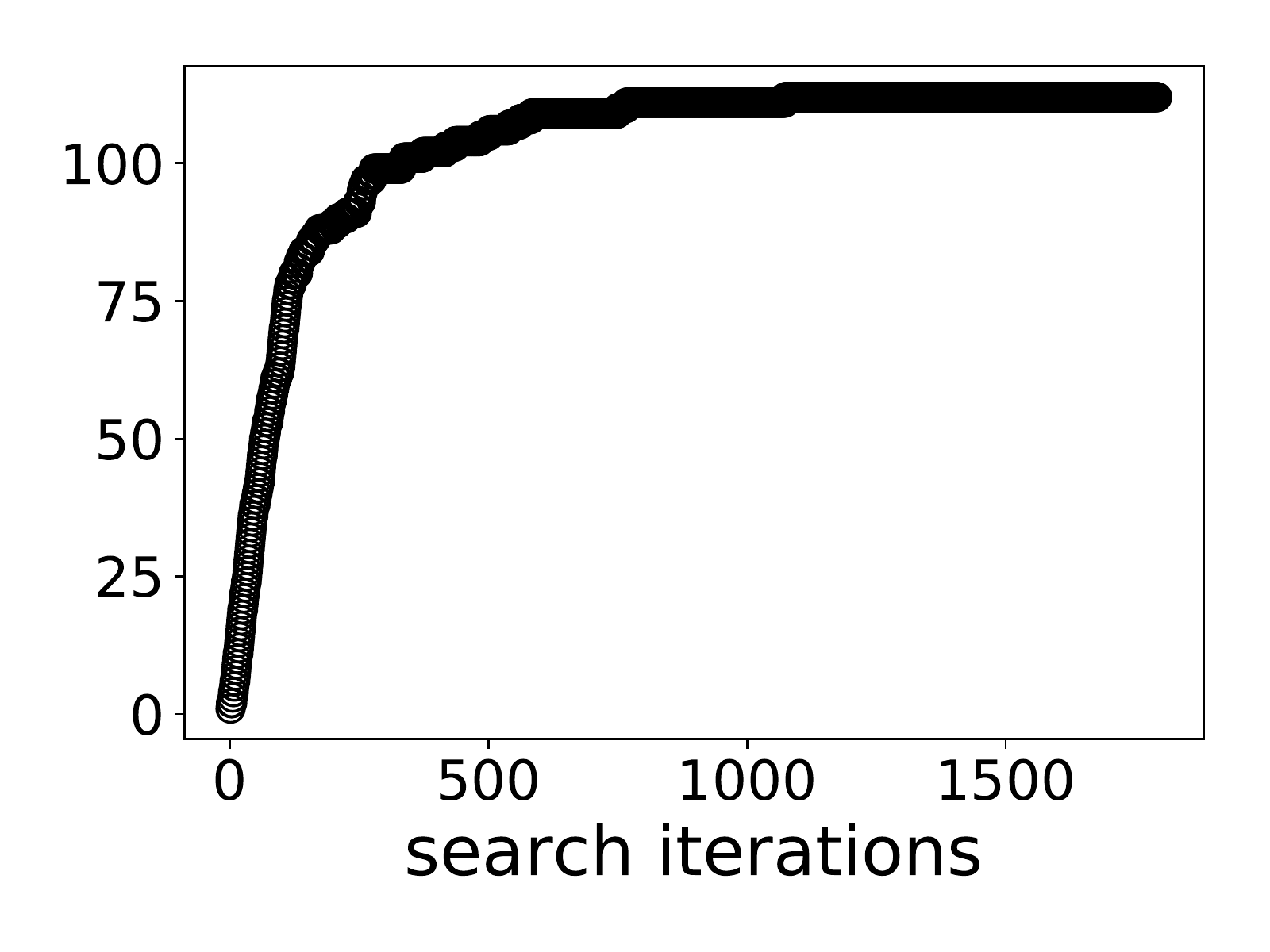} 

		\end{tabular}
		\caption{\label{fig:main}
			Experimental results for our method \ourmethod, and the baselines \bld and \bls on real-world datasets. 
			Each row represents one dataset. 
			The first column shows the values of subgraph edge similarity and density of the discovered solutions. 
			Column 1 also show the values of solutions, discovered by the baselines. 
			Columns 2 and 3 show \dns and \sml as a function of $\lambda$. 
			Column 4 shows how the number of discovered unique optimal solutions grows with the number of iterations in 
			$\lambda$-exploration.}
	\end{center}
\end{figure*}

\spara{Experimental Results.}
Figure~\ref{fig:main} shows the different characteristics of solutions discovered in the datasets during $\lambda$-exploration. 
We observe that the baselines are extremely sensitive to the values of $\gamma$: 
it is hard to find a set of $\gamma$ values that lead to distinct solutions. 
Moreover, the range and granularity of $\gamma$ depend on the datasets, 
and it is up to the end-user to decide their values.
To provide a somewhat unified comparison, we allow $\gamma$ to range  
from $0$ to $10$ with step $0.1$.

The first column in Figure~\ref{fig:main} shows the values of density and subgraph edge similarity of the solutions found. 
The solutions discovered by \ourmethod cover the space of possible values 
of similarity and density rather uniformly, providing a range of trade-offs. 
The solutions discovered by the baselines are mostly grouped around the same values and 
often dominated by solutions of \ourmethod. 
Note that the baselines successfully find the points with the largest density or similarity. 
By design, these solutions correspond to values of $\gamma = 0$, 
and they also correspond to the solutions of \ourmethod for $\lambda = \lambda_{\mathit{min}}$ and $\lambda = \lambda_{\mathit{max}}$.


Columns 2 and 3 show optimal density \dns and subgraph edge similarity \sml as functions of $\lambda$. 
As expected, larger values of $\lambda$ correspond to solutions with larger density and smaller similarity.
The range of $\lambda$ that gives unique optimal solutions is dataset-dependent 
and not uniform. 
However, due to the monotonicity property we can search for these values efficiently,
in contrast to the na\"ive search for the baselines.
The last column of Figure~\ref{fig:main} shows the efficiency of $\lambda$-exploration. 
All unique solutions are found after 200 to 1000 iterations. 

In Table~\ref{tab:time} we report the number of the unique optimal solutions and running times. 
The total running time varies from seconds to hours.
We should highlight, however, that finding a solution for a single value of $\lambda$ 
takes on average $20$ seconds for the largest dataset. 
Thus, if the search progresses fast (as shown in the last row of Figure~\ref{fig:main})
and a sufficient number of optimal solutions have been found, 
we can terminate the search.
As we discussed before, we implement $\lambda$-exploration as a BFS, 
so that at any point we have a diverse set of $\lambda$ tested.
It is worth noting that \FPalgo converges in about $5$ iterations on average, and
thus the \mincut algorithm is not run many times.

\begin{table}[!t]
	\begin{center} 
		\caption{Number of subgraphs and running time characteristics. 
		$|\pareto|$: number of discovered optimal solutions; 
		$t$(s): total time (seconds) for the search; 
		$I_{\lambda}$: number of tested values of $\lambda$; 
		$t_{\lambda}$(s): average time to test one value of $\lambda$; 
		$I_{\mathit{MC}}$: average number of \mincut problems solved for one $\lambda$ 
		(i.e., average number of iterations in \FPalgo).}		
		\label{tab:time}
		\setlength{\tabcolsep}{0pt}
		\vspace{-2mm}
		\begin{tabular*}{\textwidth}{@{\extracolsep{\fill}}lrrrrr}
			\toprule 			
			Dataset & $|\pareto|$ & $t$(s) &$I_{\lambda}$ & $t_{\lambda}$(s) & $I_{\mathit{MC}}$\\
			\midrule						
			\cs & 15 & 2.26 & 465 & 0.003 & 2.89\\ 
			\air & 74 & 314 & 2770 & 0.069 & 6.00\\
			\concel & 72 & 1015 & 3064 & 0.244 & 4.60 \\
			\gencel & 59 & 10159 & 2561 & 3.075 & 4.72\\
			\genara & 112 & 43200 & 1794 & 20.540 & 5.68\\
			\bottomrule
		\end{tabular*}
	\end{center}
\end{table}

\begin{figure}[!t]
	\begin{center}
	\includegraphics[width=0.4\columnwidth]{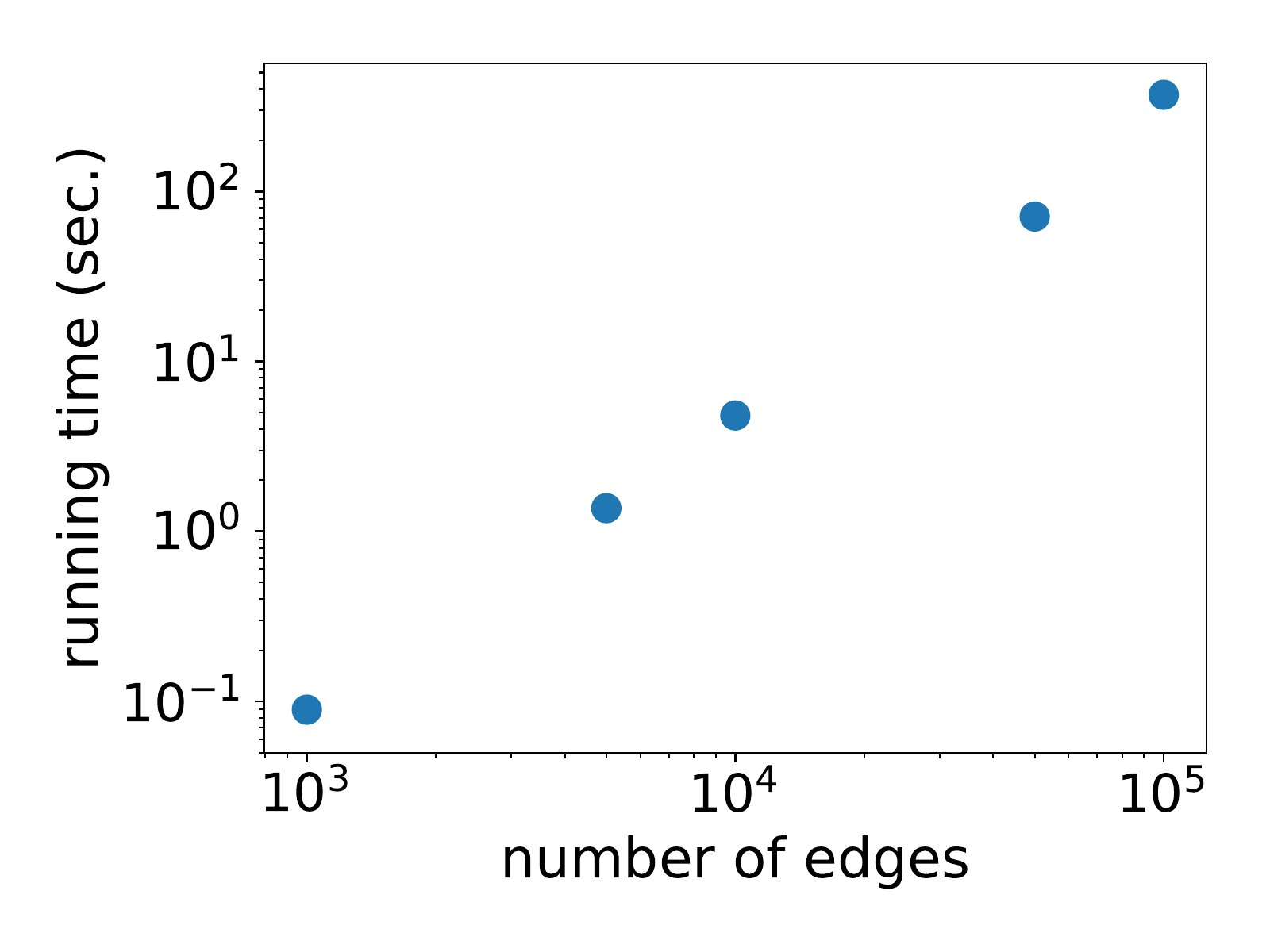} 
	\caption{Running time in seconds to calculate $10$ first solutions. The input graph is a $G_{n,m}$ random graph with $n=1000$ nodes, $m$ edges, the probability that a pair of edges has a non-zero similarity is set to $0.001$.}\label{fig:scalability}
	\end{center}
\end{figure}

\spara{Scalability.}
In order to test the scalability of \ourmethod we generate a number of random graphs with $n=1000$ nodes and varying $m$ number of edges.
We draw the graphs from a random graph model $G_{n,m}$. The similarities between the edges are random values from $(0,1]$, and the probability that a pair of edges has a non-zero similarity is set to $0.001$. We run \ourmethod until it found $10$ solutions, and the running time is reported in Figure~\ref{fig:scalability}.
It took $6$ minutes to find $10$ solutions for the largest graph with $100000$ edges.

\begin{figure*}[!h]
	\begin{center}
		\setlength{\tabcolsep}{0pt}
		\setlength{\figlength}{0.15\textheight}
		\begin{tabular}{@{\hspace{-2mm}}cccc}	
			&$\lambda_{\mathit{min}}$ (max similarity)& $\lambda_{\mathit{med}}$ (trade-off) & $\lambda_{\mathit{max}}$ (max density)\\ 
			\rotatebox{90}{\hspace*{0.7cm}lunch}			                            
			& \includegraphics[height=\figlength	]{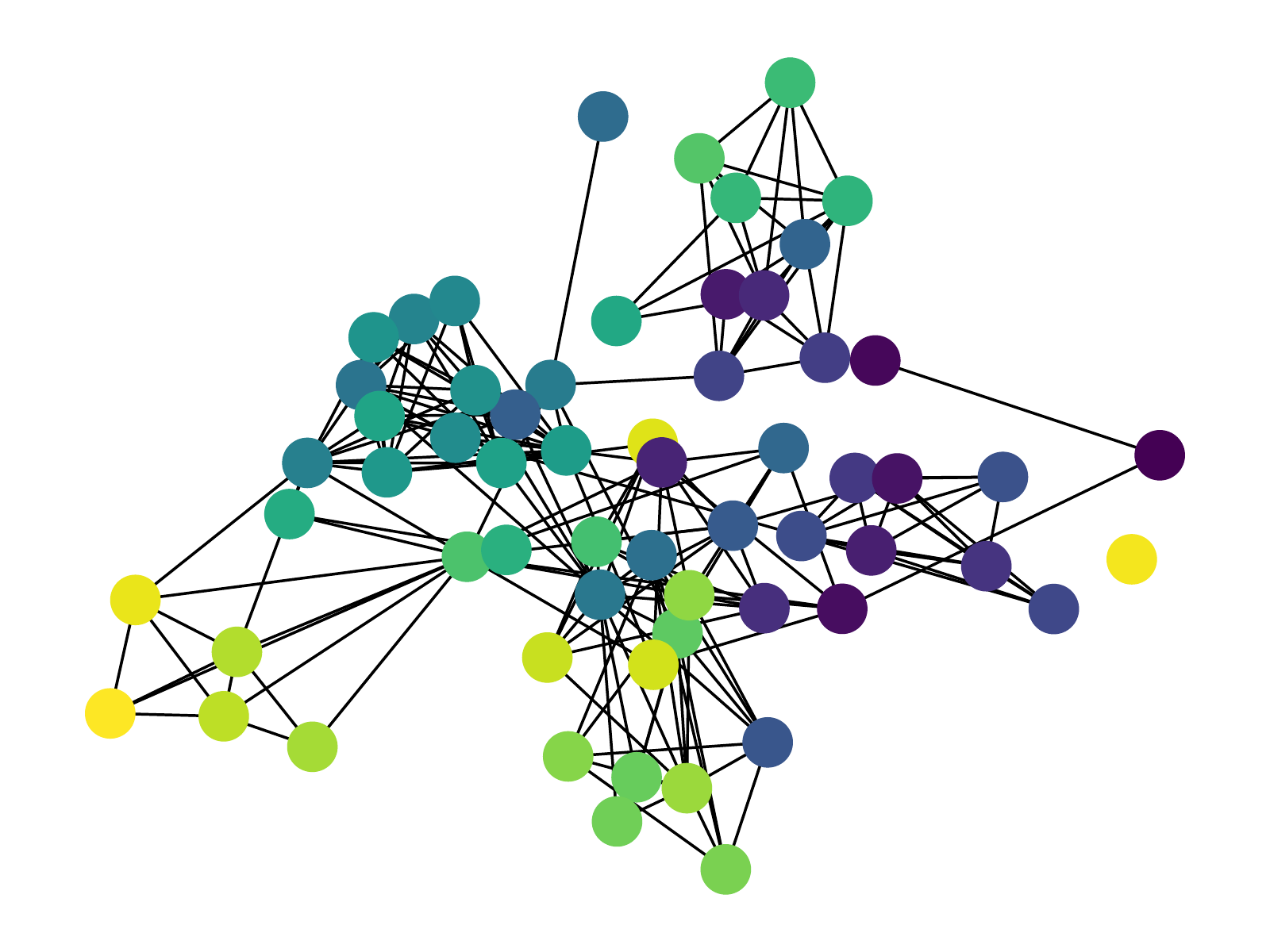} 
			& \includegraphics[height=\figlength	]{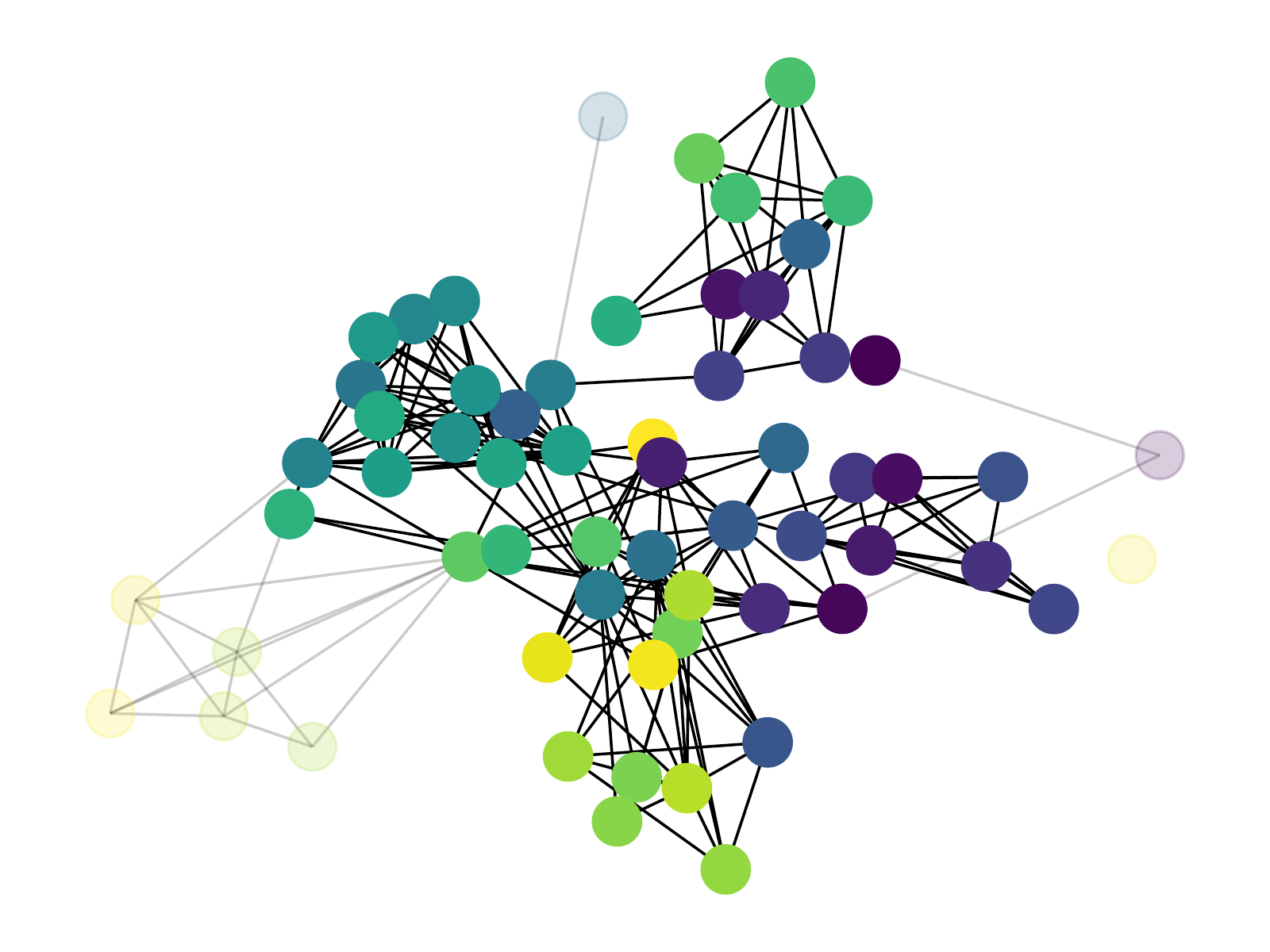} 
			& \includegraphics[height=\figlength	]{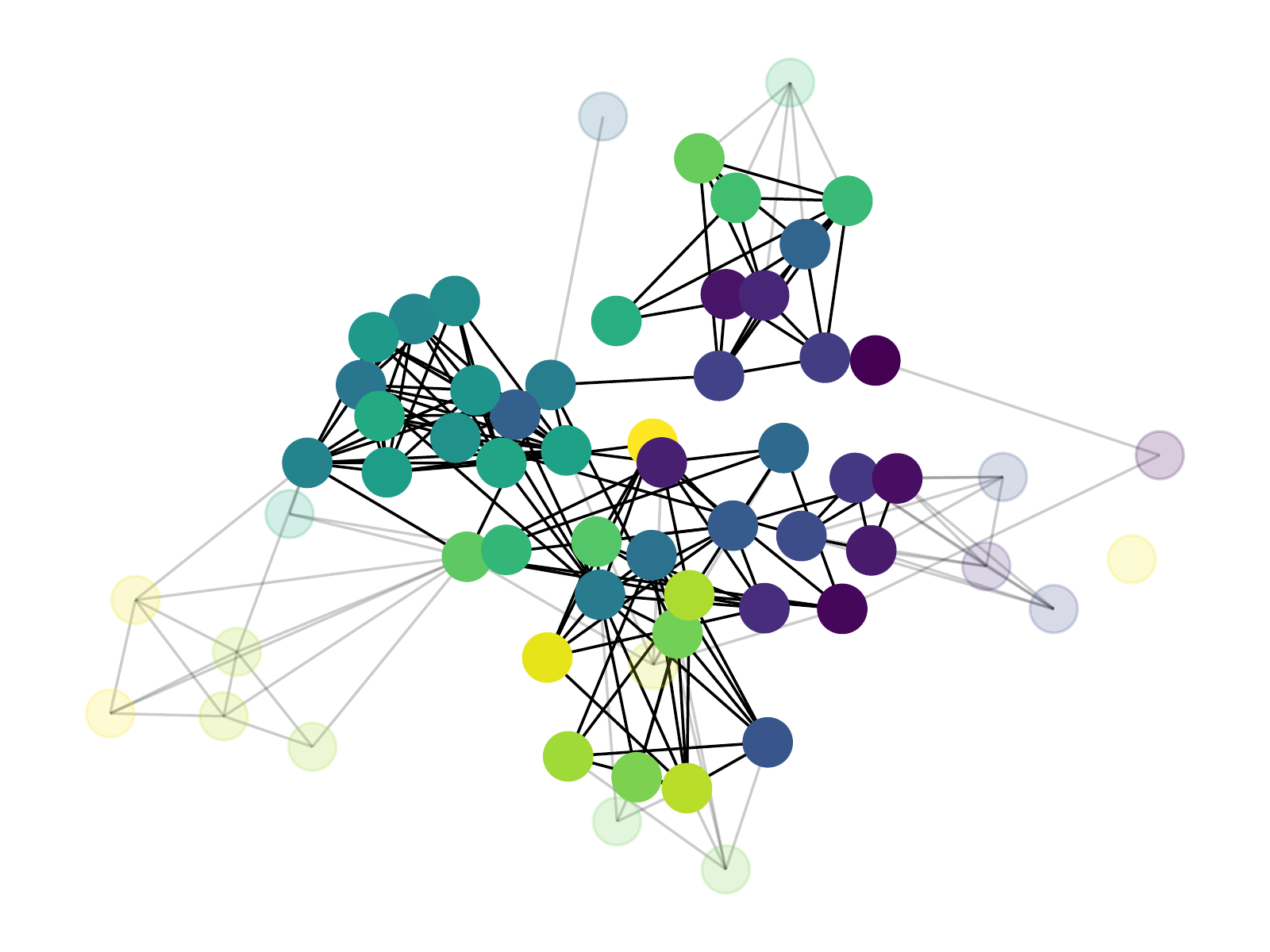} \\ 
			\rotatebox{90}{\hspace*{0.7cm}facebook} 
			& \includegraphics[height=\figlength	]{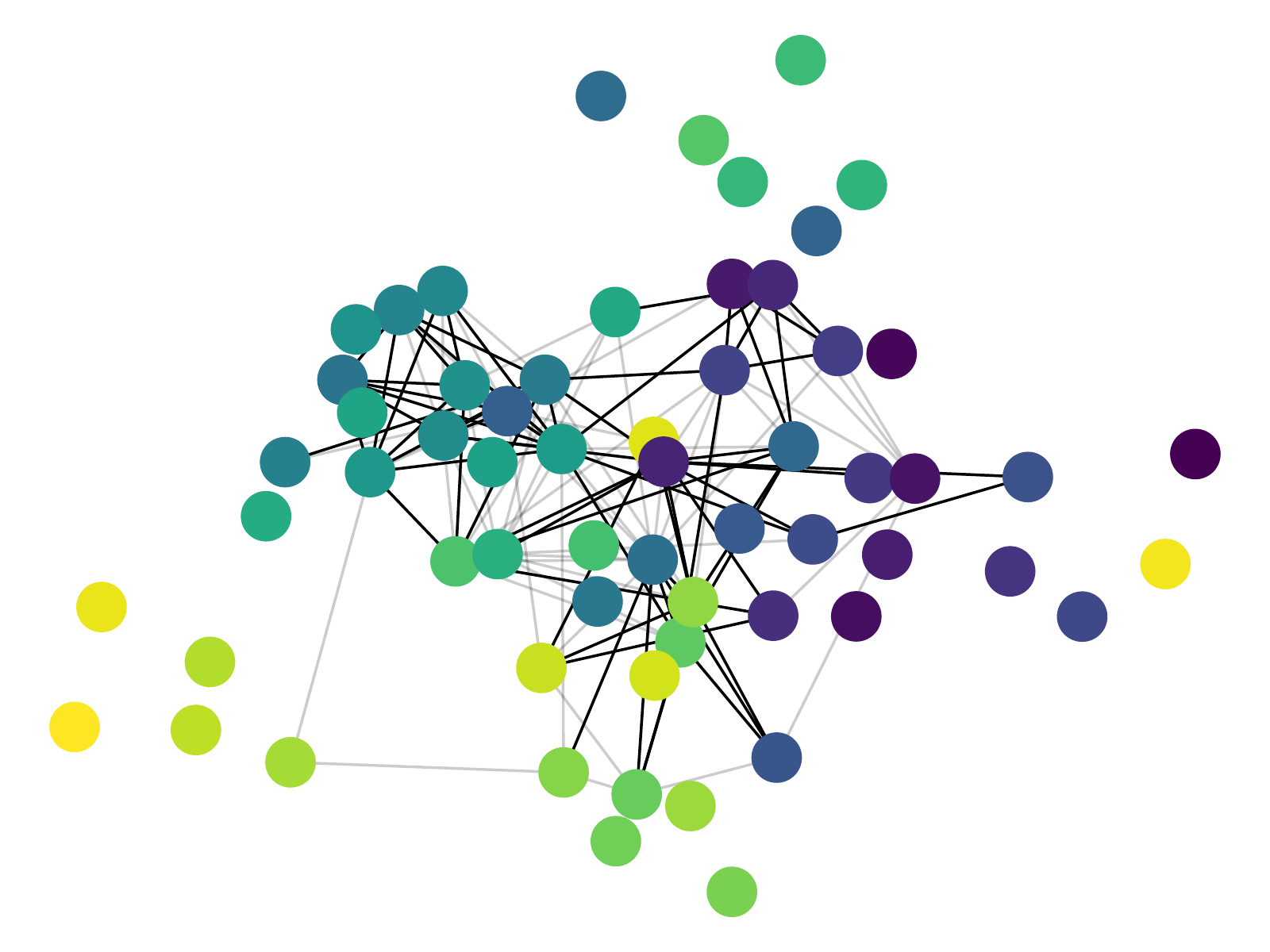} 
			& \includegraphics[height=\figlength	]{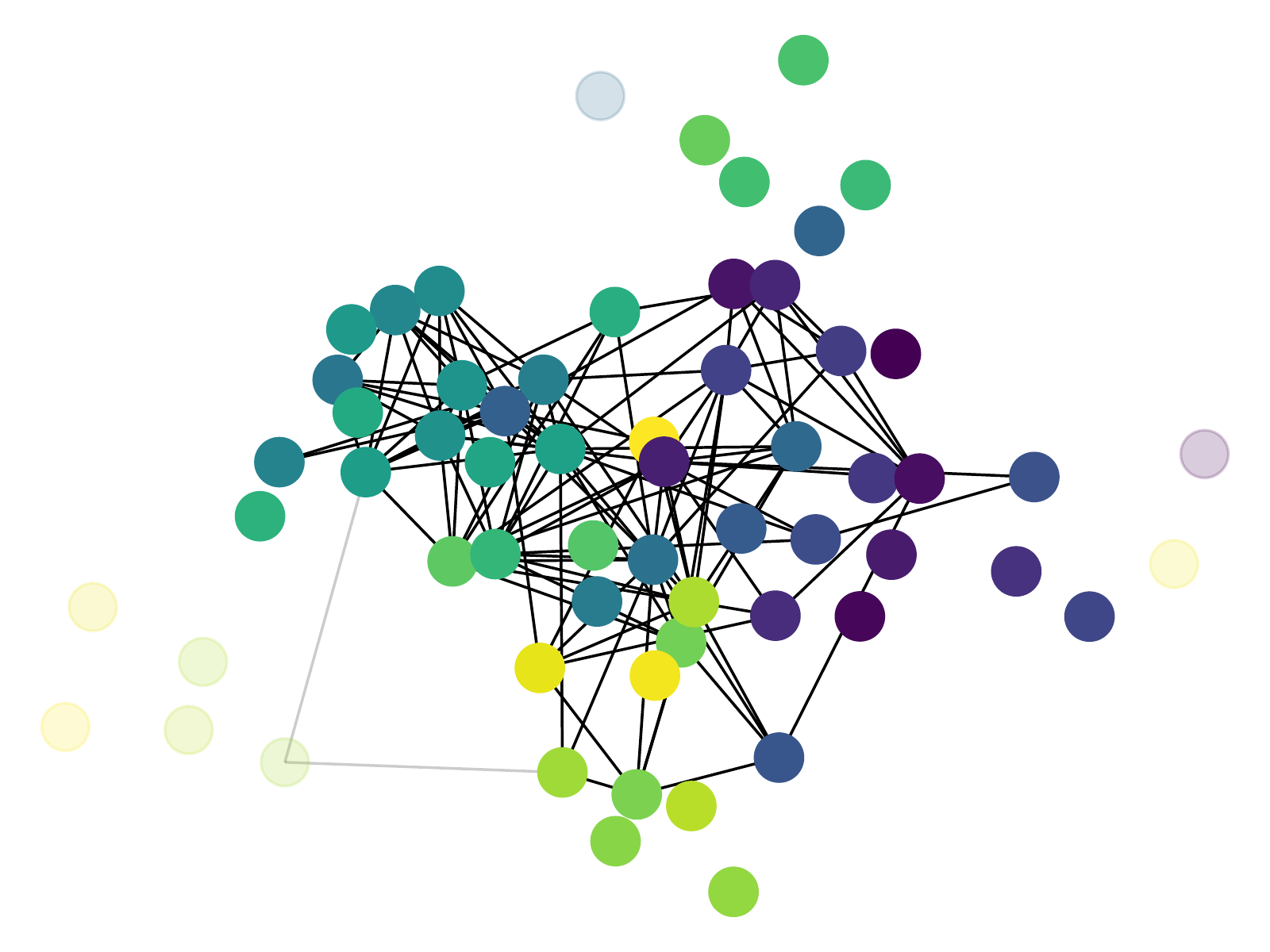} 
			& \includegraphics[height=\figlength	]{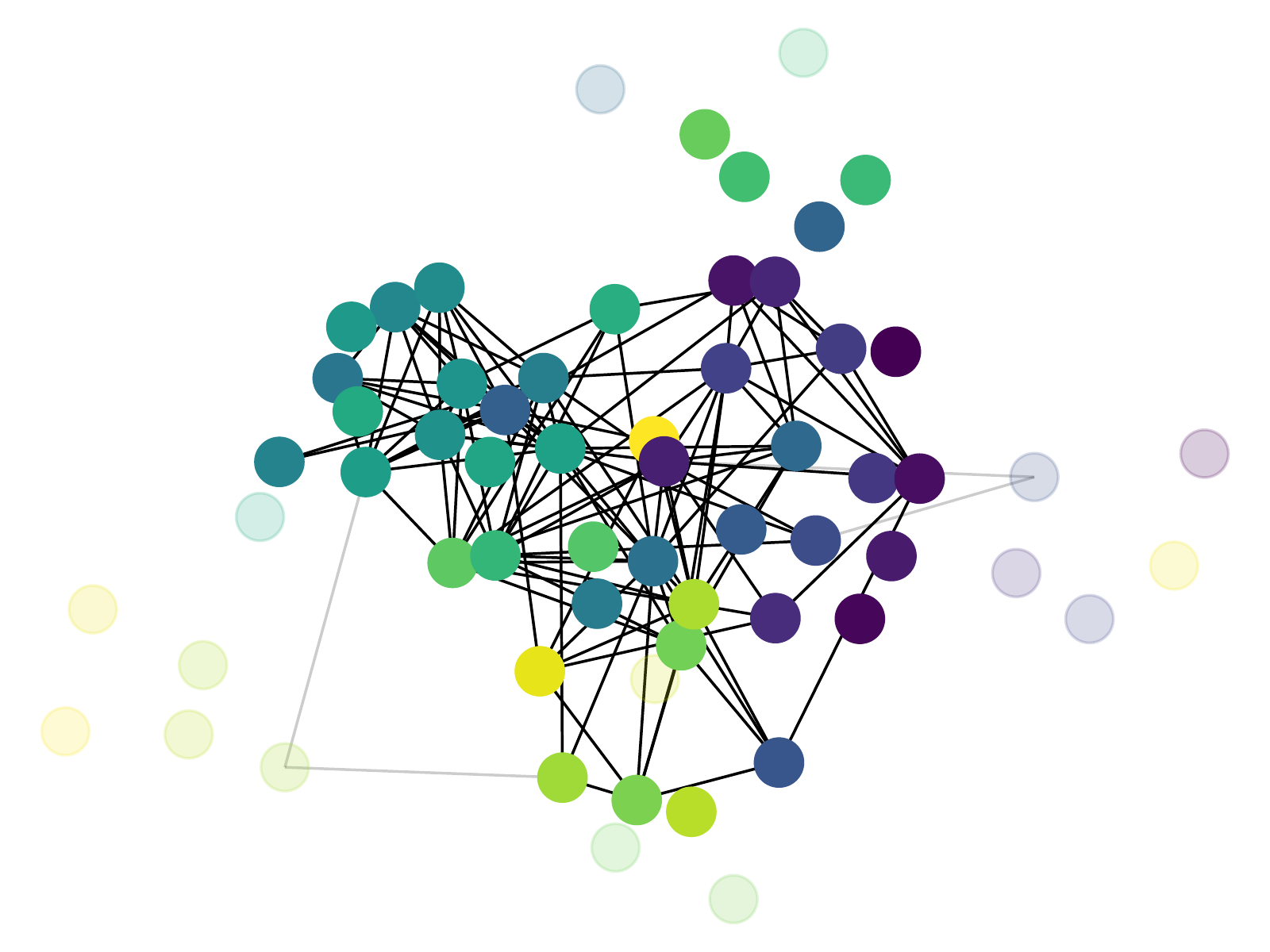}\\
			\rotatebox{90}{\hspace*{0.7cm}coauthor}
			& \includegraphics[height=\figlength	]{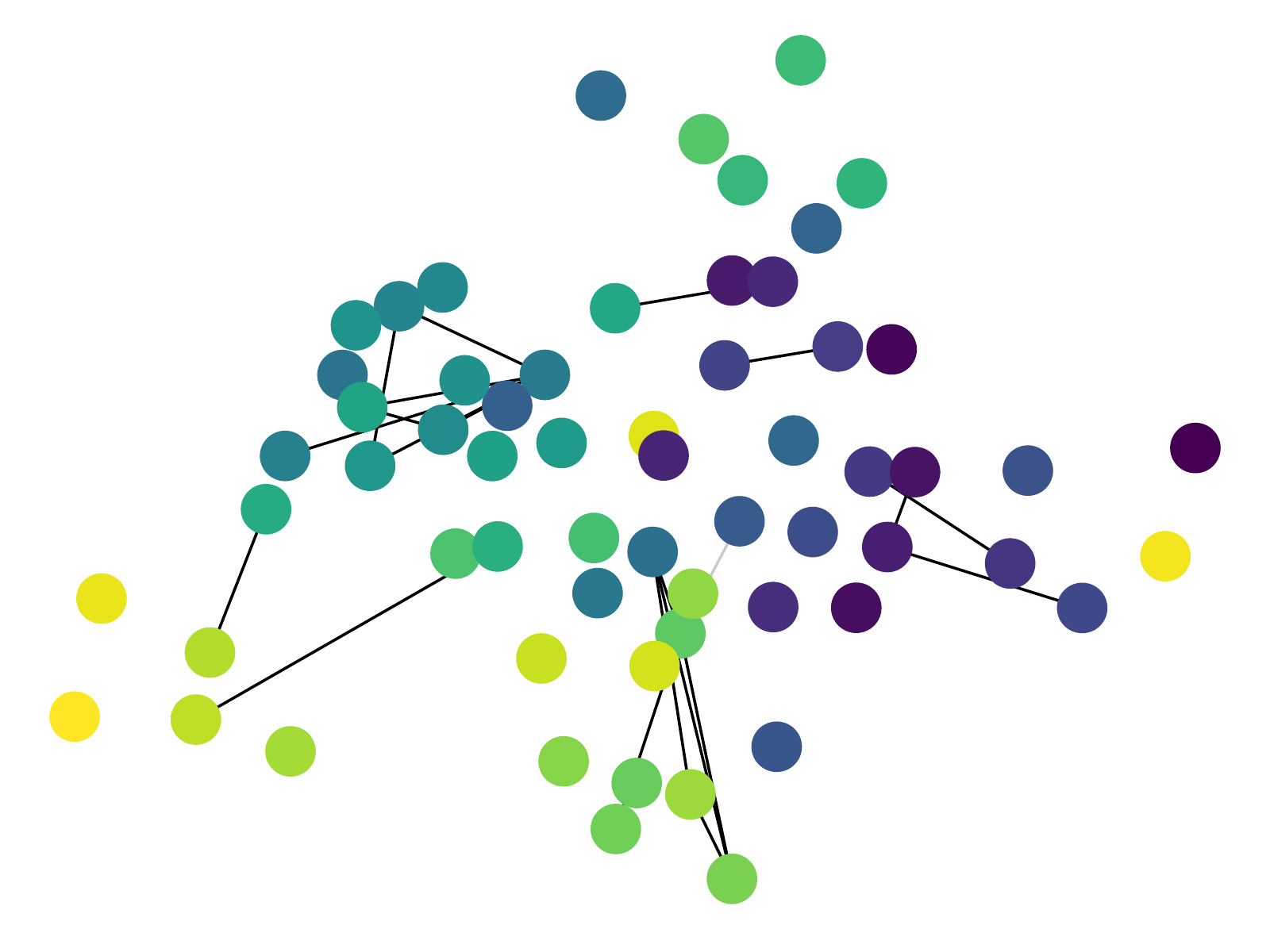} 
			& \includegraphics[height=\figlength	]{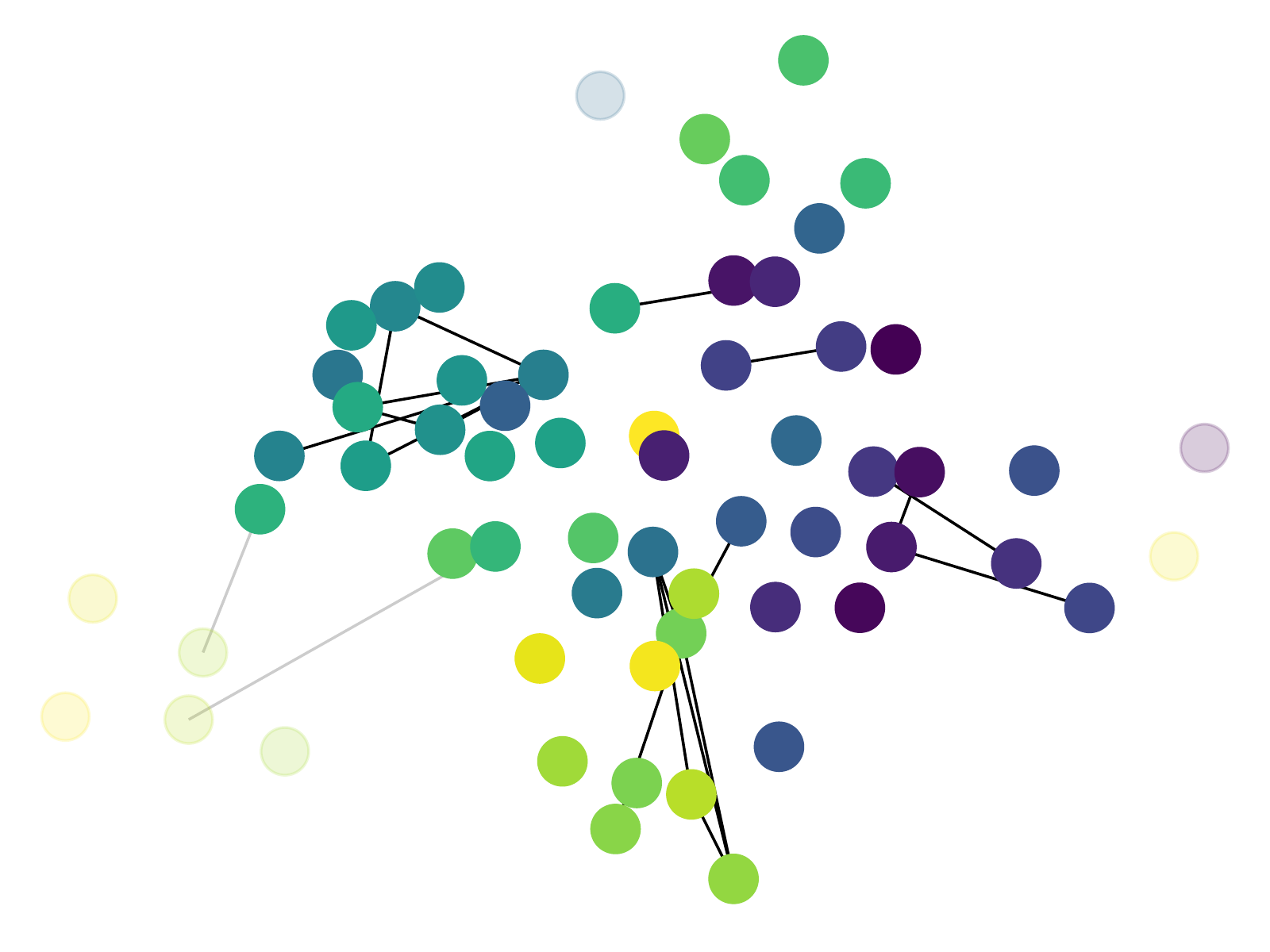} 
			& \includegraphics[height=\figlength	]{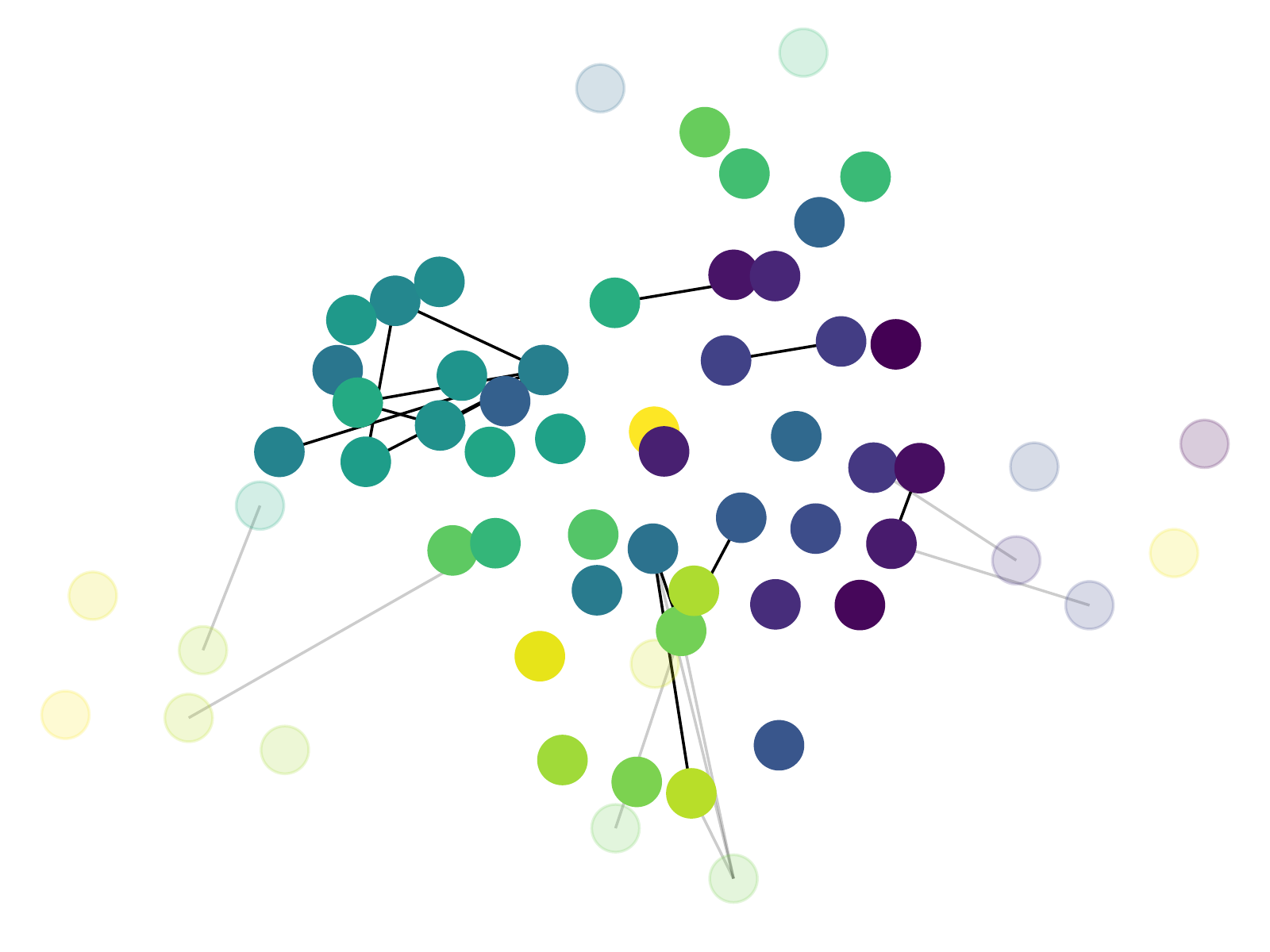}\\
			\rotatebox{90}{\hspace*{0.7cm}leisure}
			& \includegraphics[height=\figlength	]{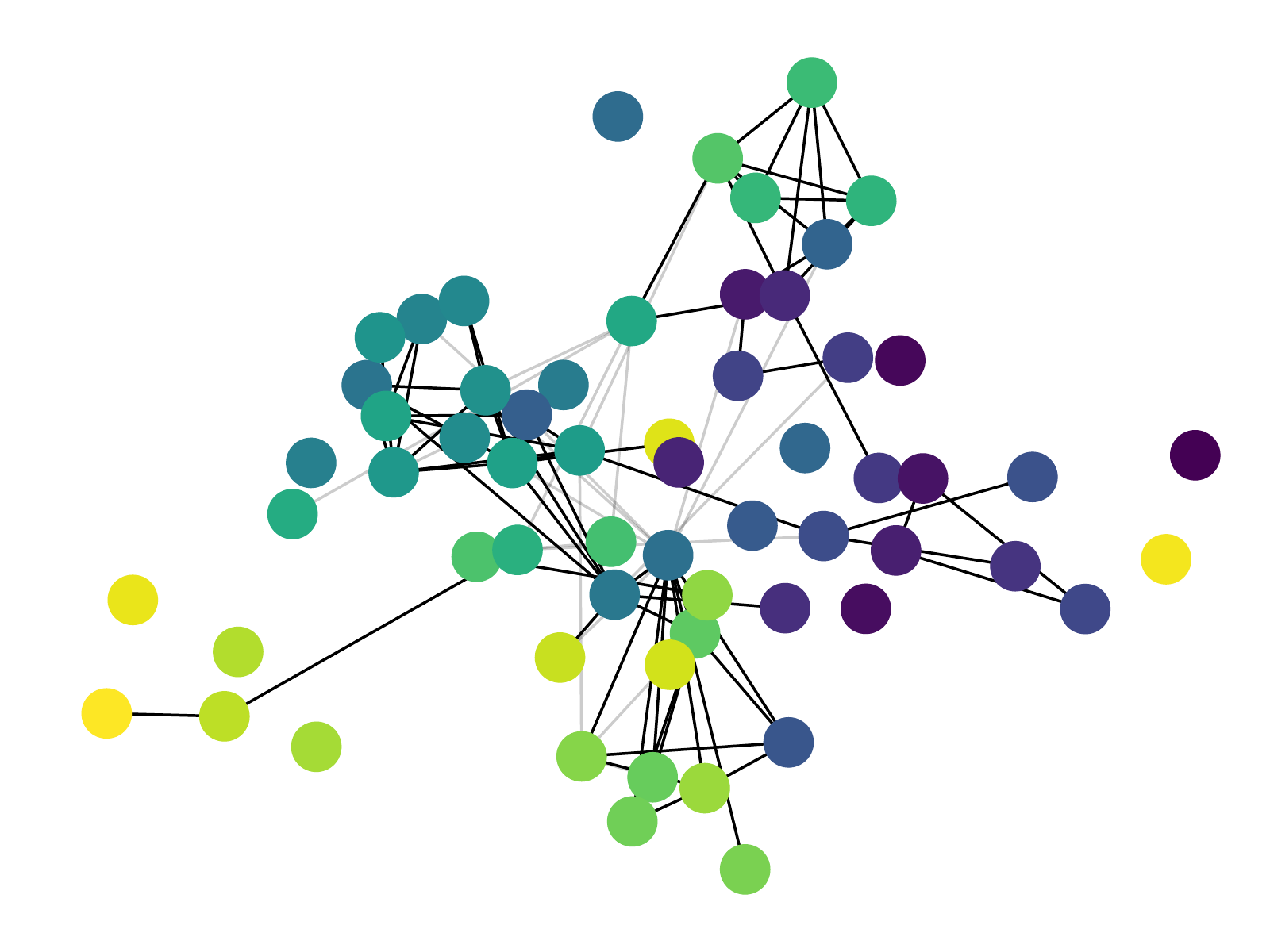} 
			& \includegraphics[height=\figlength	]{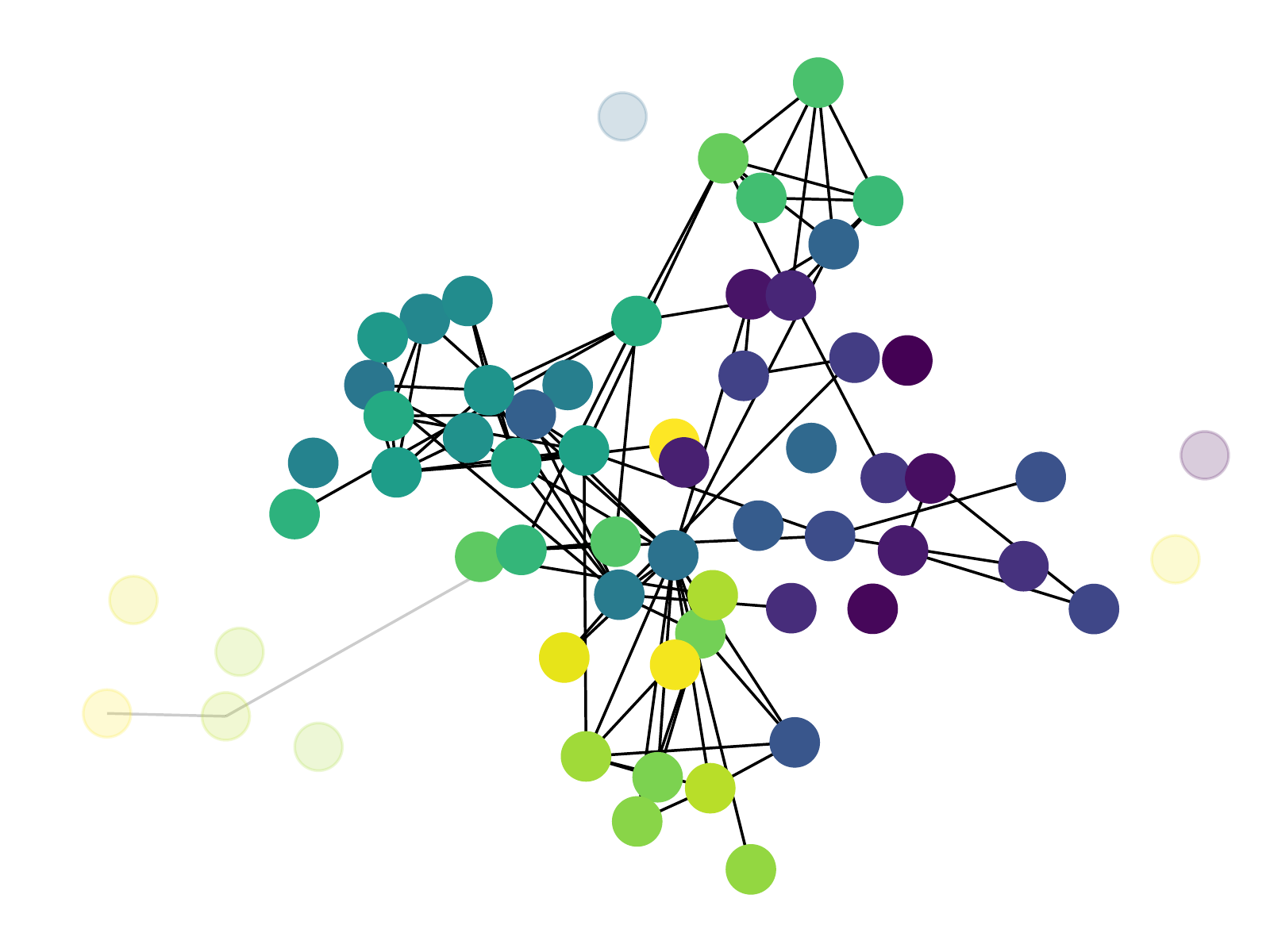} 
			& \includegraphics[height=\figlength	]{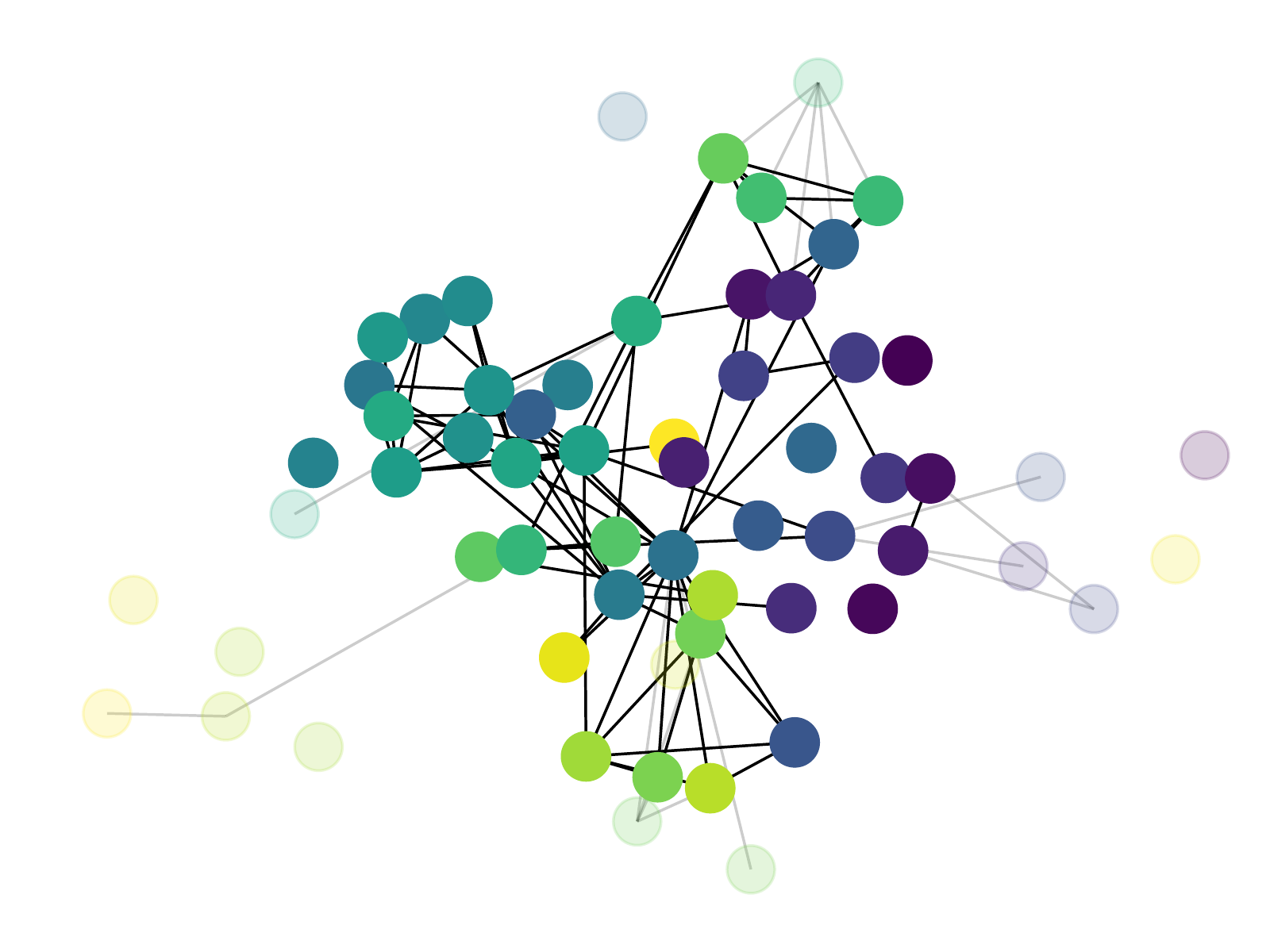} \\
			\rotatebox{90}{\hspace*{0.7cm}work} 
			& \includegraphics[height=\figlength	]{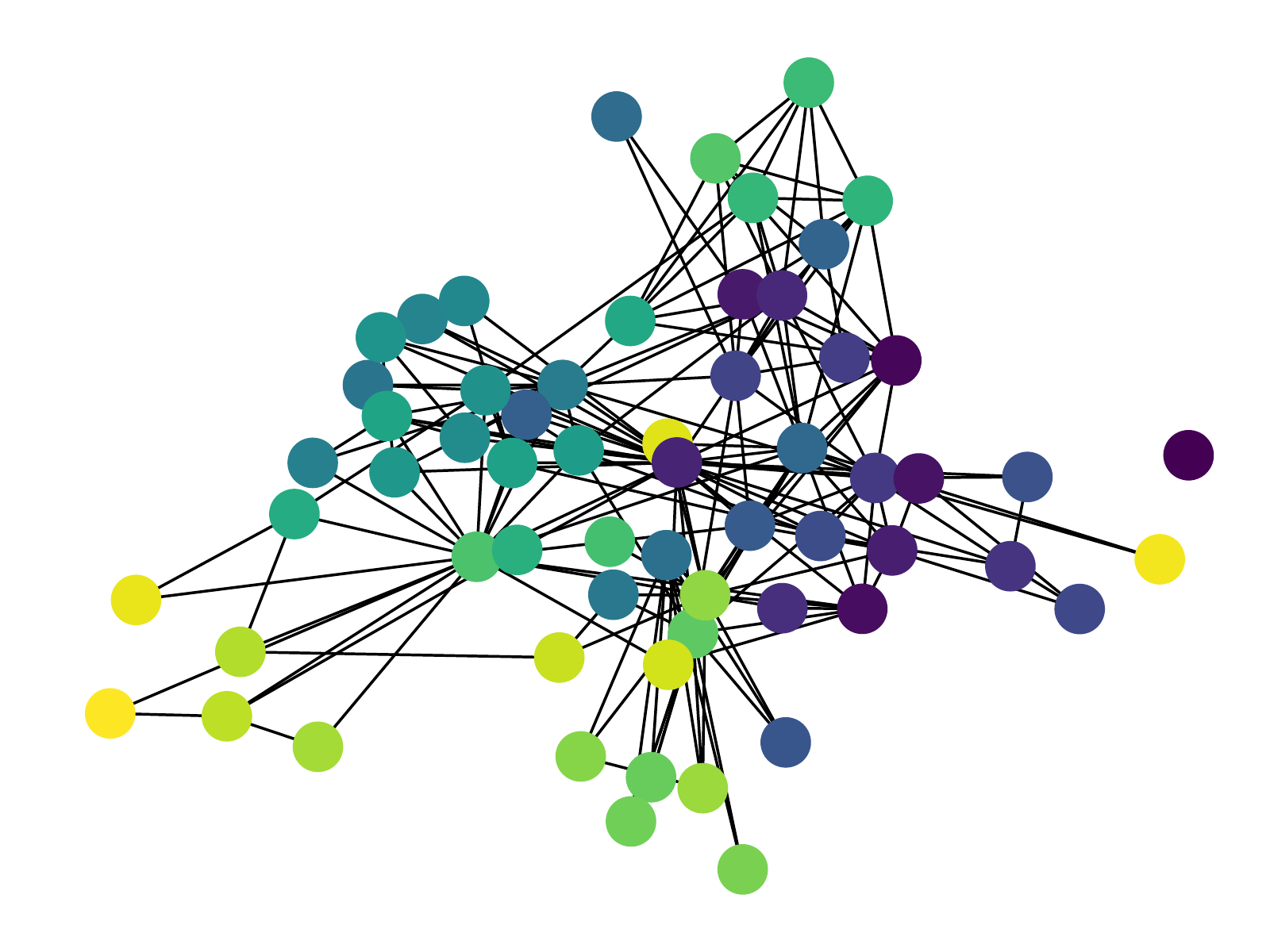} 
			& \includegraphics[height=\figlength	]{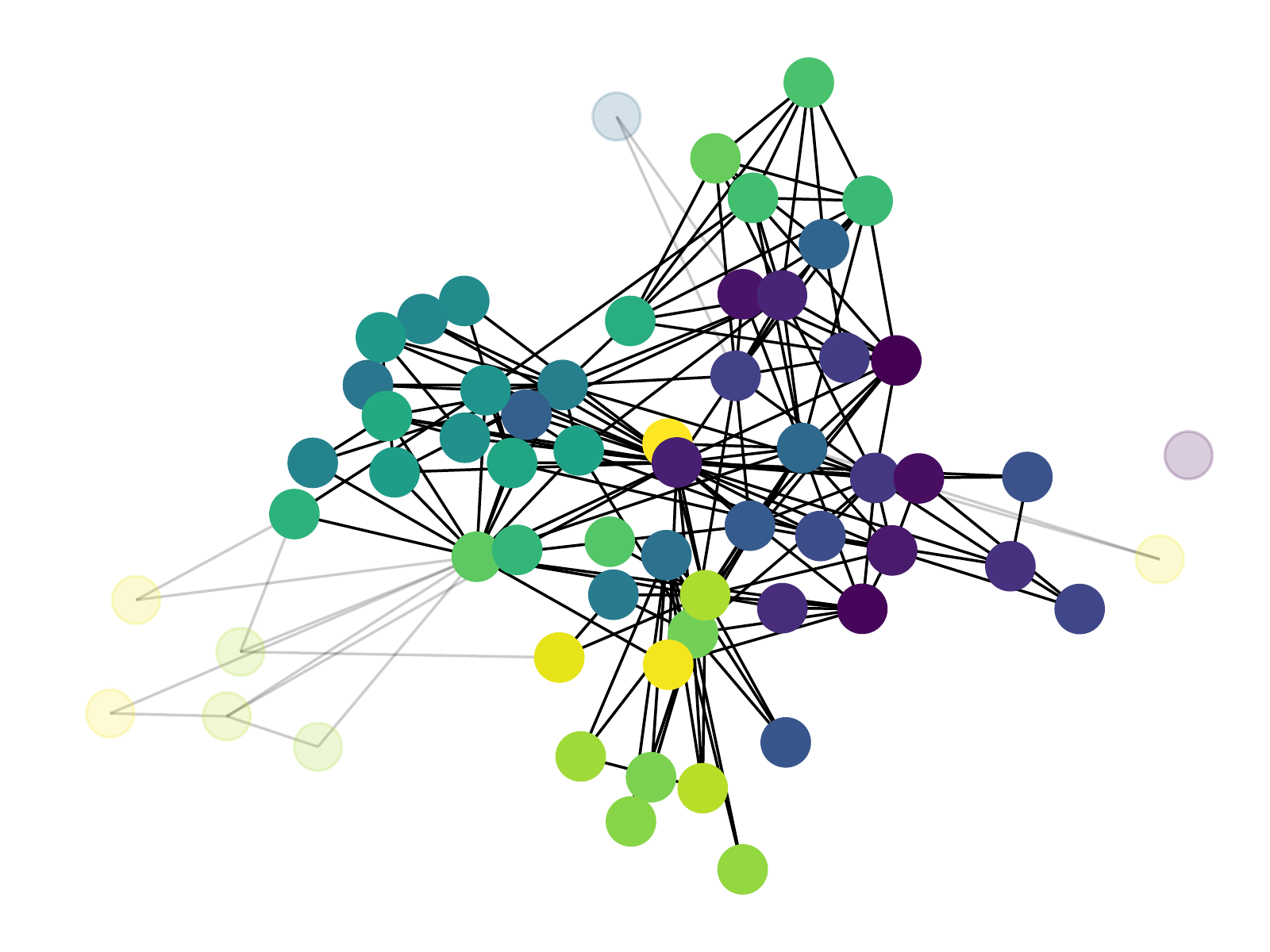} 
			& \includegraphics[height=\figlength	]{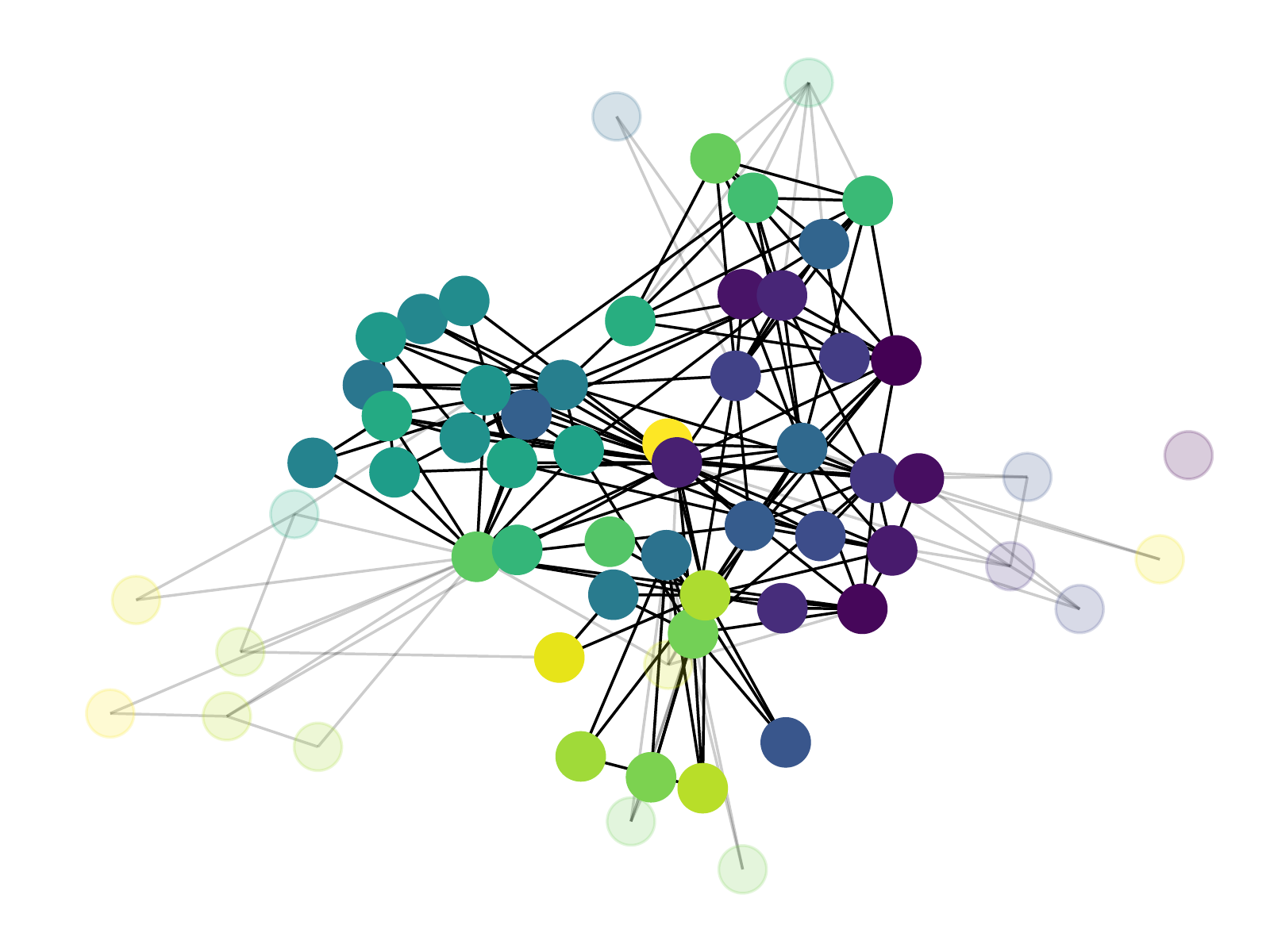} \\
		\end{tabular}
		\caption{Some solutions output by \ourmethod for \cs dataset. Each column corresponds to a solution with the smallest (left), largest (right), and median (center) value of $\lambda$. Each row corresponds to a layer in the dataset. The nodes and edges, included into the solution are shown with the bright colors, the rest of the nodes and edges on each layer are drawn transparently.}\label{fig:case}
	\end{center}
\end{figure*}

\spara{Case Study.}
We run \ourmethod on the \cs dataset.
We pick three of the solutions discovered: 
one for $\lambda_{\mathit{min}}$, 
one for $\lambda_{\mathit{max}}$, and 
one for the median value $\lambda_{\mathit{med}}$.
Recall that $\lambda_{\mathit{min}}$ gives a solution with maximum subgraph edge similarity, 
while density is ignored; 
while $\lambda_{\mathit{max}}$ gives a solution with maximum density.
ignoring edge similarity. 
Any other $\lambda_{\mathit{med}}$ should provide some balance between these extremes.
The solutions are visualized in Figure~\ref{fig:case}.

The graph maximizing the subgraph similarity ($\lambda_{\mathit{min}}$) 
includes all the edges from the layers of ``work'' and ``lunch.'' 
Since the dataset contains relationships between the employees of the same university department, 
it is intuitive that these two layers define the edge set with the largest subgraph edge similarity. 
All network nodes are included in this solution,
as all these people share similar interactions at work and lunch. 
Facebook and leisure interactions, not overlapping with ``work'' and ``lunch'', 
are excluded, as they are localized in their layers. 
The resulting graph contains 61 nodes and 289 edges, 
while the subgraph edge similarity is 59.43 and the density is~4.73.

The graph maximizing the density ($\lambda_{\mathit{max}}$) 
includes the edges of the densest subgraph from the ``work'' layer, and 
reinforces it by adding edges from other layers. 
The graph contains 45 nodes and 281 edges, and it 
is the smallest of the three. 
The subgraph edge similarity is 44.83 and the density is 6.24.

The trade-off graph ($\lambda_{\mathit{med}}$) 
selects 325 edges, more than the other two, while it has 53 nodes. 
Its subgraph similarity is 52.64 and its density 6.13. 
The graph resembles the one for $\lambda_{\mathit{max}}$, 
but adds interactions that decrease the density while increasing the subgraph similarity. 


\section{Concluding Remarks}\label{conclusion}

In this paper we study a novel graph-mining problem, 
where the goal is to find a set of edges that 
maximize the density of the edge-induced subgraph and the subgraph edge similarity.
We reformulate the problem as a non-standard Lagrangian relaxation and develop a novel efficient algorithm to solve the relaxation based on parametric minimum-cut~\cite{gallo1989fast, hochbaum2008pseudoflow}. We pro\-vide an efficient search strategy through the values of Lagrangian multipliers. The approach is evaluated on real-world datasets and compared against intuitive baselines. 

\section*{Acknowledgments}
This research was partially supported by the National Research Foundation, Singapore under its AI Singapore Programme (AISG Award No: AISG-GC-2019-001).
Aristides Gionis is supported by three Academy of Finland projects (286211, 313927, 317085),
the ERC Advanced Grant REBOUND (834862),
the EC H2020 RIA project ``SoBigData++'' (871042), and the
Wallenberg AI, Autonomous Systems and Software Program (WASP).
The funders had no role in study design, data collection and analysis, decision to publish, or preparation of the manuscript. 


\bibliographystyle{splncs04}

\newpage
\appendix
\section{Appendix}
\label{append}

Proof of Proposition~1.
\begin{proof}
	
	First, observe that an empty edge set $X_0=\emptyset$ cannot be an optimal solution of Problem~\LRES, since $\OLR(\emptyset\mid\mu) = 0$ and any one-edge set $X_1$ has strictly positive value $\OLR(X_1\mid\mu) > 0$. Thus, even with the constraint $|X|\geq 1$ in Problem~$\LRID$, every optimal solution of Problem~$\LRES$ is in the feasible set of Problem~$\LRID$.
	
	Now we start with the if-statement. Let $\esetopt$ be a solution of Problem~$\LRES$ for a fixed $\mu$. Let us show that $\esetopt$ is also a solution of Problem~$\LRID$ for $\lambda=\dns^2(\esetopt)\mu$.
	
	Since $\esetopt$ is an optimum, for all $\eset\subseteq\edges$ it holds that:  
	\begin{equation}
	\label{eq1}
	\sml(\esetopt)-\sml(\eset)\geq \mu(\dns(\eset)-\dns(\esetopt)).
	\end{equation}
	We need to show that $\OID(\esetopt\mid \lambda)\geq \OID(\eset\mid \lambda)$ for all $\eset\subseteq\edges$. This can be written as $\sml(\esetopt)-\sml(\eset)\geq \dns^2(\esetopt)\mu(-1/\dns(\eset)+1/\dns(\esetopt))$ or
	\begin{equation}
	\label{eq2}
	\sml(\esetopt)-\sml(\eset)\geq \frac{\dns(\esetopt)}{\dns(\eset)}\mu(\dns(\eset)-\dns(\esetopt)).
	\end{equation}
	Now we consider two cases:\\
	
	Case (i). Let $\dns(\esetopt)\geq\dns(\eset)$. Since $\dns(\esetopt)>0$ and $\dns(\eset)>0$, then $\mu(\dns(\eset)-\dns(\esetopt))\geq \frac{\dns(\esetopt)}{\dns(\eset)}\mu(\dns(\eset)-\dns(\esetopt))$ and from Inequality~\ref{eq1} it follows that Inequality~(\ref{eq2}) holds.\\
	Case (ii). Let $\dns(\esetopt)<\dns(\eset)$. Still $\mu(\dns(\eset)-\dns(\esetopt))\geq \frac{\dns(\esetopt)}{\dns(\eset)}\mu(\dns(\eset)-\dns(\esetopt))$ and Inequality~(\ref{eq2}) holds.
	Thus, the if-statement is true.
	
	To prove the only-if-statement, we can apply an identical argumentation, after substituting $\dns()$ with $\neginvd()$ and swapping $\mu$ with $\lambda$. This proves that if $\esetopt$ is a solution to Problem~$\LRID$ for a fixed $\lambda$, then $\esetopt$ is also a solution to Problem~$\LRES$ for
	$\mu = \neginvd^2(\esetopt)\lambda = 1/\dns^2(\esetopt)\lambda$.

	
\end{proof}

Proof of Proposition~3.
\begin{proof}
	Let us consider the minimum cut set \usetopt in \flowgr.
	The set~\usetopt in \flowgr corresponds to a set of edges $\edges(\usetopt) \equiv \usetopt\cap \uset_\edges$ 
	and a set of nodes $\nodes(\usetopt) \equiv \usetopt\cap \uset_\nodes$ in \graph.
	
	We first show that the edges $\edges(\usetopt)$ cover all the nodes $\nodes(\usetopt)$ in \graph and thus $(\nodes(\usetopt), \edges(\usetopt))$ is a valid subgraph in $\graph$.
	If a node $u_e$ belongs to $\usetopt$, 
	then the nodes $u_v$ for which $v$ is an end-point of $e$ also belong to \usetopt; 
	otherwise the weight of~\usetopt would be infinite 
	(which cannot be, as the cut $\{\src\}$ is finite). 
	Next, if all $u_e$, such that $v$ is an end-point of $e$, belong to $\usetoptc$, 
	then $u_v$ also must be in $\usetoptc$, as this decreases the cost of the cut by $\lambda$. Similarly, if $u_v \in \usetopt$ for some $v\in\nodes$ 
	then there exists $u_e$ 
	with $e\in\edges$ and $e$ being an end-point of $v$, 
	that also belongs to \usetopt; otherwise moving $u_v$ to $\usetoptc$ would reduce the cost by $\lambda$.
	This proves our claim that 
	the edges $\edges(\usetopt)$ cover all the nodes $\nodes(\usetopt)$ in \graph.
	
	Let $\edges(\usetoptc) = \edges\setminus \edges(\usetopt)$. 
	We show the equivalence of \mincut problem and the {\Qproblem}.
	The cost $\optmccost(c,\lambda)$ of the minimum cut in \flowgr is
	\[
	\optmccost(c,\lambda) = 
	\frac{1}{2}\!\!\sum_{\substack{e\in \edges(\usetopt)\\d\in \edges(\usetoptc)}}\!\!\esim(e,d) +  
	\frac{1}{2}\!\!\sum_{e\in \edges(\usetoptc)}\sum_{\substack{d\in E\\d\ne e}}\esim(e, d) - 
	c|\edges(\usetoptc)| + 
	\lambda|\nodes(\edges(\usetopt))|.
	\]
	Using the fact that
	\begin{align*}
	\frac{1}{2}\!\!\sum_{\substack{e\in \edges(\usetopt)\\d\in \edges(\usetoptc)}}\!\!\esim(e,d) +  
	\frac{1}{2}\!\!\sum_{e\in \edges(\usetoptc)}\sum_{\substack{d\in E\\d\ne e}}\esim(e, d) - 
	\sum_{\{e,d\}\in\edges^2} \!\!\esim(e, d) =
	-\!\!\sum_{\{e,d\}\in \edges(\usetopt)^2}\!\! \esim(e, d),
	\end{align*}
	and that
	\; $-c|\edges(\usetoptc)| + c|\edges| = c|\edges(\usetopt)|$ \;
	we can show that
	\begin{align*}
	&-\optmccost(c,\lambda) +\sum_{\{e,d\}\in\edges^2} \!\!\esim(e, d)- c|\edges| =\\
	& =\sum_{\{e,d\}\in \edges(\usetopt)^2}\!\! \esim(e, d) - \lambda|\nodes(\edges(\usetopt))| - 
	c|\edges(\usetopt)| 
	=  Q(\edges(\usetopt), c,\lambda).
	\end{align*}
	Since $\sum_{\{e,d\}\in\edges^2}\esim(e, d)$ and $c|\edges|$ are constants, 
	searching for the minimum $\optmccost(c,\lambda)$ is equivalent to maximizing $Q$. 
	Thus, a solution minimum cut $\usetopt$ to \mincut on \flowgr 
	provides a solution edge set $\edges(\usetopt)$ for {\Qproblem} on $\graph$.
\end{proof}

Proof of Proposition~5.
\begin{proof}
	First, we prove the monotonicity of the optimal solution value
	of \LRID.
	Let us
	consider values $\lambda_1$ and $\lambda_2$ such that $0\le\lambda_1<\lambda_2$, with 
	corresponding optimal solutions $\eset_1$ and $\eset_2$. Write $\sml_1$ and $\dns_1$ for $\sml(\eset_1)$ and $\dns(\eset_1)$, $\sml_2$ and $\dns_2$ for $\sml(\eset_2)$ and $\dns(\eset_2)$.
	Suppose that the optimal solution to \LRID is increasing for $\lambda_1$ and $\lambda_2$, i.e.,
	$\sml_1 - \lambda_1/\dns_1<\sml_2 - \lambda_2/\dns_2$. Then $\sml_2 - \lambda_2/\dns_2 < \sml_2 - \lambda_1/\dns_2$ and thus $\eset_1$ is not optimal for $\lambda_1$, and a contradiction is reached.	
	
	Next, we show the monotonicity of the optimal density. Let $0\le\lambda_1<\lambda_2$. By optimality, $\sml_1-\lambda_1/\dns_1\geq \sml_2-\lambda_1/\dns_2$ and $\sml_2-\lambda_2/\dns_2\geq \sml_1-\lambda_2/\dns_1$. Thus, $\lambda_2(1/\dns_2 -1/\dns_1)\leq\lambda_1(1/\dns_2 -1/\dns_1)$ and $(\lambda_2-\lambda_1)(\dns_1-\dns_2)\leq 0$. It follows that $\dns_1\leq \dns_2$ and this concludes the proof for optimal density.
	
	Last, we prove the optimal subgraph similarity. Let $0\le\lambda_1<\lambda_2$. By monotonicity of the cost function, $\sml_1-\lambda_1/\dns_1\geq \sml_2-\lambda_1/\dns_2$. Since the density is non-decreasing, $-\lambda_1/\dns_1\leq -\lambda_2/\dns_2$, and thus, the first inequality hold if $\sml_1\geq \sml_2$ and the optimal subgraph similarity is non-increasing.
\end{proof}

Proof of Proposition~6.
\begin{proof}
	
	Let $\sml^{min}$ and $\neginvd^{min}$ denote the smallest possible values of subgraph similarity and the inverse density for a graph, respectively, and $\sml^{max}$ and $\neginvd^{max}$ be their respective maximum values. 
	Denote the granularity of $\lambda$ as
	$\delta_{\lambda}=\min{|\lambda_1 - \lambda_2|}$, where $\lambda_1$ and $\lambda_2$ lead to solution edge sets $X_1$ and $X_2$ for $\LRID$, so that $\sml(X_1) \ne \sml(X_2)$ and $\neginvd(X_1) \ne \neginvd(X_2)$.
	Finally, we define the granularity and the range of possible similarity and inverse density values:
	
	$\delta_{\sml}=\min{|\sml(X_1) - \sml(X_2)|}$ where $X_1$ and $X_2$ are solutions for $\LRID$ with some $\lambda_1$ and $\lambda_2$ so that $\sml(X_1) \ne \sml(X_2)$.
	
	$\delta_{\neginvd}=\min{|\neginvd(X_1) - \neginvd(X_2)|}$ where $X_1$ and $X_2$ are solutions for $\LRID$ with some $\lambda_1$ and $\lambda_2$ so that $\neginvd(X_1) \ne \neginvd(X_2)$.
	
	$\Delta_{\sml}=\max{|\sml(X_1) - \sml(X_2)|}$ where $X_1$ and $X_2$ are solutions for $\LRID$ with some $\lambda_1$ and $\lambda_2$.
	
	$\Delta_{\neginvd}=\max{|\neginvd(X_1) - \neginvd(X_2)|}$ where $X_1$ and $X_2$ are solutions for $\LRID$ with some $\lambda_1$ and $\lambda_2$.
	
	Let us estimate these values. It is easy to see that
	$\delta_{\sml}\geq \sml_{min}/|E|$ and $\delta_{\neginvd}\geq 1/|E|$.
	
	The lower bound for the subgraph similarity is defined as $0$ for one-edge graphs and the upper bound occurs when all the edges in the graph have the highest pairwise similarity $s_{max}$, i.e., $\Delta_{\sml}\leq s_{max}(|E|-1)|E|/2|E| = s_{max}(|E|-1)/2$.
	Similarly, the lower bound for the negative inverse density is $-2$ when the graph is a collection of disjoint edges, and the upper bound occurs when the graph is a clique, i.e., $\Delta_{\neginvd}\leq -2/(|V|-1) + 2 \leq 2$.
	
	{\it (i)Lower bound:}
	Since the optimum subgraph similarity is a non-increasing function of $\lambda$, $\lambda_{min}$ is a value such that the solution has the maximum possible similarity regardless of the value of density on the solution edge set. If we compare the solution with $\sml^{max}$ and $\neginvd^{min}$ to another solution with the best possible values of subgraph similarity and density, the following inequality must hold for any $\lambda_{min}\leq \delta_{\sml}/\Delta_{\neginvd}$: 
	
	$\sml^{max} + \lambda_{min} \neginvd^{min}\geq \sml^{max} - \delta_{\sml} + \lambda_{min}\neginvd^{max}$. 
	
	Since $\delta_{\sml}/\Delta_{\neginvd}\geq s_{min}/2|E|$, $\lambda_{min} = s_{min}/2|E|$ is a lower bound for $\lambda$.
	
	{\it (ii)Upper bound:}
	Similarly to the lower bound, the following inequality must hold for any $\lambda_{max}\geq \Delta_{\sml}/\delta_{\neginvd}$:
	
	$\sml^{min} + \lambda_{max} \neginvd^{max}\geq \sml^{max} + \lambda_{max}(\neginvd^{max}-\delta_{\neginvd})$.
	
	Since $\Delta_{\sml}/\delta_{\neginvd}\leq s_{max}|E|^2/2$, $\lambda_{max} = s_{max}|E|^2/2$ is an upper bound for $\lambda$.
	
	{\it (iii)Granularity:}
	Let $\lambda_1$ and $\lambda_2 = \lambda_1 + \delta_{\lambda}$ with $\delta_{\lambda} > 0$. The corresponding values of subgraph similarity and density of optimal solutions for these $\lambda$ values are $\sml_2 \leq \sml_1$ and $\neginvd_2 \geq \neginvd_1$, due to monotonicity.
	
	Due to optimality, it must hold that:
	$\sml_2 + (\lambda_1 + \delta_{\lambda}) \neginvd_2 \geq \sml_1 + (\lambda_1 + \delta_{\lambda}) \neginvd_1$,\\
	and by applying $\sml_2\leq\sml_1$, we get
	$\sml_2 + (\lambda_1 + \delta_{\lambda})\neginvd_2 \geq \sml_2 + (\lambda_1 + \delta_{\lambda})\neginvd_1$.
	
	Due to optimality it must also hold that:
	$\sml_1 + \lambda_1 \neginvd_1 \geq \sml_2 + \lambda_1 \neginvd_2$.
	
	Thus, $\sml_2 + (\lambda_1 + \delta_{\lambda}) \neginvd_2\leq \sml_1 + \lambda_1\neginvd_1 + \delta_{\lambda}\neginvd_2$.
	
	As a result, $\sml_2 + \lambda_1\neginvd_1 + \delta_{\lambda}\neginvd_1\leq \sml_1 + \lambda_1\neginvd_1 + \delta_{\lambda}\neginvd_2$ and $\delta_{\lambda}\geq(\sml_1-\sml_2)/(\neginvd_2-\neginvd_1)\geq \delta_{\sml}/\Delta_{\neginvd}$.
	
	Thus, $\delta_{\lambda}=s_{min}/2|E|$ is a lower bound for $\lambda$ granularity.
	
\end{proof}

\end{document}